\documentclass{JHEP3}
\usepackage{amsmath}
\input epsf
\usepackage{epsfig}
\usepackage{amssymb}
\usepackage{graphics}
\usepackage[active]{srcltx}

\usepackage{shuffle}

\newcommand{\bea}{\begin{eqnarray}}
\newcommand{\eea}{\end{eqnarray}}
\newcommand{\bean}{\begin{eqnarray*}}
\newcommand{\eean}{\end{eqnarray*}}
\newcommand{\nn}{\nonumber \\}

\def\O #1{\overline{#1}}

\def\W #1{\widetilde{#1}}

\def\braket#1{\left\langle #1 \right\rangle}

\def\eref#1{(\ref{#1})}

\def\d{{\rm d}}

\def\a{{\alpha}}

\def\b{{\beta}}

\def\d{\partial}

\def\eps{\epsilon}

\def\vev{\braket}

\def\Spaa{\vev}

\def\Label#1{\label{#1}%
  \smash{\hbox to0pt{\raise1ex\hbox{\tiny[#1]}\hss}}}

\title{Expansion of Einstein-Yang-Mills Theory by Differential Operators}
\author{ Bo Feng$^{ab}$, Xiaodi Li$^{a}$, Kang Zhou$^{c}$
\footnote{Emails:  fengbo@zju.edu.cn, lixiaodi@zju.edu.cn, zhoukang@yzu.edu.cn. The corresponding author is
Kang Zhou.} \\
{$^a$\small Zhejiang Institute of Modern Physics, Zhejiang University, Hangzhou, 310027, P. R. China \\
$^b$ Center of Mathematical Science, Zhejiang University, Hangzhou, 310027, P. R. China\\
$^c$ Center for Gravitation and Cosmology, College of Physical Science and Technology, Yangzhou University,
Yangzhou, 225009, P.R. China}}

\date{\today}
\abstract{ The factorization form of  the integrands in the Cachazo-He-Yuan (CHY)
formalism makes
the generalized Kawai-Lewellen-Tye (KLT)  relations manifest, thus amplitudes of one
theory can be expanded in terms of the amplitudes of another theory. Although this claim
seems a rather
natural consequence
of the above structure, finding the exact expansion coefficients to express an amplitude
in terms of another amplitudes is, nonetheless, a nontrivial task despite many efforts
 devoted to it in the literature. In this paper, we propose a new strategy based in using
the differential operators introduced by Cheung, Shen and Wen, and taking advantage of the
fact these operators already relate the amplitudes of different theories. Using this new
method, expansion coefficients can be found effectively.

~~~Although the method should be general, to demonstrate the idea, we focus on the
expansion of  single trace
Einstein-Yang-Mills (sEYM) amplitudes  in the Kleiss-Kuijf (KK)-basis and
Bern-Carrasco-Johansson (BCJ)-basis of Yang-Mills theory.
Using the new method, the general recursive expansion to the
 KK-basis has been reproduced. The expansion to the BCJ-basis is
a more difficult problem. Using the new method, we have worked out the details for sEYM
 with one, two and three gravitons.
As a by-product, profound relations among two kinds of expansion coefficients,
i.e., the expansion of sEYM amplitudes to the BCJ basis of YM theory and the expansion
of any color ordered Yang-Mills amplitudes to its BCJ-basis, have been observed.

}


\begin{document}

\section{Introduction}

Witten's famous work \cite{Witten:2003nn} has opened a door to study
scattering amplitudes from angles differed from the standard Feynman
diagrams. One of such  new angles is the Cachazo-He-Yuan (CHY)
formalism \cite{Cachazo:2013gna,Cachazo:2013hca, Cachazo:2013iea,
Cachazo:2014nsa,Cachazo:2014xea}. The CHY formalism has made many
beautiful properties manifest, such as the gauge invariance, the
soft behavior
\cite{Weinberg:1965nx,Cachazo:2014fwa,Casali:2014xpa,Schwab:2014xua,Afkhami-Jeddi:2014fia,
Zlotnikov:2014sva, Kalousios:2014uva, Cachazo:2015ksa} and the
double copy structure \cite{Bern:2008qj, Bern:2010ue}, etc. Among
those important results, two of them are related to our current
work. The first one is that for all theories known now in the CHY
formalism, the weight-4 CHY-integrands can be factorized as the
product of two weight-2 ingredients, formally written as
\bea \cal{I}=\cal{I}_L\times \cal{I}_R~.~~~\label{gen-1}\eea
For each weight-2 ingredient, when multiplying by another particular
weight-2 ingredient, i.e., the color ordered Parke-Taylor factor
\bea {\rm PT}(\alpha):={1\over (z_{\alpha_1}-z_{\alpha_2})\cdots
(z_{\alpha_{n-1}}-z_{\alpha_n})(z_{\alpha_n}-z_{\alpha_1})}~,~~~\eea
they define corresponding color ordered amplitudes
$A_L(\alpha),A_R(\beta)$. Thus \eref{gen-1} naturally gives the well
known  generalized Kawai-Lewellen-Tye (KLT) \cite{Kawai:1985xq}
relations (or the double copy form)\footnote{See section $5$ of
\cite{Bern:2019prr} for a comprehensive up-to-date list of the
theories connected by double-copy relations.},
\bea A=\sum_{\sigma,\widetilde{\sigma}\in
S_{n-3}}A_L(n-1,n,\sigma,1){\cal{S}}[\sigma|\widetilde{\sigma}] A_R
( 1,\widetilde{\sigma}, n-1, n)~,~~~\label{gen-3}\eea
As pointed out in \cite{Fu:2017uzt}, the double copy form implies
immediately that the original amplitude $A$ can be expanded in terms
of  either the amplitudes $A_L$ or the amplitudes $A_R$. For
example, with a fixed $\widetilde{\sigma}$ ordering, we sum $\sigma$
over all permutations of  $S_{n-3}$  and denote the result as
\bea {\cal{C}}(\widetilde{\sigma}):=\sum_{\sigma\in
S_{n-3}}A_L(n-1,n,\sigma,1){\cal{S}}[\sigma|\widetilde{\sigma}]~,~~\label{gen-4}\eea
then the original amplitude can be rewritten as
\bea A=\sum_{\widetilde{\sigma}\in
S_{n-3}}{\cal{C}}(\widetilde{\sigma})A_R(1,\widetilde{\sigma},n-1,n)~,~~~\label{gen-5}\eea
where $\cal{C}(\widetilde{\sigma})$'s serve as the expansion
coefficients. Above short deduction seems to indicate  that
expanding amplitudes of one  theory in terms of  amplitudes of
another theory is a trivial consequence of the double copy form
\eref{gen-3}. However, such a triviality is just an illusion. When
trying to get the exact expansion coefficients ${\cal C}$, one will
find it is very difficult task. If we use the expression
\eref{gen-4} directly, one needs to do the sum over $(n-3)!$ terms.
Since analytical expressions for both parts $A_L(n-1,n,\sigma,1)$
and ${\cal{S}}[\sigma|\widetilde{\sigma}]$ are very complicated, it
is no wonder why in the practice no one uses this method except for
some special cases\footnote{In \cite{Feng:2010hd}, since all
amplitudes are the MHV amplitudes, the analytical expressions are
very simple, thus one can carry out the $(n-3)!$ sum recursively.}.
Many efforts have been devoted to avoid the mentioned technical
difficulties. In \cite{Stieberger:2016lng}, using the string theory,
the single trace Einstein-Yang-Mills (sEYM) amplitudes with just one
graviton have been expanded in terms of the color ordered Yang-Mills
(YM) amplitudes. In \cite{Schlotterer:2016cxa}, using the heterotic
string theory, evidence of the expansion of general EYM amplitudes
in terms of  YM amplitudes has also been presented, but explicit
expansions are done  only for some special cases, such as up to
three gravitons, or up to two color traces. In
\cite{Chiodaroli:2017ngp}, using the fact that expansion
coefficients ${\cal C}$'s in the Kleiss-Kuijf (KK)-basis can be
identified with the BCJ numerators of DelDuca-Dixon-Maltoni (DDM)
chain \cite{DelDuca:1999rs},
 an assumption of algebraic
expressions has been made with the proper hints from  Feynman
diagrams of DDM chain. After that,  the gauge invariance and other
properties have been imposed on the assumption, thus  expansion
coefficients up to five gravitons have been finally determined. In
\cite{Nandan:2016pya,delaCruz:2016gnm}, using the CHY-integrand and
the techniques developed and presented in \cite{Cardona:2016gon,
Bjerrum-Bohr:2016juj, Bjerrum-Bohr:2016axv, Huang:2017ydz}, such as
the cross ratio identities etc, expansion coefficients of sEYM
amplitudes up to four gravitons have been presented. In these works,
using the CHY-integrands instead of amplitudes directly is a big
simplification, since the analytical expressions of CHY-integrands
are much more easier to write down while analytical expressions of
amplitudes are very hard to imagine. However, since the cross ratio
identities have a lot of equivalent forms,  results obtained by this
way contain a lot of freedoms, thus it is very hard to observe the
general patterns from these explicit examples. A different approach
has been used in \cite{Fu:2017uzt} to attack this difficult problem:
with the proper arrangement of the known expansion coefficients for
one, two and three gravitons, gauge symmetry consideration has led
us to a conjecture having the natural iterative expansion pattern.
Thus the expansion coefficients for arbitrary number of gravitons in
the sEYM theory can be systematically worked out. With the guidance
of the conjectured pattern in \cite{Fu:2017uzt}, rigorous proof of
the conjecture has been given in \cite{Teng:2017tbo} by carefully
dealing with the CHY-integrands of sEYM theory, especially the
Pfaffian structure. Based on these results, expansions for
multi-trace EYM amplitudes as well as nonlinear sigma model (NLSM)
amplitudes have been given in \cite{Du:2017kpo, Du:2017gnh}, again
using the gauge symmetry considerations plus the manipulations of
the CHY-integrands.

The second important result coming from the CHY formalism  is that
different theories have some deep connections, where KLT-like
relations are just a reflection of them. In
\cite{Cachazo:2014xea}, using three operations, i.e., dimension
reduction, squeezing and generalized dimension reduction, various
theories can be deduced from the Einstein gravity theory. Recently,
the web of connections of different theories given in
\cite{Cachazo:2014xea} has also been reproduced in
\cite{Cheung:2017ems} from a different point of view. It is found
that one can define different kinds of  differential
operators, which  act on variables formed by
 Lorentz contractions among external momenta and
polarization vectors. When the proper combinations of differential
operators act on the on-shell amplitudes of one theory, amplitudes
of another theory will be reached. Since these two different
strategies lead to the same web picture, there must be a connection
between them. Such a connection  has been explained in
\cite{Zhou:2018wvn, Bollmann:2018edb}, by showing that when acting
proper combinations of differential operators on the CHY-integrands
of Einstein gravity theory, the corresponding CHY-integrands of
other theories have been produced.

Above two results lead naturally to the following question: since
using differential operators one can  connect different theories,
could we use them to establish expansion relations as given in
\eref{gen-5}? In this paper, we will show that indeed one can do
that. The basic idea is following: by applying differential operators
at the both sides of \eref{gen-5}, one gets differential equations
for expansion coefficients. Solving them, one obtains the wanted
coefficients. In fact, we achieve more than just finding
coefficients. We will show  that by proper choices of building
blocks, we can {\it derive} the expansion relation \eref{gen-5} and
find coefficients at the same time. More explicitly, when we act
differential operators on amplitudes of sEYM theory at the left hand
side, we land naturally to YM amplitudes at the right hand side. The
emergence of relations \eref{gen-5} comes from the web connections
between different theories through differential operators as
explained in \cite{Cheung:2017ems}.

Although our new method is general, in this paper, we will mostly
focus on the expansion of sEYM amplitudes. The reason is that as
shown in \cite{Cachazo:2014xea, Cheung:2017ems}, by acting
differential operators on amplitudes of the sEYM theory\footnote{
The CHY-integrand of gravity theory is defined by the Pfaffian of a
matrix, which, when expanding, becomes the sum of integrands of sEYM
theory with different gluon structures. }, one can get amplitudes of
all other theories in the web (see also
\cite{Zhou:2018wvn,Bollmann:2018edb}). Thus if we can solve the
expansion of sEYM amplitudes using the new method, such a method
should be applicable to all other theories. Furthermore, when we
talk about the expansion, we need to be careful about two different
kinds of expansions. The factorization form of CHY-integrands leads
naturally the expansion form \eref{gen-5}, where the independent
basis is $(n-3)!$ color ordered amplitudes with three legs fixed.
This basis is called Bern-Carrasco-Johansson (BCJ) basis. As we will
show, with different choices of building blocks, our new method will
give either the expansion to the BCJ-basis, or the expansion to the
Kleiss-Kuijf (KK)-basis, which has $(n-2)!$ color ordered amplitudes
with only two legs fixed (for example, the first and the last legs).
Results reviewed in above references (such as
\cite{Stieberger:2016lng,Schlotterer:2016cxa, Chiodaroli:2017ngp,
Nandan:2016pya, delaCruz:2016gnm, Fu:2017uzt,Teng:2017tbo,
Du:2017kpo, Du:2017gnh} ) are mostly expanded in the KK-basis.
Although using the Bern-Carrasco-Johansson (BCJ) relations among
KK-basis
\cite{Bern:2008qj,BjerrumBohr:2009rd,Stieberger:2009hq,Feng:2010my,
Chen:2011jxa} one can connect these two kinds of expansions, such a
connection is not so easy to get as we will show in this paper.
Another motivation of our current paper is to expand sEYM amplitudes
directly in the BCJ-basis without relying on expansion in the
KK-basis.

Having above clarifications, the tasks of the paper are following
two: (1) First  we re-derive  expansions in the KK-basis using only
differential operators; (2) Secondly, we derive expansion in the
BCJ-basis directly. Our derivations will deepen our understanding of
the intricate connections between on-shell amplitudes of different
theories. For example, we will see how the expansion coefficients of
sEYM amplitudes in \eref{gen-5} and the expansion coefficients (see
\eref{BCJ-exp-2}) of arbitrary color ordered YM amplitudes in its
BCJ basis are related to each other.

The plan of the paper is following. In section $\S2$, we present
some backgrounds needed for later discussions. They include the
expansion of arbitrary color ordered YM amplitude in its BCJ-basis,
the recursive expansion of sEYM amplitudes in the KK-basis of YM
amplitudes, and the differential operators used heavily in this
paper. In section $\S3$, we re-derive the recursive expansion of
sEYM amplitudes in the  KK-basis of YM amplitudes using the proposed
new method, i.e., using differential operators. As emphasized in
previous paragraphes, results in this section show that differential
operator method is an independent systematical new method to
determine the expansion. In section $\S4$, we apply the same method
to Einstein gravity theory and Born-Infeld theory. In section $\S5$,
starting from the expansion of sEYM amplitudes in the KK-basis of YM
amplitudes, with careful algebraic manipulations, we translate the
KK-basis  to the BCJ-basis, thus  get the expansion in the
BCJ-basis. In section $\S6$, following the similar idea of $\S3$,
but with a different choice of building blocks, we derive the
expansion of sEYM amplitudes in the BCJ-basis with one, two and
three gravitons by applying differential operators directly.  In
section $\S7$, some discussions have been given. Some technical
calculations are collected in the Appendix. In the appendix A, we
show terms with index cycle structure will drop out from the
expansion. In the appendix B, we discuss the structure of functions
when imposing manifestly gauge invariant conditions.  These
structures
 provide the building blocks used in the expansion in the BCJ-basis. In the
appendix C, some omitted calculation details in $\S6$ have been
given here.

\section{Some backgrounds}

In this section, we will review some known results. They  will serve
as the backgrounds of our discussions in the whole paper. This
section contains three parts. In the first part, we review the
expansion of arbitrary color ordering YM amplitudes to its
BCJ-basis. Especially we have rewritten the expansion coefficients
in a compact form comparing to these given in \cite{Bern:2008qj,
Chen:2011jxa}. In the second part, we  review the recursive
expansion of sEYM amplitudes in the KK-basis of YM amplitudes given
in \cite{Fu:2017uzt, Teng:2017tbo, Du:2017gnh}. The result will be
reproduced by using differential operators proposed in this paper.
In the third part, we  review differential operators introduced in
\cite{Cheung:2017ems}. They are our main tools. As a simple
application of these tools, three simple but nontrivial new
relations for EYM amplitudes have been derived  in \eref{color-rev},
\eref{pho-dec-gen} and \eref{gener-KK-rela}.

\subsection{The expansion of color ordered YM amplitudes to its BCJ basis}

It is  well known that all color ordered Yang-Mills amplitudes are
not independent of each other. There are two kinds of relations
among them. The first kind is the Kleiss-Kuijf (KK) relation
\cite{Kleiss:1988ne}, which is given by
\bea A(1,\a, n,\b)=\sum_{\shuffle} (-)^{n_\b} A(1, \a \shuffle \b^T,
n)~~~~\label{KK-relation} \eea
where $\a,\b$ are two ordered subsets of external gluons and $n_\b$
is the number of elements in the subset $\b$, while $\b^T$ means the
reversing of the ordered subset $\b$. The sum is over all possible
shuffles of two ordered subsets. All shuffles can be obtained
recursively as
\bea & & \a\shuffle \emptyset =\a,~~~~\emptyset \shuffle \b=\b,\nn
& &  \{\a_1,...,\a_m\}\shuffle\{\b_1,...,\b_k\} =\{\a_1,
\{\a_2,...,\a_m\}\shuffle \b\}+\{\b_1,
\a\shuffle\{\b_2,...,\b_k\}\}~.~~\label{shuffle}\eea
Using the KK-relations \eref{KK-relation}, one can expand arbitrary
color ordered Yang-Mills amplitudes in the basis, where two legs are
fixed at the two ends (hence we will call such a basis the "KK
basis").

The second kind is the  Bern-Carrasco-Johansson (BCJ) relations
\cite{Bern:2008qj}. To present these relations compactly, let us
define some notations first. Given two ordered sets
$\Xi=\{\xi_1,\xi_2,...,\xi_n\}$ and  $\b=\{\b_1,...,\b_r\}$ where
the set $\b$ is the subset of $\Xi$, for a given element $p\in \Xi$
with its position $K$ in $\Xi$ (i.e., $\xi_K=p$),  we define:
\begin{itemize}

\item (1) {\bf The momentum $X_p$:} It is given by
\bea X_p=\sum_{i=1}^{K-1} k_{\xi_i}~~~~\label{Xp-def}\eea
, i.e., the sum of momenta of these elements, which are on the
left hand side of $p$ in the set $\Xi$.

\item (2) {\bf The momentum $Y_p$:} It is given by
\bea Y_p=\sum_{i=1, \xi_i\not\in \b}^{K-1}
k_{\xi_i}~~~~\label{Yp-def}\eea
, i.e., the sum of momenta of these elements, which are on the
left hand side of $p$ in the set $\Xi$, but do not belong to the
set $\b$.

\item (3) {\bf The momenta $W^{(L,L)}_{p}, W^{(L,R)}_p, W^{(R,L)}_p,W^{(R,R)}_p$:}
To define them, we require $p\in \b$,  thus $p$ split the
set $\b$ into two subsets $\b_{p}^{L}$ and $\b_{p}^{R}$, i.e., the
collections of elements on the left hand side of $p$ and on the
right hand side of $p$ respectively. Now we define
\bea W^{(L,L)}_{p} & = & \sum_{i=1, \xi\not\in \b_{p}^{R}}^{K-1}
k_{\xi_i}~~~~\label{WLL-def}\eea
, i.e., $W^{(L,L)}_{p}$ is the sum of momenta of these elements,
which are  on the left hand side of $p$ in the set $\Xi$, but do
not belong to  the subset $\b_{p}^{R}$. In other words,
$W^{(L,L)}_{p}$ is the sum of momenta of these elements at the
left hand side of $p$ in the set $\Xi$ satisfying either the
condition that they do not belong to the set $\b$ or the
condition that they belong to the subset $\b_{p}^{L}$. This is
 the reason we call it $W^{(L,L)}_{p}$, where the first $L$ is
for these elements belonging to $\Xi\backslash\b$, and the
second $L$ is for these elements belonging to $\b_{p}^{L}$.
Similarly we define
\bea W^{(L,R)}_{p} & = & \sum_{i=1, \xi\not\in \b_{p}^{L}}^{K-1}
k_{\xi_i}~,~~~\label{WLR-def}\\
 W^{(R,L)}_{p} & = & \sum_{i=K+1, \xi\not\in \b_{p}^{R}}^{n}
k_{\xi_i}~,~~~\label{WRL-def}\\
 W^{(R,R)}_{p} & = & \sum_{i=K+1, \xi\not\in \b_{p}^{L}}^{n}
k_{\xi_i}~.~~~\label{WRR-def}\eea
\end{itemize}

After above preparations, let us begin with the (generalized)
fundamental BCJ relation \cite{Chen:2011jxa}. If we divide the set
$\{2,...,n-1\}$ into two ordered subsets $\a=\{a_1,...,\a_m\}$ and
$\b=\{\b_1,...,\b_t\}$ (so $m+t=n-2$), the relations are given by
\bea \sum_{\shuffle} (\sum_{i=1}^t k_{\b_i}\cdot X_{\b_i}  ) A(1,
\a\shuffle \b,n)=0~.~~\label{generalized-BCJ} \eea
Using these relations, one can easily solve
\bea A_{n+1}^{\rm YM}(1,p,\{2,\ldots,n-1\},n)={-(k_p\cdot X_p)\over
(k_p\cdot k_1)}~ A_{n+1}^{\rm
YM}(1,2,\{3,\ldots,n-1\}\shuffle\{p\},n)~~~\label{Fund-BCJ-deform-1}\eea
for $\Xi=\{1,\{2,...,n-1\}\shuffle\{p\},n\}$ and $\b=\{p\}$, and
\bea & & A^{\rm YM}_{n+2}(1,p,q, \{2,...,n-1\},n) \nn
& = &  {(k_p\cdot k_1+k_q\cdot (Y_q+k_p))(k_p\cdot (Y_p+k_q))\over
{\cal K}_{1pq} {\cal K}_{1p}}  ~A_{n+1}^{\rm
YM}(1,2,\{3,\ldots,n-1\}\shuffle\{q,p\},n)\nn
& & +{(k_p\cdot (Y_p- k_1))(k_q\cdot (Y_q+k_p))\over {\cal K}_{1pq}
{\cal K}_{1p}}~A^{\rm YM}_{n+2}(1,2,
\{3,...,n-1\}\shuffle\{p,q\},n)~~~\label{G-BCJ-b=2-3}\eea
for  $\Xi=\{1,\{2,...,n-1\}\shuffle\{p,q\},n\}$ and $\b=\{p,q\}$.
For simplicity, we have omitted the $\sum_\shuffle$ on the right
hand side when its meaning is obvious and defined
\bea {\cal K}_{\a}=\sum_{i< j;i,j\in \a} k_i\cdot
k_j~~~\label{K-contract}\eea
for any set $\a$.

As shown in \cite{Chen:2011jxa}, by recursively using the relations
\eref{generalized-BCJ}, one can expand arbitrary color ordered
Yang-Mills amplitudes in the BCJ-basis, i.e., those color ordered
Yang-Mills amplitudes with three legs fixed at some particular
positions (for example, at the first, second and the last
positions), as
\bea A_n (1,\b_1,...,\b_r,2,\a_1,...,\a_{n-r-3},n) & = &
\sum_{\{\xi\} \in \{\b\} \shuffle_{\cal P} \{\a\}} {\cal
C}_{\{\a\},\{\b\};\{\xi\}}
A_n(1,2,\{\xi\},n)~.~~\label{BCJ-exp-1}\eea
The expansion coefficients (we will call them the "BCJ
coefficients") are first conjectured in \cite{Bern:2008qj} and then
proved in \cite{Chen:2011jxa}. Using our new defined notations, they
are given by
\bea {\cal C}_{\{\a\},\{\b\};\{\xi\}} & = & \prod_{k=1}^r {{\cal
F}_{\b_k}(\{\a\},\{\b\};\{\xi\}) \over {\cal
K}_{1\b_1...\b_k}}~,~~\label{BCJ-exp-2} \eea
where
\bea {\cal F}_{\b_k}(\{\a\},\{\b\};\{\xi\})& = &
\theta(\xi_{\b_{k}}-\xi_{k-1})\left\{k_{\b_k}\cdot
W^{(R,R)}_{\b_k}+\theta(\xi_{\b_{k+1}}-\xi_{\b_k}) {\cal
K}_{1\b_1...\b_k} \right\}\nn & & +
\theta(\xi_{\b_{k-1}}-\xi_{k})\left\{-k_{\b_k}\cdot
(W^{(L,R)}_{\b_k}-k_1)-\theta(\xi_{\b_{k}}-\xi_{\b_{k+1}}) {\cal
K}_{1\b_1...\b_k}\right\} ~~~\label{BCJ-exp-3}\eea
with $W$'s  defined in \eref{WRR-def} and \eref{WLR-def} and
$\theta(x)=1$ when $x>0$ and $\theta(x)=0$, otherwise\footnote{It is
worth to emphasize that in the definition of \eref{BCJ-exp-3}, there
are three orderings $\{\a\},\{\b\}$ and $\{\xi\}$. To define $W$'s we
need both of them, i.e., $\b=\{\b_1,...,\b_r\}$ and
$\Xi=\{1,2,\xi,n\}$.
 To define $\theta(x)$ only $\{\xi\}$ is needed.} . Results \eref{BCJ-exp-1},
\eref{BCJ-exp-2} and \eref{BCJ-exp-3} are a little bit complicated
and we give some explanations:
\begin{itemize}

\item In \eref{BCJ-exp-1}, we have defined a new notation $\shuffle_{\cal P}$,
the "partial ordered shuffle", which is the sum over all
    permutations that maintain only the relative order of the
    subset $\{\a\}$ (i.e., the ordering of elements of the
    subset $\{\b\}$ in $\{\xi\}$ can be arbitrary).

\item   The expression \eref{BCJ-exp-2} tells us that for each
element $\b_k\in \b$, there is an associate kinematic factor.
The BCJ coefficients are the product of these factors.

\item For the set $\{\xi\}=\{\xi_1,...,\xi_{n-3}\}$, we use the $\xi_{\b_k}$
to denote the position of the element $\b_{k}$ in $\xi$. Also,
we define the auxiliary elements $\b_0,\b_{r+1}$ with positions
$\xi_{\b_0}\equiv +\infty$ and $\xi_{\b_{r+1}}=0$, i.e., $\b_0$
is always  treated at the rightmost position and $\b_{r+1}$ is
always treated at the leftmost position. With this
understanding, the $\theta(x)$'s are well defined for all $\b_k$
with $k=1,2,...,r$.

\end{itemize}

Having explained the meanings of \eref{BCJ-exp-1}, \eref{BCJ-exp-2}
and \eref{BCJ-exp-3}, let us apply them to one example:
\bea & &
A^{\rm YM}_{n+2}(1,p,q,r, \{2,...,n-1\},n)\nn
& =& \sum_{\rho\in S_3} {\cal C}[\{p,q,r\}; \rho\{p,q,r\}] A^{\rm
YM}_{n+3}(1,
2,\{3,...,n-1\}\shuffle\{\rho\{p,q,r\}\},n)~~~~\label{3legs-BCJ}\eea
where the sum is over all permutations $\rho$ of three legs
$\{p,q,r\}$. Coefficients in \eref{3legs-BCJ} involve two ordered
lists: the first one $\{p,q,r\}$ is the remembering of the original
ordering on the left hand side, while the second one $\rho\{p,q,r\}$
is  the ordering in the expansion on the right hand side. For these
six coefficients with orderings  $\rho\{p,q,r\}$, we have:
\begin{itemize}

\item {\bf The order $\{p,q,r\}$:}  For $p$,  $\xi_{0}=+\infty>\xi_{p}$ and $\xi_p<\xi_q$,
so we  have
\bean & & -k_{p}\cdot (W^{(L,R)}_{p}-k_1)=-k_{p}\cdot
(Y_{p}-k_1)~.\eean
For $q$, $\xi_p<\xi_q$ and $\xi_q<\xi_r$, so we have
\bean k_{q}\cdot W^{(R,R)}_{q}+ {\cal K}_{1pq}=-k_{q}\cdot X_q+
{\cal K}_{1pq}=-k_{q}\cdot (Y_q+k_p)+ {\cal K}_{1pq}\eean
where momentum conservation has been used. For $r$,
$\xi_q<\xi_r$ and $\xi_r>0$, so we have
\bean k_{r}\cdot W^{(R,R)}_{r}=-k_r\cdot X_r~.\eean
This ordering is a special case since it is the same ordering
comparing to the original ordering. This case is easy to
generalize to a set with arbitrary length $m$ and we have
following results. For $\b_r$ is not the first or the last
elements of $\b$, the corresponding kinematic factor is
\bea & & k_{\b_r}\cdot W^{(R,R)}_{\b_r}+ {\cal
K}_{1\b_1...\b_r}=  -(k_{\b_r}\cdot X_{\b_r}- {\cal
K}_{1\b_1...\b_r})~.~~~~~\label{same-order-1}\eea
For the first element, the corresponding kinematic factor  is
\bea & & -k_{\b_1}\cdot (W^{(L,R)}_{\b_1}-k_1)=-k_{\b_1}\cdot
(Y_{\b_1}-k_1)~.~~~~~\label{same-order-2}\eea
For the last element, the corresponding kinematic factor  is
\bea & & k_{\b_m}\cdot W^{(R,R)}_{\b_m}=k_{\b_m}\cdot \O
Y_{\b_m}= -k_{\b_m}\cdot X_{\b_m}~.~~~~~\label{same-order-3}\eea

\item {\bf The order $\{p,r,q\}$:}
For $p$,  $\xi_{0}=+\infty>\xi_{p}$ and $\xi_p<\xi_q$, so we
have $-k_{p}\cdot (W^{(L,R)}_{p}-k_1)=-k_{p}\cdot (Y_{p}-k_1)$.
 For $q$,  $\xi_p<\xi_q$ and $\xi_q>\xi_r$, so
we have $ k_{q}\cdot W^{(R,R)}_{q}=k_{q}\cdot \O Y_q=-k_{q}\cdot
X_q$. For $r$, $\xi_q>\xi_r$ and $\xi_r>0$, so we have
$-k_{r}\cdot (W^{(L,R)}_{r}-k_1)-{\cal K}_{1pqr}=-k_r\cdot
(Y_r-k_1) -{\cal K}_{1pqr}$.

\item {\bf The order $\{q,p,r\}$:}
For $p$,  $\xi_{0}=+\infty>\xi_{p}$ and $\xi_p>\xi_q$, so we
have $-k_{p}\cdot (W^{(L,R)}_{p}-k_1)-{\cal
 K}_{1p}=-k_p\cdot X_p$. For $q$, $\xi_p>\xi_q$ but
$\xi_q<\xi_r$, so we  have $ -k_{q}\cdot
(W^{(L,R)}_{q}-k_1)=-k_{q}\cdot (Y_{q}-k_1)$.  For $r$,
$\xi_q<\xi_r$ and $\xi_r>0$, so we have $k_r\cdot
W^{(R,R)}_r=-k_r\cdot X_r$.

\item {\bf The order $\{q,r,p\}$:}
For $p$,  $\xi_{0}=+\infty>\xi_{p}$ and $\xi_p>\xi_q$, so we
have $-k_{p}\cdot (W^{(L,R)}_{p}-k_1)-{\cal K}_{1p}=-k_p\cdot
 X_p$. For $q$,  $\xi_p>\xi_q$ and $\xi_q<\xi_r$, so we have  $
-k_{q}\cdot (W^{(L,R)}_{q}-k_1)=-k_{q}\cdot (Y_{q}-k_1)$. For
$r$, $\xi_q<\xi_r$ and $\xi_r>0$, so we have $k_r\cdot
 W^{(R,R)}_r=-k_r\cdot (X_r+k_p)$.

\item {\bf The order $\{r,p,q\}$:}
For $p$, we have $\xi_{0}=+\infty>\xi_{p}$ and $\xi_p<\xi_q$, so
we will have $-k_{p}\cdot (W^{(L,R)}_{p}-k_1)=-k_{p}\cdot
 (X_{p}-k_1)$. For $q$, we have $\xi_p<\xi_q$ and $\xi_q>\xi_r$,
so it is $ k_{q}\cdot W^{(R,R)}_{q}= -k_q\cdot X_q$. For $r$,
$\xi_q>\xi_r$ and $\xi_r>0$, so it is $-k_r\cdot
 (W^{(L,R)}_r-k_1)-{\cal K}_{1pqr}=-k_r\cdot ( Y_r-k_1)-{\cal
 K}_{1pqr}$.

\item {\bf The order $\{r,q,p\}$:}
This case is also special, since the ordering $\{r,q,p\}$ is the
reversing of  the original one $\{p,q,r\}$. In general, for the
ordering $\b^T$,   each element $\b_r$ has the corresponding
kinematic factor
\bea -k_{\b_r}\cdot (W^{(L,R)}_{\b_r}-k_1)-{\cal
K}_{1\b_1...\b_r} =-k_{\b_r}\cdot (X_{\b_r}-k_1)-{\cal
K}_{1\b_1...\b_r}~~~~~~\label{reverse-order-1}\eea

\end{itemize}
For later convenience, we collect these six coefficients as
following:
\bea {\cal C}[\{p,q,r\};\{p,q,r\}] &= &  {  -(k_p\cdot ( Y_p-
k_1))\over {\cal K}_{1p}}\times {({\cal K}_{1pq}-k_q\cdot X_q)\over
{\cal K}_{1pq}}\times {-(k_r\cdot X_r)\over {\cal K}_{1pqr}}\nn
{\cal C}[\{p,q,r\};\{p,r,q\} ] & = &   {- (k_p\cdot ( Y_p-
k_1))\over {\cal K}_{1p}} \times { -(k_q\cdot X_q)\over {\cal
K}_{1pq}} \times{ -k_r\cdot (Y_r-k_1)-{\cal K}_{1pqr} \over {\cal
K}_{1pqr}} \nn
{\cal C}[\{p,q,r\};\{q,p,r\} ] & = &  { -k_p\cdot X_p\over {\cal
K}_{1p}} \times { -(k_q\cdot (Y_q-k_1))\over {\cal K}_{1pq}} \times{
-k_r\cdot X_r\over {\cal K}_{1pqr}} \nn
{\cal C}[\{p,q,r\};\{q,r,p\} ]& = &  { -k_p\cdot X_p\over {\cal
K}_{1p}} \times { -(k_q\cdot (Y_q-k_1))\over {\cal K}_{1pq}} \times{
-k_r\cdot (X_r+k_p)\over {\cal K}_{1pqr}} \nn
{\cal C}[\{p,q,r\};\{r,p,q\} ]& = &  { -(k_p\cdot (X_p-k_1))\over
{\cal K}_{1p}} \times {-k_q\cdot X_q \over {\cal K}_{1pq}} \times{
-k_r\cdot (Y_r-k_1)-{\cal K}_{1pqr}\over {\cal K}_{1pqr}} \nn
{\cal C}[\{p,q,r\};\{r,q,p\} ]& = &  { -k_p\cdot X_p\over {\cal
K}_{1p}} \times { -k_q\cdot (X_q-k_1)-{\cal K}_{1pq} \over {\cal
K}_{1pq}} \times{ -k_r\cdot (Y_r-k_1)-{\cal K}_{1pqr}\over {\cal
K}_{1pqr}} ~~~~~~~\label{BCJ-pqr-collect}\eea

Before ending this subsection, let us make an observation. For color
ordered YM amplitudes, we have two basis: one is the KK-basis with
$(n-2)!$ elements and another one is the BCJ-basis with $(n-3)!$
elements. As it has been seen in many examples, when we expand in
the  KK-basis, the expansion coefficients can be properly chosen
 to be polynomial functions, while when we expand in the BCJ-basis, the
 expansion
coefficients will be rational functions in general. It is a simple
but very useful observation. The reason for the polynomial functions
rather than rational functions is clear: we must have some poles in these
coefficients to match up pole structures  at the both sides of the
expansion.

\subsection{Expansion of sEYM amplitudes by YM amplitudes}

As explained in the introduction,  in this paper we will focus on
the expansion of sEYM amplitudes in terms of YM amplitudes. As
mentioned before, the expansion  in the KK-basis has been
conjectured first by gauge symmetry consideration\cite{Fu:2017uzt}
and then proved using the CHY integrands in \cite{Teng:2017tbo}. In
this subsection, we recall the recursive expansion of sEYM
amplitudes $A_{|\mathsf{H}|}(1,2\ldots r\,\Vert\,\mathsf{H})$ given
in ~\cite{Fu:2017uzt,Teng:2017tbo,Du:2017gnh}
\bea  A^{\rm EYM}_{r,\mathsf{H}|}(1,2\ldots
r\,\Vert\,\mathsf{H})=\sum_{\pmb{h}\,|\,\mathsf{\W h}
=\mathsf{H}\backslash h_a}\Big[C_{h_a}(\pmb{h})A^{\rm
EYM}_{r+|\mathsf{H}|-|\mathsf{\W h}|,|\mathsf{\W h}|} (1,\{2\ldots
r-1\}\shuffle\{\pmb{h},h_a\},r\,\Vert\,\mathsf{\W h})\Big]\,,
~~~~\label{eq:singletrace}\eea
where the $\mathsf{H}$ is the list of gravitons. The sum in
\eref{eq:singletrace} is over all partial ordered splitting of the
list $\mathsf{H}\backslash h_a$, i.e., $\mathsf{H}\backslash
h_a=\pmb{h}\bigcup\mathsf{\W h}$, where different orderings of the
subset $\pmb{h}$ are treated to be different splitting while the
orderings of $\mathsf{\W h}$ do not matter. Suppose
${\pmb{h}}=\{i_s,i_{s-1}\ldots i_1\}$, the coefficient
$C_{h_a}(\pmb{h})$ can be written as:
\bea C_{h_a}(\pmb{h})=\epsilon_{h_a}\cdot f_{i_1}\cdot
f_{i_{2}}\ldots f_{i_{s-1}}\cdot f_{i_s}\cdot
Y_{i_s}\,,~~~\label{C-single-exp} \eea
where
$(f_i)^{\mu\nu}=(k_i)^{\mu}(\epsilon_i)^{\nu}-(k_i)^{\nu}(\epsilon_i)^{\mu}$
is the field strength,  and when $\pmb{h}=\emptyset$,
$C_{h_a}(\emptyset)=\epsilon_{h_a}\cdot Y_{h_a}$. The definition of
$Y_q$ \eref{Yp-def} is according to the set $\Xi=\{11,\{2\ldots
r-1\}\shuffle\{\pmb{h},h_a\},r\}$ and the set $\b=\{\pmb{h},h_a\}$.

The expression \eqref{eq:singletrace} is manifestly invariant under
all permutations and gauge transformations of gravitons in the set
$\mathsf{H}\backslash h_a$. The $h_a$ is special in the expansion
and will be called as \emph{the fiducial graviton}. The full
permutation and gauge invariances for all gravitons, although not
explicitly, are guaranteed by the generalized BCJ
relations~\cite{Chen:2011jxa} (see explicit proof given in
\cite{Hou:2018bwm} ). When we use \eref{eq:singletrace} recursively,
we can expand any sEYM amplitude in the KK-basis of YM amplitudes
with \emph{polynomial coefficients}. In this paper, we will only use
differential operators to reproduce the recursive expansion formula
\eref{eq:singletrace}, thus provide a new strategy to find the
expansion.

Before ending this subsection, let us give an important remark.
Because of the generalized BCJ-relations \eref{generalized-BCJ}, even
imposing the polynomial conditions, the coefficients
$C_{h_a}(\pmb{h})$'s in \eref{C-single-exp} can not be unique, since
one can always add these zero combinations. Such a freedom can
easily be fixed by making some choices for expansion coefficients of
some elements in KK-basis.

\subsection{Differential operators}

In \cite{Cheung:2017ems}, to study connections of amplitudes of
different theories, some kinds of differential operators have been
defined. The first kind is the trace operator ${\cal T}_{ij}$
\bea {\cal
T}_{ij}\equiv\partial_{\epsilon_i\epsilon_j}~,~~~\label{trace} \eea
where $\eps_i\eps_j\equiv \eps_i\cdot \eps_j$ is the Lorentz
invariant contraction (similar understanding holds for all operators
in this paper). The second kind is the insertion operator  defined
by
\bea {\cal
T}_{ikj}\equiv\partial_{k_i\epsilon_k}-\partial_{k_j\epsilon_k}\,.~~~\label{insertion}
\eea
As pointed out in  \cite{Cheung:2017ems}, ${\cal T}_{ikj}$ itself is
not a gauge invariant operator, but when it acts on those objects
obtained after acting a trace operator, it is effectively gauge
invariant. Insertion operators will be extensively used in our
paper. When it acts on the sEYM amplitudes, the operator ${\cal
T}_{ikj}$ has the physical meaning of inserting the graviton $k$
between $i$ and $j$ when $i,j$ are nearby in a trace.  If $i,j$ are
not adjacent in a trace, for example, ${\cal T}_{ik(i+2)}$, we can
write it as
\bea {\cal T}_{ik(i+2)}={\cal T}_{ik(i+1)}+{\cal
T}_{(i+1)k(i+2)}~,~~\label{Inser-not-nearby}\eea
thus the physical meaning is clear. If $i,j$ are not in same trace,
the physical meaning is not clear. The third kind of useful
operators is the gauge invariant differential operator
\bea {\cal W}_{i}\equiv \sum_{v} (k_i
v)\d_{v\eps_i}~~~~\label{Gauge-oper}\eea
where the sum is over all Lorentz contractions with $\eps_i$. When
it is applied, $\eps_i$ is  effectively replaced by $k_i$  in any
expression. Thus if an expression is gauge invariant, the action of
${\cal W}_{i}$ will give zero result.  The fourth kind of useful
operators is longitudinal operators defined via
\bea {\cal L}_i\equiv\sum_{j\neq i}k_i
k_j\partial_{k_j\epsilon_i}\,, ~~~\label{Li-def}\eea
and the related operators
\bea {\cal L}_{ij}\equiv-k_i k_j\partial_{\epsilon_i\epsilon_j}\,.
~~~\label{Lij-def}\eea
The roles of ${\cal L}_i$ and ${\cal L}_{ij}$ are to add derivative
interactions in vertexes while getting rid of polarization vectors.
Using them, we can define the operators ${\cal L}$ and ${\cal \W L}$
as \cite{Cheung:2017ems}
\bea {\cal L}\equiv\prod_i{\cal L}_i\,,~~~~~{\cal \W L}\equiv
\sum_{\rho\in {\rm pair}} \prod_{i,j\in\rho}{\cal
L}_{ij}.~~~\label{LT} \eea
At the algebraic level, the actions of ${\cal L}$ and ${\cal \W L}$
are different. However, if we consider the combination ${\cal
L}\cdot{\cal T}_{ab}\,{\bf Pf}'\Psi$, and let subscripts of ${\cal
L}_i$'s and ${\cal L}_{ij}$'s run through all legs in
$\{1,2,\cdots,n\}\setminus\{a,b\}$, the effects of ${\cal L}$ and
${\cal \W L}$ are same and give a result which has a meaningful
explanation. We will see it later.

As the simple applications of these kinds of differential operators,
 we  derive three generalized relations for
general EYM amplitudes. For simplicity we will use the sEYM
amplitudes to demonstrate the idea. The first relation is the
generalized color ordered reversed relation of  EYM amplitudes:
\bea \boxed{~{ A}^{\rm EYM}(1,2,\cdots,(n-1),n;\{h_g\}) =(-)^n{
A}^{\rm EYM}(n,(n-1),\cdots,2,1;\{h_g\})\,}~,~~~~\label{color-rev}
\eea
which is  reduced to the familiar color ordered reversed relation
for pure YM amplitudes when $\{ h_g\}=\emptyset$. To derive this, we
use the fact that the ordering $(1,2,\cdots,(n-1),n)$ of sEYM
amplitudes can be created by following combination of  operators
\bea {\cal T}[1,2,\cdots,(n-1),n]={\cal T}_{(n-2)(n-1)n}\cdots{\cal
T}_{23n}{\cal T}_{12n}{\cal T}_{1n}\,,~~~~\label{oper-color} \eea
acting on the amplitudes of  gravity theory.  Using the fact that
${\cal T}_{1n}={\cal T}_{n1}$ and ${\cal T}_{ikj}=-{\cal T}_{jki}$,
we get
\bea {\cal T}_{(n-2)(n-1)n}\cdots{\cal T}_{23n}{\cal T}_{12n}{\cal
T}_{1n} &=&(-)^{n-2}{\cal T}_{n(n-1)(n-2)}\cdots{\cal T}_{n32}{\cal
T}_{n21}{\cal T}_{n1}\nn &=&(-)^{n-2}{\cal T}[n,(n-1),\cdots,2,1]\,.
\eea
Substituting it to \eref{oper-color}, we find
\bea {\cal T}[1,2,\cdots,(n-1),n]&=&(-)^n{\cal
T}[n,(n-1),\cdots,2,1]\,.~~~~\label{oper-color-rev} \eea
Then the relation \eref{color-rev} can be trivially obtained by
applying \eref{oper-color-rev} on the gravity amplitude ${
A}^{G}(\{1,\cdots,n\}\cup\{h_g\})$.

The second relation is  the generalized photon decoupling relation
of EYM amplitudes. By definition, we have

\bea {\cal T}_{1hn}&=&\partial_{\epsilon_h\cdot
k_1}-\partial_{\epsilon_h\cdot k_n}\nn
&=&\partial_{\epsilon_h\cdot k_1}-\partial_{\epsilon_h\cdot k_2}
+\partial_{\epsilon_h\cdot k_2}-\partial_{\epsilon_h\cdot
k_3}+\cdots +\partial_{\epsilon_h\cdot
k_{n-1}}-\partial_{\epsilon_h\cdot k_n}\nn &=&{\cal T}_{1h2}+{\cal
T}_{2h3}+\cdots+{\cal T}_{(n-1)hn}\,. \eea
Putting them together gives
\bea {\cal T}_{1h2}+{\cal T}_{2h3}+\cdots+{\cal T}_{(n-1)hn}+{\cal
T}_{nh1}=0\,. ~~~\label{inser-n-rel}\eea
Applying this identity on ${A}^{\rm EYM}(1,\cdots,n;h\cup\{h_i\})$
leads to
\bea \boxed{~\sum_{\shuffle}\,{A}^{\rm
EYM}(1,h\shuffle\{2,\cdots,n\};\{h_i\})=0\,}~,~~~~\label{pho-dec-gen}
\eea
which is the wanted generalized photon decoupling relation for EYM
amplitudes.

The third relation is the generalized KK relation, which is given by
\bea \boxed{~ { A}^{\rm EYM}(1,\alpha,n,\beta;\{h_g\})
=\sum_{\shuffle}(-)^{n_\beta}{ A}^{\rm
EYM}(1,\alpha\shuffle\beta^{T},n;\{h_g\})\,}~.~~~~\label{gener-KK-rela}
\eea
First we want to point out that when $\a$ is the empty subset, it
reduces to the generalized color-ordered reserved relation
\eref{color-rev}. The proof of
the relation \eref{gener-KK-rela} is extremely similar to the proof
of \eref{color-rev}. Suppose $\beta=\{b_1,b_2,\cdots,b_r\}$, we have
\bea { A}^{\rm EYM}(1,\alpha,n,\beta;\{h_g\})&=&{ A}^{\rm
EYM}(\alpha,n,\beta,1;\{h_g\})\nn &=&{\cal T}_{nb_11}{\cal
T}_{b_1b_21}\cdots{\cal T}_{b_{r-1}b_r1}{ A}^{\rm
EYM}(\alpha,n,1;\{h_g\}\cup\beta)\nn &=&(-)^{n_\beta}{\cal
T}_{1b_1n}{\cal T}_{1b_2b_1}\cdots{\cal T}_{1b_rb_{r-1}}{ A}^{\rm
EYM}(1,\alpha,n;\{h_g\}\cup\beta)\,.~~~~\label{deri-gener-KK} \eea
In the third step, the relation ${\cal T}_{ikj}=-{\cal T}_{jki}$ and
the cyclic symmetry have been used. To understanding the meaning of
last line in \eref{deri-gener-KK}, recalling that all insertation
operators are commutative and by the explanation of
\eref{Inser-not-nearby},  the operator ${\cal T}_{1b_1n}$ inserts
$b_1$ at any position between $1$ and $n$, thus turns ${ A}^{\rm
EYM}(1,\alpha,n;\{h_g\}\cup\beta)$ to ${ A}^{\rm
EYM}(1,\alpha\shuffle b_1,n;\{h_g\}\cup \beta\backslash \b_1)$. The
next action of ${\cal T}_{1b_2b_1}$ gives ${ A}^{\rm
EYM}(1,\alpha\shuffle \{\b_2,b_1\},n;\{h_g\}\cup\beta
\backslash\{b_1,\b_2\})$. Continuing the action until the last one,
 the last line in \eref{deri-gener-KK} gives $(-)^{n_\b}{
A}^{\rm EYM}(1,\alpha\shuffle\beta^T,n;\{h_g\})$, thus the
generalized KK relation \eref{gener-KK-rela} is obtained.

\section{Expansion in the  KK-basis by differential operators}

In this section, we will propose a new strategy to derive the
expansion of sEYM amplitudes in the KK-basis of YM amplitudes by
differential operators given in \cite{Cheung:2017ems}. Before doing
so, we make some remarks:
\begin{itemize}

\item (1) Although it is well known that sEYM amplitudes can be expanded
in terms of YM amplitudes, we don't need to assume this in
the beginning. From the presentation in this section, we will
see that assuming the factorization of polarization tensors
$\epsilon_i^{\mu\nu}$ of gravitons into two polarization vectors $\eps$
and $\W \eps$ of gluons (i.e.,
$\epsilon_i^{\mu\nu}=\epsilon_i^{\mu}\tilde{\epsilon}_i^{\nu}$),
the web connections given in \cite{Cheung:2017ems} will
naturally lead to  the expansion of sEYM amplitudes in terms of
YM amplitudes.

\item (2) Although in general, the expansion coefficients in the KK-basis can be
rational functions as in \cite{Nandan:2016pya,delaCruz:2016gnm},
one can arrange them properly using the generalized BCJ
relations \eref{generalized-BCJ} to get polynomial coefficients
as written in \eref{eq:singletrace}. If we impose the condition that the expansion coefficients are polynomial, the KK-basis can be treated as independent up
to the generalized BCJ-relations
\eref{generalized-BCJ}\footnote{If we use the language of
algebraic geometry, generalized BCJ relations
\eref{generalized-BCJ} will generate an ideal in the polynomial
ring of $(n-2)!$ variables. KK-basis will be the independent
basis in the quotient ring.}. Under this understanding, we see that the expansion coefficients in the KK-basis have the freedom caused by the generalized BCJ-relations.

\item (3) Since we have assumed the factorization of polarization
tensors of gravitons, we consider  building blocks for the
polarization vectors $\eps$. For a given graviton, the expansion
coefficients must be the function of Lorentz invariant
combinations $(k_i\cdot k_j)$, $(\eps_i\cdot k_j)$, as well as
$(\eps_i\cdot \eps_j)$~\footnote{Here we have excluded the
contraction $\eps_i\cdot \W\eps_i$. In other words, we have
assumed that two polarization vectors coming from the same
graviton should not contract with each other. This is a very
natural assumption and could be seen from the traditional
Feynman rules of gravitons.}. Among them, $(\eps_i\cdot k_j)$
and $(\eps_i\cdot \eps_j)$ are the building blocks for the
expansion when applying the differential operators. One reason
using these building blocks is that we can transform the
differential equations of expansion  to linear algebraic
equations of unknown coefficients of building blocks. One will
see that such a choice of building blocks will naturally lead to
the expansion in the KK-basis.

\item (4) The mass dimension of expansion coefficients is $|H|$, the number of gravitons.

\item (5) For the expansion of sEYM amplitudes
${A}^{\text{EYM}}_{n,m}(1,\cdots,n;\{h_1,\cdots,h_m\})$, the
KK-basis is not general since the ordering of $\{1,2,\cdots,
n\}$ will be kept for each amplitude of the basis with nonzero
coefficient. This fact can easily be understood from the pole
structures of both sides and be proved using the insertion
operators. Keeping only these KK-basis with the ordering
$\{1,2,\cdots, n\}$ has also partially fixed the freedoms caused
by the generalized BCJ relations \eref{generalized-BCJ}.

\end{itemize}

Above five points come from  general considerations. Based on them,
the structure of this section is following. In the first subsection,
we consider the sEYM amplitudes with only one graviton. The
calculations in this part are straightforward.  In the second
subsection, we consider the sEYM amplitudes with two gravitons. This
part is very important because it contains the recursive strategy used in
the third subsection for general situations. Based on the idea in
the second part, recursive expansion for sEYM amplitudes with
arbitrary number of gravitons has been presented in the third
subsection. Thus we have shown that using differential operators, we
could indeed derive the expansion  of sEYM amplitudes in the
KK-basis of YM amplitudes as given in \eref{eq:singletrace}.

\subsection{The case with one graviton}

According to the principles of Lorentz invariance and momentum
conservation, the  sEYM amplitude with one graviton
$A^{\text{EYM}}_{n,1}(1,\cdots,n;h_1)$ can always be written as
\begin{align}
 A^{\text{EYM}}_{n,1}(1,\cdots,n;h_1)=\sum_{i=1}^{n-1}(\epsilon_{h_1}\cdot k_i)\W B_i,
\end{align}
where $(\epsilon_{h_1}\cdot k_i)$'s\footnote{The
$\epsilon_{h_1}\cdot k_n$ is not independent by the momentum
conservation.} constitute a linearly independent building blocks
with coefficients $\W B_i$ being functions of
$\tilde{\epsilon}_{h_1}$ and $(k_i\cdot k_j)$. By our experiences,
to simplify the calculation  we can choose a new basis of building
blocks and the expansion is
\begin{align}
 A^{\text{EYM}}_{n,1}(1,\cdots,n;h_1)=\sum_{i=1}^{n-1}(\epsilon_{h_1}\cdot K_i)B_i,~~~\label{1g-exp-1}
\end{align}
with $K_i=\sum_{j=1}^{i}k_j$, $i=1,...,n-1$.

Now we use differential operators to determine  $B_i$'s.  To do so,
we apply insertion operators $\mathcal{T}_{ah_1(a+1)}$ with
$a=1,\cdots,n-1$ to the expansion \eref{1g-exp-1}. Since
$\mathcal{T}_{ah_1(a+1)}=\partial_{\epsilon_{h_1}\cdot
k_a}-\partial_{\epsilon_{h_1}\cdot k_{a+1}}$ doesn't act on $B_i$,
we have
\begin{align}
 \mathcal{T}_{ah_1(a+1)}A^{\text{EYM}}_{n,1}(1,\cdots,n;h_1)
 =& A^{\text{YM}}_{n+1}(1,\cdots,a,h_1,a+1,\cdots,n-1)  \notag\\
 =& \sum_{i=1}^{n-1}\left\{ \mathcal{T}_{ah_1(a+1)}(\epsilon_{h_1}\cdot K_i)\right\} B_i
 = \sum_{i=1}^{n-1} \delta_{a,i} B_i
 = B_a. ~~~\label{1g-exp-2}
\end{align}
Above manipulations show that after applying differential operators to
the building blocks, we  obtain  linear algebraic equations for
these $B_i$'s. With the smart choice of building blocks, these
linear equations are very easy to solve
\bea B_i=A^{\text{YM}}_{n+1}(1,\cdots,i,h_1,i+1,\cdots,n-1)~.
~~~\label{1g-exp-3}\eea
Here is a subtlety for the solution \eref{1g-exp-3}. As we have
remarked, the KK-basis is independent only up to the generalized BCJ
relations \eref{generalized-BCJ}. Thus in principle, one can add
these terms at the right hand side of \eref{1g-exp-3}. But with the
polynomial condition  as well as  the mass dimension condition of
coefficients, one can see that \eref{1g-exp-3} is the only allowed
solution. Similar phenomena will appear in later subsections and we
will not discuss them further. Putting the solutions back, we get
\begin{align}
A^{\text{EYM}}_{n,1}(1,\cdots,n;h_1)
=& \sum_{i=1}^{n-1}(\epsilon_{h_1}\cdot K_i) A^{\text{YM}}_{n+1}(1,\cdots,i,h_1,i+1,\cdots,n-1) \notag\\
=& \sum_{\shuffle} (\epsilon_{h_1}\cdot Y_{h_1}) A^{\text{YM}}_{n+1}(1,\{2,\cdots,n-1\}
\shuffle\{h_1\},n),~~~\label{1g-exp-4}
\end{align}
where in the second line we have used the definition of $\shuffle$
and $Y_{h_1}$ (see \eref{Yp-def}).

Although this example is trivial, it reveals the essence of our
strategy: (1) The expansion in the KK-basis of YM amplitudes is the
natural consequence of web connections established in
\cite{Cheung:2017ems} without relying other information such as the
double copy form \eref{gen-3}; (2) The differential operators act on
building blocks only, and we get  linear algebraic equations for
unknown coefficients, which are much easier to solve than
differential equations. This transmutation reflects  the important
role of building blocks.

\subsection{The case with two gravitons}

Now we move to the less trivial case $A^{\text{EYM}}_{n,2}$. This
example will show  how to generalize our strategy to arbitrary
amplitudes $A^{\text{EYM}}_{n,m}$. As in previous subsection, we
expand $A^{\text{EYM}}_{n,2}$ according to the building blocks of
$\eps_{h_1}$ as
\begin{align}
 A^{\text{EYM}}_{n,2}(1,\cdots,n;h_1,h_2)
 = \sum_{i=1}^{n-1}(\epsilon_{h_1}\cdot K_i)B_i+(\epsilon_{h_1}
 \cdot k_{h_2})(\epsilon_{h_2}\cdot D_{h_2})+(\epsilon_{h_1}\cdot
 \epsilon_{h_2})E_{h_2}\,. \label{KKtwo}
\end{align}
 To determine $B_i$, we use insertion
operators $\mathcal{T}_{ah_1(a+1)}$ with $a=1,\cdots,n-1$. After
acting on both sides, we arrive
\begin{align}
\mathcal{T}_{ah_1(a+1)}A^{\text{EYM}}_{n,2}(1,\cdots,n;h_1,h_2)
= A^{\text{EYM}}_{n+1,1}(1,\cdots,a,h_1,a+1,\cdots,n;h_2)
= B_a,
\end{align}
Putting the solutions back, (\ref{KKtwo}) becomes
\begin{align}
 A^{\text{EYM}}_{n,2}(1,\cdots,n;h_1,h_2)
 =& \sum_{\shuffle}(\epsilon_{h_1}\cdot Y_{h_1}) A^{\text{EYM}}_{n+1,1}(1,\{2,\cdots,n-1\}\shuffle\{h_1\},n;h_2) \notag\\
   & +(\epsilon_{h_1}\cdot k_{h_2})(\epsilon_{h_2}\cdot D_{h_2})+(\epsilon_{h_1}\cdot \epsilon_{h_2})E_{h_2}. \label{KKtwo1}
\end{align}
To determine the remaining two variables $D_{h_2}, E_{h_2}$, we use
the gauge invariance of the graviton $h_2$~\footnote{The gauge
invariance of graviton ${h_i}$ means the Ward's identity, i.e., the
amplitude vanishes under the replacement $\epsilon_{h_i}\to
k_{h_i}$.}.There are two approachs. The first approach is that
since the first term at the RHS of \eref{KKtwo1} is ${h_2}$ gauge
invariant already, after $\epsilon_{h_2}\to k_{h_2}$ in the
remaining two terms, we get
\bea (\epsilon_{h_1}\cdot k_{h_2})(k_{h_2}\cdot {
D}_{h_2})+(\epsilon_{h_1}\cdot k_{h_2}) E_{h_2}=0 \
~~\Longrightarrow (k_{h_2}\cdot { D}_{h_2})+ E_{h_2}=0\,. \eea
The second approach is to  consider the commutation relation of
insertion operators and gauge invariant operators (see
\eref{insertion} and \eref{Gauge-oper})\footnote{Although for this
simple example, the first approach is simpler, but when trying to
generalize to general cases, the second approach is more suitable.}
\bea [{\cal T}_{ijk},{\cal W}_{l}]=\delta_{il}{\cal
T}_{ij}-\delta_{kl}{\cal T}_{jk}\,, \eea
where ${\cal T}_{jk}$ is the trace operator \eref{trace}. For
current case, we apply the commutation relation
$[\mathcal{T}_{h_2h_1n},\mathcal{W}_{h_2}]=\mathcal{T}_{h_2h_1}$  to
both sides of \eref{KKtwo1} and get
\bea &  &
[\mathcal{T}_{h_2h_1n},\mathcal{W}_{h_2}]A^{\text{EYM}}_{n,2}(1,\cdots,n;h_1,h_2)
=
-\mathcal{W}_{h_2}\mathcal{T}_{h_2h_1n}A^{\text{EYM}}_{n,2}(1,\cdots,n;h_1,h_2)
= -(k_{h_2}\cdot D_{h_2}) \nn
 & &\mathcal{T}_{h_2h_1}A^{\text{EYM}}_{n,2}(1,\cdots,n;h_1,h_2)
=E_{h_2}~. \eea
No matter which approach is used, now (\ref{KKtwo1}) becomes
\begin{align}
A^{\text{EYM}}_{n,2}(1,\cdots,n;h_1,h_2)
 =& \sum_{\shuffle}(\epsilon_{h_1}\cdot Y_{h_1}) A^{\text{EYM}}_{n+1,1}(1,\{2,\cdots,n-1\}\shuffle\{h_1\},n;h_2)
  +(\epsilon_{h_1}\cdot f_{h_2}\cdot D_{h_2})~,
\end{align}
where the gauge invariance of $h_2$ is manifest. To determine the
last variable $D_{h_2}$, we expand it according to the building
blocks of $\eps_{h_2}$ as
\begin{align}
A^{\text{EYM}}_{n,2}(1,\cdots,n;h_1,h_2)
 =& \sum_{\shuffle}(\epsilon_{h_1}\cdot Y_{h_1}) A^{\text{EYM}}_{n+1,1}(1,\{2,\cdots,n-1\}\shuffle\{h_1\},n;h_2) \notag\\
 &+\sum_{i=1}^{n-1} (\epsilon_{h_1}\cdot f_{h_2}\cdot K_i)H_{i}+(\epsilon_{h_1}
 \cdot f_{h_2}\cdot k_{h_1}) H_{h_1}~.  \label{KKtwo2}
\end{align}
To determine  the $H_i$'s, we apply the combinations
$\mathcal{T}_{ah_2(a+1)}\mathcal{T}_{h_2h_1(a+1)},~a=1,...,n-1$ of
two insertion operators, and get
\begin{align}
&\mathcal{T}_{ah_2(a+1)}\mathcal{T}_{h_2h_1(a+1)}A^{\text{EYM}}_{n,2}(1,\cdots,n;h_1,h_2)
= A^{\text{YM}}_{n+2}(1,\cdots,a,h_2,h_1,a+1,n) \notag\\
=& \mathcal{T}_{ah_2(a+1)}\Big\{ -\sum_{i=a+1}^{n-1} A^{\text{EYM}}_{n+1,1}(1,\cdots,i,h_1,i+1,\cdots,n;h_2)
  +\sum_{i=1}^{n-1} (\epsilon_{h_2}\cdot K_i)H_{i}+(\epsilon_{h_2}\cdot k_{h_1}) H_{h_1}  \Big\} \notag\\
  =& -\sum_{i=a+1}^{n-1} A^{\text{YM}}_{n+2}(1,\cdots,a,h_2,a+1,\cdots,i,h_1,i+1,
  \cdots,n) +H_{a}.
\end{align}
From it we can solve
\bea
H_a=\sum_{\shuffle}A^{\text{YM}}_{n+2}(1,\cdots,a,h_2,\{a+1,\cdots,n-1\}\shuffle\{h_1\},n)~,\eea
so the expansion becomes
\begin{align}
A^{\text{EYM}}_{n,2}(1,\cdots,n;h_1,h_2)
 =& \sum_{\shuffle}(\epsilon_{h_1}\cdot Y_{h_1}) A^{\text{EYM}}_{n+1,1}(1,\{2,\cdots,n-1\}\shuffle\{h_1\},n;h_2) +(\epsilon_{h_1}\cdot f_{h_2}\cdot k_{h_1}) H_{h_1}\notag\\
 &+ \sum_{\shuffle}(\epsilon_{h_1}\cdot f_{h_2}\cdot Y_{h_2})A^{\text{YM}}_{n+2}(1,\{2,\cdots,n-1\}\shuffle\{h_2,h_1\},n). \label{KKtwo2.1}
\end{align}

Now only the $H_{h_1}$ is unknown. This one is  manifestly gauge
invariant and symmetric because $(\epsilon_{h_1}\cdot f_{h_2}\cdot
k_{h_1}) H_{h_1}=\frac{1}{2}\text{tr}(f_{h_1}f_{h_2})H_{h_1}$. To
fix it, we use  the special operator
$\mathcal{T}_{jh_1h_2}\mathcal{T}_{jh_2h_1}$.  While acting on the
right hand side of (\ref{KKtwo2.1}) it gives
\begin{align}
 R=&\mathcal{T}_{jh_1h_2}\mathcal{T}_{jh_2h_1} \Big\{ \sum_{\shuffle}(\epsilon_{h_1}\cdot Y_{h_1}) A^{\text{EYM}}_{n+1,1}(1,\{2,\cdots,n-1\}\shuffle\{h_1\},n;h_2) \notag\\
 &+ \sum_{\shuffle}(\epsilon_{h_1}\cdot f_{h_2}\cdot Y_{h_2})A^{\text{YM}}_{n+2}(1,\{2,\cdots,n-1\}\shuffle\{h_2,h_1\},n)
 +(\epsilon_{h_1}\cdot f_{h_2}\cdot k_{h_1}) H_{h_1}\Big\}   = H_{h_1},
\end{align}
when acting on the left hand side, the physical meaning of
"insertion" is not clear at the level of amplitudes. However, as
shown in \cite{Zhou:2018wvn, Bollmann:2018edb}, its action at the
level of CHY-integrands is still clear.  Doing it leads to
\begin{align}
 L=&\mathcal{T}_{jh_1h_2}\mathcal{T}_{jh_2h_1}A^{\text{EYM}}_{n,2}(1,\cdots,n;h_1,h_2) \notag\\
 =&  \int d\mu \text{PT}(1,2,\cdots,n)  \Big\{ \mathcal{T}_{jh_1h_2}
 \mathcal{T}_{jh_2h_1} \text{Pf} \Psi_H  \Big\} \text{Pf}' \Psi~.
\end{align}
To continue, we use the Pfaffian expansion given in
\cite{Lam:2016tlk,He:2016iqi}  and keep only terms contributing
under these operators. The related terms\footnote{More details,
please see the Appendix \ref{cycle-index}.} are
$(\Psi_{(h_1)}\Psi_{(h_2)}-\Psi_{(h_1h_2)})$ with
\bea \Psi_{(h_1)}=C_{h_1h_1}=-\sum_{a\ne h_1}
\frac{\epsilon_{h_1}\cdot k_a}{\sigma_{h_1a}},~~
\Psi_{(h_2)}=C_{h_2h_2}=-\sum_{a\ne h_2} \frac{\epsilon_{h_2}\cdot
k_a}{\sigma_{h_2a}},~~~ \Psi_{(h_1h_2)}=\frac{\text{tr}(f_{h_1}\cdot
f_{h_2})}{2\sigma_{h_1h_2}\sigma_{h_2h_1}}~.\eea
Using them, we find
\begin{align}
\mathcal{T}_{jh_1h_2}\mathcal{T}_{jh_2h_1} \text{Pf}(\Psi_H)
&= \mathcal{T}_{jh_1h_2}\mathcal{T}_{jh_2h_1}\Big\{ \sum_{a\ne h_1} \frac{\epsilon_{h_1}\cdot k_a}{\sigma_{h_1a}} \sum_{b\ne h_2} \frac{\epsilon_{h_2}\cdot k_b}{\sigma_{h_2b}}-\frac{\text{tr}(f_{h_1}\cdot f_{h_2})}{2\sigma_{h_1h_2}\sigma_{h_2h_1}} \Big\}  \notag\\
&=(\frac{1}{\sigma_{h_1j}}-\frac{1}{\sigma_{h_1h_2}})  (\frac{1}{\sigma_{h_2j}}-\frac{1}{\sigma_{h_2h_1}})-\frac{1}{\sigma_{h_1h_2}\sigma_{h_2h_1}} \notag\\
&=\frac{\sigma_{jh_2}}{\sigma_{h_1j}\sigma_{h_1h_2}} \frac{\sigma_{jh_1}}{\sigma_{h_2j}\sigma_{h_2h_1}}-\frac{1}{\sigma_{h_1h_2}\sigma_{h_2h_1}} \notag\\
&=0~.
\end{align}
Since the left-hand side is zero, we find $H_{h_1}=0$.

Assembling all pieces together,  we obtain the expansion of
$A^{\text{EYM}}_{n,2}$ as
\begin{align}
A^{\text{EYM}}_{n,2}(1,\cdots,n;h_1,h_2)
 =& \sum_{\shuffle}(\epsilon_{h_1}\cdot Y_{h_1}) A^{\text{EYM}}_{n+1,1}(1,\{2,\cdots,n-1\}\shuffle\{h_1\},n;h_2) \notag\\
 &+ \sum_{\shuffle}(\epsilon_{h_1}\cdot f_{h_2}\cdot Y_{h_2})A^{\text{YM}}_{n+2}(1,\{2,\cdots,n-1\}\shuffle\{h_2,h_1\},n).~~~\label{li-2g-exp}
\end{align}
It is a recursive expansion of $A^{\text{EYM}}_{n,2}$ by
$A^{\text{EYM}}_{n+1,1}$ and $A^{\text{YM}}_{n+2}$. We can get the
complete expansion of $A^{\text{EYM}}_{n,2}$ in terms of  purely YM
amplitudes by putting back the known  expansion of
$A^{\text{EYM}}_{n+1,1}$ in \eref{1g-exp-4}.

From above calculations we see again that the expansion of sEYM
amplitudes to the KK-basis of YM amplitudes is the natural
consequence of web connections established in  \cite{Cheung:2017ems}
by differential operators. Furthermore, we see that to  find the
expansion, we could do it step by step. At the first step,  we
organize expansion according to the building blocks of
$\epsilon_{h_1}$ and then using insertion and gauge invariant
differential operators to determine some coefficients. At the second
step, we organize the remaining unknown variables according to the
building blocks of the polarization vector of  next gravitons, and
then solve some variables again. Repeating the procedure, we will
finally determine the expansion. In the next subsection, we will see
more details about the iterative operations.

A nontrivial thing in above derivation is the possible appearance
of the term $(\epsilon_{h_1}\cdot f_{h_2}\cdot k_{h_1})
H_{h_1}=\frac{1}{2}\text{tr}(f_{h_1}f_{h_2})H_{h_1}$. It satisfies
all general conditions, such as the multi-linearity of $\eps_{h_i}$,
gauge invariance and polynomial, etc.  To show its vanishing, we
need to consider the action of differential operators at the level
of CHY-integrands instead of at the level of amplitudes directly.
The vanishing result has been carefully proved in the Appendix
\ref{cycle-index}, which may imply some interesting things at the
level of sEYM amplitudes.

\subsection{The case with $m$ gravitons}

Having outlined our derivation strategy in the previous subsection,
we consider the expansion of arbitrary sEYM amplitudes
$A^{\text{EYM}}_{n,m}$. Let us begin with  the expansion according
to the building blocks of graviton $h_1$
\begin{align}
A_{n,m}(1,2,\cdots,n;\{h_1,\cdots,h_m\})
= \sum_{i=1}^{n-1}(\epsilon_{h_1}\cdot K_i)B_i+\sum_{j=2}^{m}(\epsilon_{h_1}\cdot k_{h_j})
(\epsilon_{h_j}\cdot D_{h_j})+\sum_{j=2}^{m}(\epsilon_{h_1}\cdot \epsilon_{h_j})E_{h_j}~.  \label{KKm}
\end{align}
At the first step
 we apply the insertion operators $\mathcal{T}_{ah_1(a+1)}$ with $a=1,\cdots,n-1$ to (\ref{KKm}) and get
\begin{align}
\mathcal{T}_{ah_1(a+1)}A_{n,m}(1,2,\cdots,n;\mathbf{h})
=A_{n+1,m-1}(1,2,\cdots,a,h_1,a+1,\cdots,n;\mathbf{h}/\{h_1\})
=B_a,
\end{align}
where we use $\mathbf{h}$ to denote the set of gravitons
$\{h_1,\cdots,h_m\}$. Thus (\ref{KKm}) becomes
\begin{align}
A_{n,m}(1,2,\cdots,n;\mathbf{h})
=& \sum_{\shuffle}(\epsilon_{h_1}\cdot Y_{h_1})A_{n+1,m-1}(1,\{2,\cdots,n-1\}\shuffle\{h_1\},n;\mathbf{h}/\{h_1\}) \notag\\
&+\sum_{j=2}^{m}(\epsilon_{h_1}\cdot k_{h_j})(\epsilon_{h_j}\cdot D_{h_j})+\sum_{j=2}^{m}(\epsilon_{h_1}\cdot \epsilon_{h_j})E_{h_j}.  \label{KKm1}
\end{align}
At the second step, using the gauge invariance conditions, i.e., the
commutation relations
$[\mathcal{T}_{h_ah_1h},\mathcal{W}_{h_a}]=\mathcal{T}_{h_ah_1}$
with $a=2,\cdots,m$ to (\ref{KKm1}), we get
\begin{align}
&[\mathcal{T}_{h_ah_1h},\mathcal{W}_{h_a}]A_{n,m}(1,2,\cdots,n;\mathbf{h})
=-(k_{h_a}\cdot D_{h_a})
=\mathcal{T}_{h_ah_1}A_{n,m}(1,2,\cdots,n;\mathbf{h})
=E_{h_a},
\end{align}
thus (\ref{KKm1}) becomes
\bea A_{n,m}(1,2,\cdots,n;\mathbf{h}) =
\sum_{\shuffle}(\epsilon_{h_1}\cdot
Y_{h_1})A_{n+1,m-1}(1,\{2,\cdots,n-1\}\shuffle\{h_1\},n;
\mathbf{h}/\{h_1\}) +\sum_{j=2}^{m}(\epsilon_{h_1}\cdot f_{h_j}\cdot
D_{h_j}). ~~~~~\label{KKm2} \eea
One needs to notice that  $D_{h_j}$'s are required to be gauge
invariant for gravitons $h\neq h_1, h_j$\footnote{It is because
$A_{n+1,m-1}(1,\{2,\cdots,n-1\}\shuffle\{h_1\},n;\mathbf{h}/\{h_1\})$
is gauge invariant for the remaining gravitons and
$(\epsilon_{h_1}\cdot f_{h_j}\cdot D_{h_j})$'s are linearly
independent.}.

To determine the $D_{h_j}$, we need to expand $(\epsilon_{h_j}\cdot
D_{h_j})$ in \eref{KKm1} further according to $h_j$'s building
blocks
\begin{align}
(\epsilon_{h_j}\cdot D_{h_j})
=&\sum_{j_2=2,j_2\ne j}^{m}(\epsilon_{h_j}\cdot k_{h_{j_2}})(\epsilon_{h_{j_2}}\cdot D_{h_jh_{j_2}})+\sum_{j_2=2,j_2\ne j}^{m}(\epsilon_{h_j}\cdot \epsilon_{h_{j_2}}) E_{h_jh_{j_2}} \notag\\
&+\sum_{i=1}^{n-1} (\epsilon_{h_j}\cdot K_i)B_{h_ji}+ (\epsilon_{h_j}\cdot k_{h_1}) B_{h_jh_1}. \label{KKme}
\end{align}
Because of the gauge invariance of $D_{h_j}$, we can apply the
commutation relations
$[\mathcal{T}_{h_{a}h_jn},\mathcal{W}_{h_{a}}]=\mathcal{T}_{h_{a}h_j}$
(with $a=2,\cdots,m,$ but $a\ne j$) to (\ref{KKme}) and get
$E_{h_jh_{j_2}}=-(k_{h_{j_2}}\cdot D_{h_jh_{j_2}})$. Then
(\ref{KKme}) becomes
\begin{align}
(\epsilon_{h_j}\cdot D_{h_j})
=&\sum_{j_2=2,j_2\ne j}^{m}(\epsilon_{h_j}\cdot f_{h_{j_2}}\cdot D_{h_jh_{j_2}})+\sum_{i=1}^{n-1} (\epsilon_{h_j}\cdot K_i)B_{h_ji}+ (\epsilon_{h_j}\cdot k_{h_1})B_{h_jh_1},
\end{align}
and (\ref{KKm2}) becomes
\begin{align}
A_{n,m}(1,2,\cdots,n;\mathbf{h})
=& \sum_{\shuffle}(\epsilon_{h_1}\cdot Y_{h_1})A_{n+1,m-1}(1,\{2,\cdots,n-1\}\shuffle\{h_1\},n;\mathbf{h}/\{h_1\})  \notag\\
&+\sum_{j=2}^{m}\sum_{j_2=2,j_2\ne j}^{m}(\epsilon_{h_1}\cdot f_{h_j}\cdot f_{h_{j_2}}\cdot D_{h_jh_{j_2}})
+\sum_{j=2}^{m}\sum_{i=1}^{n-1} (\epsilon_{h_1}\cdot f_{h_j}\cdot K_{i})B_{h_ji}  \notag\\
&+\sum_{j=2}^{m}(\epsilon_{h_1}\cdot f_{h_j}\cdot k_{h_1})B_{h_jh_1}.  \label{KKm3}
\end{align}

To determine $B_{h_ji}$'s we  apply the products of two insertion
operators, i.e., $\mathcal{T}_{ah_j(a+1)}\mathcal{T}_{h_jh_1(a+1)}$
with $a=1,\cdots,n-1$ to (\ref{KKm3})
\begin{align}
&\mathcal{T}_{ah_j(a+1)}\mathcal{T}_{h_jh_1(a+1)}A_{n,m}(1,2,\cdots,n;\mathbf{h})
=A_{n+2,m-2}(1,2,\cdots,a,h_j,h_1,a+1,\cdots,n;\mathbf{h}/\{h_1,h_j\}) \notag\\
=& -\sum_{i=a+1}^m A_{n+2,m-2}(1,2,\cdots,a,h_j,a+1,i,h_1,i+1,\cdots,n;\mathbf{h}/\{h_1,h_j\})
+B_{h_ja},
\end{align}
thus obtain
$B_{h_ja}=A_{n+2,m-2}(1,2,\cdots,a,h_j,\{a+1,\cdots,n-1\}\shuffle\{h_1\},n;\mathbf{h}/\{h_1,h_j\})$.
Next, doing similar thing (i.e., by acting proper operators
$\mathcal{T}_{kh_1h_j}\mathcal{T}_{kh_jh_1}$ on the CHY-integrand)
like in the previous subsection, one can show the vanishing of
$B_{h_jh_1}$. Thus (\ref{KKm3}) becomes
\begin{align}
A_{n,m}(1,2,\cdots,n;\mathbf{h})
=& \sum_{\shuffle}(\epsilon_{h_1}\cdot Y_{h_1})A_{n+1,m-1}(1,\{2,\cdots,n-1\}\shuffle\{h_1\},n;\mathbf{h}/\{h_1\})  \notag\\
&+\sum_{j=2}^{m}\sum_{\shuffle} (\epsilon_{h_1}\cdot f_{h_j}\cdot Y_{h_j})A_{n+2,m-2}(1,\{2,\cdots,n-1\}\shuffle\{h_j,h_1\},n;\mathbf{h}/\{h_1,h_j\}) \notag\\
&+\sum_{j=2}^{m}\sum_{j_2=2,j_2\ne j}^{m}(\epsilon_{h_1}\cdot f_{h_j}\cdot f_{h_{j_2}}\cdot D_{h_jh_{j_2}}) .  \label{KKm4}
\end{align}
where  $D_{h_jh_{j_2}}$'s are also gauge invariant for remaining
gravitons.

Now we  compare (\ref{KKm2}) and (\ref{KKm4}). We see that to get
(\ref{KKm4}) from (\ref{KKm2}), we need just do the replacement
\bea D_{h_j}^{\mu} \to
\sum_{\shuffle}Y_{h_j}^{\mu}A_{n+2,m-2}(1,\{2,\cdots,n-1\}\shuffle\{h_j,h_1\},n;\mathbf{h}/\{h_1,h_j\})
+\sum_{j_2=2,j_2\ne j}^{m} (f_{h_{j_2}}\cdot D_{h_jh_{j_2}})^{\mu}.
\eea
This replacement is derived by using proper products of insertions
operators, which act either at the amplitude level or at the
CHY-integrand level as shown in above. Now the pattern is clear. To
expand $(\epsilon_{h_{j_2}}\cdot D_{h_jh_{j_2}})$, we act on with
products of three insertion operators and find the result is
equivalent to the replacement
\begin{align}
 D_{h_jh_{j_2}}^{\mu}\rightarrow& \sum_{\shuffle}Y_{h_{j_2}}^{\mu}A_{n+3,m-3}(1,\{2,\cdots,n-1\}\shuffle\{h_{j_2},h_j,h_1\},n;\mathbf{h}/\{h_1,h_j,h_{j_2}\}) \notag\\
&+\sum_{j_3=2,j_3\ne j,j_2}^{m} (f_{h_{j_3}}\cdot D_{h_jh_{j_2}h_{j_3}})^{\mu}~.
\end{align}
Continuing the procedure  until the coefficient $D$ does not contain
any polarization vector $\eps$ (so the iteration stops), we obtain
the complete recursive expansion of $A^{\text{EYM}}_{n,m}$ as
\begin{align}
&A_{n,m}(1,2,\cdots,n;\mathbf{h}) \notag\\
=& \sum_{\shuffle}(\epsilon_{h_1}\cdot Y_{h_1})A_{n+1,m-1}(1,\{2,\cdots,n-1\}\shuffle\{h_1\},n;\mathbf{h}/\{h_1\})  \notag\\
&+\sum_{j=2}^{m} (\epsilon_{h_1}\cdot f_{h_j}\cdot Y_{h_j})A_{n+2,m-2}(1,\{2,\cdots,n-1\}\shuffle\{h_j,h_1\},n;\mathbf{h}/\{h_1,h_j\}) \notag\\
&+\sum_{j=2}^{m}\sum_{j_2=2,j_2\ne j}^{m}\sum_{\shuffle} (\epsilon_{h_1}\cdot f_{h_j}\cdot f_{h_{j_2}}\cdot Y_{h_{j_2}})
   A_{n+3,m-3}(1,\{2,\cdots,n-1\}\shuffle\{h_{j_2},h_j,h_1\},n;\mathbf{h}/\{h_1,h_j,h_{j_2}\}) \notag\\
&+\cdots  \notag\\
&+\sum_{j=2}^{m}\sum_{j_2=2,j_2\ne j}^m\cdots \sum_{j_m\ne j,j_2,\cdots,j_{m-1}} \sum_{\shuffle}(\epsilon_{h_1}f_{h_j}\cdots f_{h_{j_m}}Y_{h_{j_m}})
  A_{n+m}(1,\{2,\cdots,n-1\}\shuffle\{h_{j_m},h_{j_{m-1}},\cdots,h_1\},n).
\end{align}
In above derivation, an important thing is that at each step, we
need to show building blocks like $(\epsilon_{h_1}\cdot f_{h_j}\cdot
f_{h_{j_2}}\cdot k_{h_1})$, $(\epsilon_{h_1}\cdot f_{h_j}\cdot
f_{h_{j_2}}\cdot k_{h_j})$ and $(\epsilon_{h_1}\cdot f_{h_j}\cdot
f_{h_{j_2}}\cdot k_{h_{j_2}})$ with $j=2,\cdots,m$ having vanished
coefficients. These building blocks belong to the "index circle
structure" and in the Appendix \ref{cycle-index} we will prove their
coefficients are zero.

Now we have reached our first goal, i.e., to expand sEYM amplitudes
in the KK-basis of YM amplitudes,  by using differential operators.
Compared to other methods appearing in the literatures, our method
provides a different angle. In the whole procedure, the use of
building blocks is very crucial. By using them, we have translated the
problem to the solving of linear algebraic equations. The solutions
are naturally the KK-basis of YM amplitudes. In other words, instead
of assuming it, we have derived the fact that  sEYM amplitudes can
be expanded in terms of  YM amplitudes.

\section{Further applications}

As explained in the introduction, the expansion of sEYM amplitudes
in the KK-basis of YM amplitudes is  the primary example. In this
section, we will demonstrate this claim by two new examples. In the
first subsection, we will try to derive the expansion of gravity
amplitudes using differential operators. Using the same strategy and
results from the previous section, the expansion can be written down
straightforwardly. In the second subsection, by using the
observation that all other amplitudes can be obtained from acting on
gravity amplitudes with proper differential operators
\cite{Cachazo:2014xea, Cheung:2017ems}, we derive the expansion of
amplitudes of  the Born-Infeld theory.

\subsection{Expansion of gravity amplitudes in the  KK basis of YM amplitudes}

Let us consider the gravity amplitude $ A^{\rm G}$ with $m$
gravitons. Expanding it according to the building blocks of $h_1$,
we write
\bea  A^{\rm G}_m(\{h_i\})&=&\sum_{f=2}^{m-1}\,(\epsilon_{h_1}\cdot
k_{h_f})(\epsilon_{h_f}\cdot {\cal B}_f)
+\sum_{g=2}^m\,(\epsilon_{h_1}\cdot \epsilon_{h_g}){\cal
D}_g\,,~~~~\label{exp-G-KK-mh1} \eea
where  $k_{h_m}$ is eliminated using the  momentum conservation. To
determine ${\cal D}_g$, one can use the trace operator ${\cal
T}_{h_1h_q}$, which turns the gravity amplitude to the sEYM
amplitude with just two gluons. Applying it on the LHS of
\eref{exp-G-KK-mh1}, it gives
\bea {\cal T}_{h_1h_q}A^{\rm G}_m(\{h_i\})
=A^{\rm EYM}_{2,m-2}(h_q,h_1;\{h_i\}/\{h_1,h_q\})\,,
\eea
while acting on the RHS it gives
\bea
T_{h_1h_q}\Big(\sum_{g=1}^m\,(\epsilon_{h_1}\cdot
\epsilon_{h_g}){\cal D}_g\Big)={\cal D}_q\,. \eea
Comparing two sides, we solve
\bea {\cal D}_q= A^{\rm EYM}_{2,m-2}(h_q,h_1;\{h_i\}/\{h_1,h_q\})\,. \eea

The vector ${\cal B}_f^\mu$ can be determined from ${\cal D}_f$ via
the gauge invariance of graviton ${h_f}$. Let us consider the
commutation relation
\bea [{\cal T}_{h_fh_1h_m},{\cal
W}_{h_f}]={\cal T}_{h_1h_f}\,. \eea
Applying this relation on \eref{exp-G-KK-mh1} gives
\bea -k_{h_q}\cdot {\cal B}_q={\cal D}_q= A^{\rm
EYM}_{2,m-2}(h_q,h_1;\{h_i\}/\{h_1,h_q\})\,,~~~~\label{g-bq} \eea
where the gauge invariant condition ${\cal W}_{h_f} A^{\rm
G}_m(\{h_i\})=0$ has been used. Thus we get\footnote{We want to
point out that the expansion form \eref{exp-G-KK-mh2} has also
appeared  in \cite{Fu:2017uzt} based on the Pfaffian expansion given
in \cite{Lam:2016tlk}.}
\bea A^{\rm G}_m(\{h_i\}) &=&\sum_{f=2}^{m-1}\,(\epsilon_{h_1}\cdot
f_{h_f}\cdot {\cal B}_f) +(\epsilon_{h_1}\cdot \epsilon_{h_m})A^{\rm
EYM}_{2,m-2}(h_m,h_1;\{h_i\}/\{h_1,h_m\})\,.~~~~\label{exp-G-KK-mh2}
\eea
To determine ${\cal B}_f^\mu$'s from the relations \eref{g-bq},  we
just need to  find  the expanded formula of sEYM amplitudes $A^{\rm
EYM}_{2,m-2}(h_q,h_1;\{h_g\}/\{h_1,h_q\})$, which have been solved
in previous section using the differential operators and are given
in \eref{eq:singletrace}. A special feature of current situation is
that there are no gluons $\{2,...,r-1\}$ in \eref{eq:singletrace}
and we are left with only $\{\pmb{h},h_a\}$. Because of this fact,
coefficient given in \eref{C-single-exp} will always have
$Y^\mu_{i_s}=k^\mu_{h_q}$ in our case, which is exactly the
combination $-k_{h_q}\cdot {\cal B}_q$ in \eref{g-bq}. In other
words, using \eref{eq:singletrace} and \eref{C-single-exp}, after
removing the $Y^\mu_{i_s}$, we get $B_{f}^\mu$ immediately.

It seems that the set of amplitudes appeared in above expansion is larger than
the KK-basis, since the appearance of the ordering
$(h_q,\pmb{h},h_a,h_1)$ in \eref{eq:singletrace}. However, when we
expand the sEYM amplitudes, the fiducial graviton $h_a$ can be
chosen arbitrarily. Thus we can always choose $h_a=h_m$ in
\eref{eq:singletrace}. Then one can use the cyclic symmetry to
re-write any ordering $(h_q,\pmb{h},h_m,h_1)$ as
$(h_1,h_q,\pmb{h},h_m)$. On the other hand, the ${\cal D}_m$ term $
A^{\rm EYM}_{2,m-2}(h_m,h_1;\{h_i\}/\{h_1,h_m\})$ equals to $ A^{\rm
EYM}_{2,m-2}(h_1,h_m;\{h_i\}/\{h_1,h_m\})$. Thus, every amplitude
appearing in the expansion has the ordering $(h_1,\cdots,h_m)$ and
belongs to the chosen KK-basis.

As a simple example, let us consider the expansion of $3$-graviton
amplitude $A^{\rm G}(\{h_1,h_2,h_3\})$. Using
\eref{exp-G-KK-mh2}, \eref{eq:singletrace} and \eref{C-single-exp}
we get
\bea A^{\rm G}_{3}(\{h_1,h_2,h_3\})
&=&-(\epsilon_{h_1}\cdot f_{h_2}\cdot\epsilon_{h_3}) A^{\rm YM}_3(h_2,h_3,h_1)
+(\epsilon_{h_1}\cdot \epsilon_{h_3}) A^{\rm EYM}_{2,1}(h_3,h_1;h_2)\nn
&=&\Big((\epsilon_{h_1}\cdot
\epsilon_{h_3})(\epsilon_{h_2}\cdot
k_{h_1})-(\epsilon_{h_1}\cdot f_{h_2}\cdot\epsilon_{h_3})\Big)
 A^{\rm YM}_{3}(h_1,h_2,h_3)\,.~~~~\label{exp-G-KK-exam}
\eea
The coefficient $\Big((\epsilon_{h_1}\cdot
\epsilon_{h_3})(\epsilon_{h_2}\cdot
k_{h_1})-(\epsilon_{h_1}\cdot f_{h_2}\cdot\epsilon_{h_3})\Big)$
is just the $3$-point YM amplitude $ A^{\rm
YM}_3(h_1,h_2,h_3)$, thus the KLT relation
$A^{\rm G}_3(\{h_1,h_2,h_3\})=
A^{\rm YM}_3(h_1,h_2,h_3) A^{\rm
YM}_3(h_1,h_2,h_3)$ has been reproduced.

\subsection{Expansion of  Born-Infeld amplitudes in the  KK basis of YM amplitudes}

As pointed out in \cite{Cheung:2017ems}, starting from the expansion
of gravity amplitudes
\bea  A^{\rm G}_m(\{g\})=\sum_{\shuffle}\,C(\shuffle) A^ {\rm
YM}(1,2\shuffle
3\shuffle\cdots\shuffle(m-2)\shuffle(m-1),m)\,,~~~\label{gr-ex} \eea
expansions of amplitudes of different theories can be obtained by
applying different combinations of  differential operators on both
sides of \eref{gr-ex} simultaneously. The action of operators  can
be divided to two types. If the operators ${\cal O}^{\W\epsilon}$
are defined via the polarization vectors $\W\epsilon_g$, both $
A^{\rm G}_m(\{g\})$ and $ A^{\rm YM}_m
(1,2\shuffle3\shuffle\cdots\shuffle(m-2)\shuffle(m-1),m)$ will be
transformed to amplitudes of other theories, while expansion
coefficients are not modified. For example, one can use this
procedure to expand YM amplitudes in terms of  bi-adjoint scalar
amplitudes, with the same coefficients. The second type is more
interesting, where the operators ${\cal O}^{\epsilon}$ are defined
via the polarization vectors $\epsilon_g$. While it turns the
gravity amplitudes to amplitudes of other theories on the LHS, it
modifies only expansion coefficients on the RHS. Using this
observation we will expand amplitudes of Born-Infeld(BI) theory in
the KK-basis of YM amplitudes in this subsection.

The $n$-point BI amplitudes can be created from the gravity ones by
applying differential operators as
\bea  A^{\rm BI}_n(\{g\})={\cal L}{\cal T}_{ab} A^{\rm G}_n(\{g\})\,,
\eea
where $\{g\}=\{1,\cdots,n\}$. The operator ${\cal L}$ is defined in
\eref{LT} as  ${\cal L}=\prod_i\,{\cal L}_i$ with $i$ running over
$\{g\}/\{a,b\}$ and ${\cal L}_i\equiv \sum_j\,(k_i\cdot
k_j)\partial_{\epsilon_i k_j}$ for $j\in\{g\}/i$.
 With the
observation
\bea {\cal T}_{ab} A^{\rm
G}_n(\{g\})= A^{\rm
EYM}_{2,n-2}(a,b;\{g\}/\{a,b\})\,, \eea
we have
\bea  A^{\rm BI}_n(\{g\})={\cal L}{A}^{\rm
EYM}_{2,n-2}(a,b;\{g\}/\{a,b\})\,.~~~~\label{exp-BI-1} \eea
Without loss of generality, let us choose $a=1$ and $b=n$. From
results in the previous subsection, we know how to expand the
amplitude $ A^{\rm EYM}_{2,n-2}(1,n;\{g\}/\{1,n\})$ in terms of YM
ones as
\bea  A^{\rm
EYM}_{2,n-2}(1,n;\{g\}/\{1,n\})=\sum_{\shuffle}\,C(\shuffle)
 A^{\rm YM}_{n}(1,2\shuffle 3\shuffle\cdots\shuffle
(n-1),n)\,.~~~~\label{exp-BI-2} \eea
Applying the operator ${\cal L}$ on the LHS of \eref{exp-BI-2} gives
the BI amplitude $ A^{\rm BI}_n(\{g\})$. For the RHS, ${\cal L}$
acts only on coefficients $C(\shuffle)$ since YM amplitudes do not
carry the polarization vectors $\epsilon$. The operator ${\cal L}$
turns all $(\epsilon_i\cdot k_j)$'s to $(k_i\cdot k_j)$'s, therefore
only terms with the form $\prod_i\,(\epsilon_i\cdot K_i)$ can
survive under the action of ${\cal L}$ with $K_i$'s being the sum of some
external momenta. In $C_{n-2,2}(\shuffle)$, such a part is given by
$\prod_i\,(\epsilon_i\cdot X_i)$. Carrying it out, the RHS gives
\bea & &{\cal
L}\Big(\sum_{\shuffle}\,C(\shuffle)
 A^{\rm YM}_n(1,2\shuffle 3\shuffle\cdots\shuffle
(n-1),n)\Big)\nn &=&\sum_{\shuffle}\,\Big(\prod_i\,(k_i\cdot
X_i)\Big)  A^{\rm YM}_n(1,2\shuffle
3\shuffle\cdots\shuffle (n-1),n)\,. \eea
Comparing the LHS with the RHS, the expansion of BI amplitudes is
obtained as
\bea \boxed{~ A^{\rm
BI}_n(\{g\})=\sum_{\shuffle}\,\Big(\prod_{i=2}^{n-1}\,(k_i\cdot
X_i)\Big) A^{\rm YM}_n(1,2\shuffle 3\shuffle\cdots\shuffle
(n-1),n)\,}~~~~~\label{exp-BI-3} \eea
with the very  compact expansion coefficients.

The expansion of BI amplitudes can also be derived via another
definition of the operator ${\cal L}$,
\bea {\cal L}=\sum_{\rho\in{\rm
pair}}\,\prod_{\{i,j\}\in\rho}\,{\cal L}_{ij}\,, \eea
where ${\cal L}_{ij}\equiv -(k_i\cdot
k_j)\partial_{\epsilon_i\epsilon_j}$ and the pairs $\{i,j\}$ run
over the set $\{g\}/\{1,n\}$. Applying this definition on the RHS of
\eref{exp-BI-2}, the survived terms are these in which each
polarization vector $\epsilon_i$ is contracted with another
polarization vector $\epsilon_j$. Using the results
\eref{eq:singletrace} and \eref{C-single-exp}, such a part is found
to be
\bea & &\sum_{|{\rm Or}_k|_{\rm
even}}\,\sum_{\shuffle}\,\Big(\prod_{k=1}^t\,(-)^{{|{\rm
Or}_k|\over2}}M_k(\shuffle)\Big) A^{\rm YM}_n(1,{\rm Or}_1\shuffle
{\rm Or}_2\shuffle\cdots\shuffle{\rm Or}_t,n)\,,~~~~\label{ee-term}
\eea
where ${\rm Or}_1\cup {\rm Or}_2\cup\cdots{\rm
Or}_t=\{2,\cdots,n-1\}$, and the summation $\sum_{|{\rm Or}_k|_{\rm
even}}$ is over all possible ordered splitting that the lengths of
all subsets are even. Furthermore, $M_k(\shuffle)$ for length-$r$
subset ${\rm Or}_k=\{\gamma_1,\gamma_2,\cdots \gamma_r\}$ is defined
as\footnote{The definition of $Z_i$ can be found in
\cite{Fu:2017uzt}. Since it has not been used  much in this paper,
we have not  reviewed it in the section two.}
\bea M_k(\shuffle)=(\epsilon_{\gamma_r}\cdot
\epsilon_{\gamma_{r-1}})(k_{\gamma_{r-1}}\cdot k_{\gamma_{r-2}})
(\epsilon_{\gamma_{r-2}}\cdot\epsilon_{\gamma_{r-3}})\cdots
(k_{\gamma_3}\cdot k_{\gamma_2})(\epsilon_{\gamma_2}\cdot
\epsilon_{\gamma_1}) (k_{\gamma_1}\cdot Z_{\gamma_1})\,. \eea
Acting ${\cal L}$ on $M_k$, it gives
\bea N_k(\shuffle)=(-)^{{|{\rm Or}_k|\over2}}(k_{\gamma_r}\cdot
k_{\gamma_{r-1}})(k_{\gamma_{r-1}}\cdot k_{\gamma_{r-2}})
(k_{\gamma_{r-2}}\cdot k_{\gamma_{r-3}})\cdots (k_{\gamma_3}\cdot
k_{\gamma_2})(k_{\gamma_2}\cdot k_{\gamma_1}) (k_{\gamma_1}\cdot
Z_{\gamma_1})\,. \eea
Then we get
\bea  A^{\rm {BI}}_n(\{g\})&=&\sum_{|{\rm
{Or}}_k|_{\rm{even}}}~\sum_{\shuffle}~\Big(\prod_{k=1}^t
N_k(\shuffle)\Big)A^{\rm{YM}}_n(1,{\rm {Or}}_1\shuffle {\rm
{Or}}_2\shuffle\cdots\shuffle{\rm {Or}}_t,n) .~~~~\label{exp-BI-4}
\eea

These two expressions \eref{exp-BI-3} and \eref{exp-BI-4} should be
equal if both  are  correct. Let us verify such an equality for the
simplest $4$-point example. For $4$-point case, \eref{exp-BI-3}
gives
\bea  A^{\rm BI}_4(\{1,2,3,4\})= (k_2\cdot k_1)(k_3\cdot (k_1+k_2))
A^{\rm YM}_4(1,2,3,4) +(k_2\cdot (k_1+k_3))(k_3\cdot k_1) A^{\rm
YM}_4(1,3,2,4)\,,~~~~\label{exp-BI-exa1} \eea
while \eref{exp-BI-4} gives
\bea  A^{\rm BI}_4(\{1,2,3,4\})=(k_2\cdot k_3)(k_3\cdot k_1) A^{\rm
YM}_4(1,3,2,4)\,,~~~\label{exp-BI-exa2} \eea
with the chosen order $2\prec 3$. These two results are equal to
each other only after using BCJ relations \eref{BCJ-exp-1}. The
example shows that although \eref{exp-BI-4} looks more complicated,
it gives shorter expressions. As an interesting remark, if we
reverse the logic, i.e., assuming the equivalence between two
differential operators,  we will arrive the equivalence of results
\eref{exp-BI-3} and \eref{exp-BI-4}. From them, we can derive the
generalized fundamental BCJ relations \eref{generalized-BCJ}.

\section{Expansion of sEYM amplitudes in the  BCJ-basis of YM amplitudes}

As pointed in the introduction, the expansion of sEYM amplitudes in
terms of YM amplitudes can be divided to two types: the expansion in
the KK-basis and the expansion in the BCJ-basis. Up to now,
expansion in the
 literatures are in the KK-basis, where the expansion coefficients
are much simpler. In this paper, we will initiate the investigation
of expanding in the BCJ-basis. There are two different approaches we
could think of. In the first approach, we start from the expansion
in the KK-basis \eref{eq:singletrace} and then transform the
KK-basis  to the BCJ-basis using \eref{BCJ-exp-1}. This approach is
straightforward, but as shown in this section, the algebraic
manipulations are not so trivial. In the second approach, encouraged
by results in the previous sections, we will use differential
operators. Both approaches will be explored in this paper. In this
section, we will take the first approach, while in the next section,
we will take the second approach. As emphasized in the introduction,
we are looking for a more efficient method to find the expansion
coefficients, which now will be a more complicated rational
functions.  Results obtained in this section will be
 compared with results found  in the next section.

\subsection{Case with one graviton}

We start with  the sEYM amplitudes with just a graviton. The
expansion in the KK-basis is given by
\bea A^{\rm EYM}_{n,1}(1,2,\ldots,n;p)=\sum_{\shuffle}
(\epsilon_p\cdot Y_p) A_{n+1}^{\rm
YM}(1,\{2,\ldots,n-1\}\shuffle\{p\},n)~.~~~\label{EYM1gra}\eea
 To expand in the BCJ-basis, we
use a trick, i.e., the gauge invariance of $p$, so we have
\bea 0&= & (k_p\cdot Y_p) A_{n+1}^{\rm
YM}(1,2,\{3,\ldots,n-1\}\shuffle\{p\},n) +(k_p\cdot k_1)
A_{n+1}^{\rm YM}(1,p,\{2,\ldots,n-1\},n)~. \eea
Solving the equation and putting it back, we have
\bea  A^{\rm EYM}_{n,1}(1,2,\ldots,n;p) & = & (\epsilon_p\cdot
Y_p)
A_{n+1}^{\rm YM}(1,2,\{3,\ldots,n-1\}\shuffle\{p\},n)\nn
& & -(\eps_p\cdot k_1){(k_p\cdot Y_p)\over (k_p\cdot k_1)}
A_{n+1}^{\rm YM}(1,2,\{3,\ldots,n-1\}\shuffle\{p\},n)\nn
& = &  {(\epsilon_p\cdot
Y_p) (k_p\cdot k_1)- (\eps_p\cdot k_1)(k_p\cdot Y_p)\over (k_p\cdot k_1)}A_{n+1}^{\rm YM}(1,2,\{3,\ldots,n-1\}\shuffle\{p\},n)~.\eea
Using the definition of field strength, we arrive
\bea  A^{\rm EYM}_{n,1}(1,2,\ldots,n;p) & = &  {k_1\cdot f_p\cdot
Y_p\over (k_p\cdot k_1)}A_{n+1}^{\rm YM}
(1,2,\{3,\ldots,n-1\}\shuffle\{p\},n)~. ~~~\label{My-1g-exp}\eea
The expansion coefficient's are rational functions with pole
$(k_p\cdot k_1)$, which is crucial to match up the structure of
physical poles at both sides.  The observation implies that the
expansion coefficients in the BCJ-basis will be, in general, the
rational functions with proper pole structures.

\subsection{Case with two gravitons}
Using the recursive expansion  \eref{C-single-exp}, we could write
down
\bea A^{\rm EYM}_{n,2}(1,2,\ldots,n;p,q) &=&\sum_{\shuffle}
(\epsilon_p\cdot
Y_p)A^{\rm EYM}_{n+1,1}(1,\{2,\ldots,n-1\}\shuffle\{p\},n;q)\nonumber\\
&&+\sum_{\shuffle} (\epsilon_p\cdot f_q\cdot Y_q)A_{n+2}^{\rm
YM}(1,\{2,\ldots,n-1\}\shuffle\{q,p\},n)~.~~~\label{TG2-manifest-1}\eea
The  gauge invariance of $p$ leads to the following equation
\bean 0&= &  (k_p\cdot
Y_p)A^{\rm EYM}_{n+1,1}(1,\{2,\ldots,n-1\}\shuffle\{p\},n;q)+(k_p\cdot f_q\cdot Y_q)A_{n+2}^{\rm
YM}(1,\{2,\ldots,n-1\}\shuffle\{q,p\},n)\nn
& = & (k_p\cdot
Y_p)A^{\rm EYM}_{n+1,1}(1,2,\{3,\ldots,n-1\}\shuffle\{p\},n;q)+(k_p\cdot
k_1)A^{\rm EYM}_{n+1,1}(1,p,\{2,\ldots,n-1\},n;q)\nn
& & + (k_p\cdot f_q\cdot Y_q)A_{n+2}^{\rm
YM}(1,\{2,\ldots,n-1\}\shuffle\{q,p\},n)~.\eean
Solving the $A^{\rm EYM}_{n+1,1}(1,p,\{2,\ldots,n-1\},n;q)$ and
putting it back, we get
\bea  A^{\rm EYM}_{n,2}(1,2,\ldots,n;p,q) & = & {(k_1\cdot f_p\cdot
Y_p)\over {\cal K}_{1p}}A^{\rm
EYM}_{n+1,1}(1,2,\{3,\ldots,n-1\}\shuffle\{p\},n;q) \nn & & +
{(k_1\cdot f_p\cdot f_q\cdot Y_q) \over {\cal K}_{1p}}A_{n+2}^{\rm
YM}(1,\{2,\ldots,n-1\}\shuffle\{q,p\},n)~.~~~~~~\label{TG2-manifest-2}\eea
Although we have not reached the expansion in the BCJ-basis, the
gauge invariance of all gravitons is manifest because the only
appearance of $f_p, f_q$.

For the first term of \eref{TG2-manifest-2}, we use the result
\eref{My-1g-exp} to reach
\bea T_1& = & {(k_1\cdot f_p\cdot Y_p)\over {\cal K}_{1p}}
{(k_1\cdot f_q\cdot X_q)\over {\cal K}_{1q}}A_{n+2}^{\rm
YM}(1,2,\{\{3,\ldots,n-1\}\shuffle\{p\}\}\shuffle\{q\},n)~,\eea
where it is important to notice that  since for $q$, the $p$ is just
a gluon, thus we must use the $X_q$ instead of the $Y_q$. Expanding
further to different orderings, we get
\bea T_1
& = & {(k_1\cdot f_p\cdot
Y_p)\over {\cal K}_{1p}} {(k_1\cdot f_q\cdot Y_q)\over {\cal K}_{1q}}A_{n+2}^{\rm YM}(1,2,\{3,\ldots,n-1\}\shuffle\{q,p\},n)\nn
& & + {(k_1\cdot f_p\cdot
Y_p)\over {\cal K}_{1p}} {(k_1\cdot f_q\cdot (Y_q+k_p))\over {\cal K}_{1q}}A_{n+2}^{\rm YM}(1,2,\{3,\ldots,n-1\}\shuffle\{p,q\},n)~.\eea
For the second term in \eref{TG2-manifest-2}, we split it to
\bea T_2 & = & {(k_1\cdot f_p\cdot f_q\cdot k_1) \over {\cal
K}_{1p}}A_{n+2}^{\rm YM}(1,q,p,\{2,\ldots,n-1\},n)\nn
& & + {(k_1\cdot f_p\cdot f_q\cdot k_1) \over {\cal K}_{1p}}A_{n+2}^{\rm
YM}(1,q,2,\{3,\ldots,n-1\}\shuffle\{p\},n)\nn
& & + {(k_1\cdot f_p\cdot f_q\cdot Y_q) \over {\cal K}_{1p}}A_{n+2}^{\rm
YM}(1,2,\{3,\ldots,n-1\}\shuffle\{q,p\},n)~.\eea
For the first term of $T_2$, we use the result \eref{G-BCJ-b=2-3} to
get
\bean T_{2,1}& =& {(k_1\cdot f_p\cdot f_q\cdot k_1) \over {\cal
K}_{1p}}\left\{ {(k_q\cdot k_1+k_p\cdot (Y_p+k_q))(k_q\cdot
(Y_q+k_p))\over {\cal K}_{1pq} {\cal K}_{1q}}  ~A_{n+1}^{\rm
YM}(1,2,\{3,\ldots,n-1\}\shuffle\{p,q\},n)\right.\nn
& & \left.+{(k_q\cdot (Y_q- k_1))(k_p\cdot (Y_p+k_q))\over {\cal K}_{1pq} {\cal K}_{1q}}~A^{\rm YM}_{n+2}(1,2,
\{3,...,n-1\}\shuffle\{q,p\},n)\right\}~.\eean
For the second term of $T_2$, we use \eref{Fund-BCJ-deform-1} to get
\bean T_{2,2}&= & {(k_1\cdot f_p\cdot f_q\cdot k_1) \over {\cal
K}_{1p}}{-(k_q\cdot X_q)\over (k_q\cdot k_1)}~ A_{n+2}^{\rm
YM}(1,2,\{\{3,\ldots,n-1\}\shuffle\{p\}\}\shuffle\{q\},n)\nn
&= & - {(k_1\cdot f_p\cdot f_q\cdot k_1) \over {\cal K}_{1p}}{(k_q\cdot Y_q)\over (k_q\cdot k_1)}~ A_{n+2}^{\rm
YM}(1,2,\{3,\ldots,n-1\}\shuffle\{q,p\},n)\nn
& &- {(k_1\cdot f_p\cdot f_q\cdot k_1) \over {\cal K}_{1p}}{(k_q\cdot (Y_q+k_p))\over (k_q\cdot k_1)}~ A_{n+2}^{\rm
YM}(1,2,\{3,\ldots,n-1\}\shuffle\{p,q\},n)~.\eean
Putting $T_{2,1}, T_{2,2}$ back, we get
\bea T_2&= & {(k_1\cdot f_p\cdot f_q\cdot k_1) [(k_q\cdot (Y_q- k_1))(k_p\cdot (Y_p+k_q))-(k_q\cdot Y_q) {\cal K}_{1pq}]\over {\cal K}_{1pq} {\cal K}_{1q}{\cal K}_{1p}}~A_{n+2}^{\rm
YM}(1,2,\{3,\ldots,n-1\}\shuffle\{q,p\},n) \nn
& & +{(k_1\cdot f_p\cdot f_q\cdot k_1)(k_p\cdot (Y_p-k_1))(k_q\cdot (Y_q+k_p)) \over {\cal K}_{1pq} {\cal K}_{1q}{\cal K}_{1p}}~ A_{n+2}^{\rm
YM}(1,2,\{3,\ldots,n-1\}\shuffle\{p,q\},n) \nn
& & + {(k_1\cdot f_p\cdot f_q\cdot Y_q) \over {\cal K}_{1p}}A_{n+2}^{\rm
YM}(1,2,\{3,\ldots,n-1\}\shuffle\{q,p\},n)~.\eea
Adding $T_1, T_2$ together, we have
\bea A^{\rm EYM}_{n,2}(1,2,\ldots,n;p,q)&= &\sum_{\shuffle} C(\{p,q\},\shuffle)
A_{n+2}^{\rm YM}(1,2,\{3,\ldots,n-1\}\shuffle\{p,q\},n)\nn
& & +\sum_{\shuffle} C(\{q,p\},\shuffle)A_{n+2}^{\rm
YM}(1,2,\{3,\ldots,n-1\}\shuffle\{q,p\},n)~~~ \label{EYM-2g-Exp}\eea
with coefficients
\bea C(\{p,q\},\shuffle) & = &  \left\{{(k_1\cdot f_p\cdot
Y_p)\over {\cal K}_{1p}} {(k_1\cdot f_q\cdot (Y_q+k_p))\over {\cal
K}_{1q}} +{(k_1\cdot f_p\cdot f_q\cdot k_1)(k_p\cdot
(Y_p-k_1))(k_q\cdot (Y_q+k_p)) \over {\cal K}_{1pq} {\cal
K}_{1q}{\cal K}_{1p}}\right\}\nn
C(\{q,p\},\shuffle)& = & \left\{{(k_1\cdot f_p\cdot Y_p)\over
{\cal K}_{1p}} {(k_1\cdot f_q\cdot Y_q)\over {\cal K}_{1q}} +
{(k_1\cdot f_p\cdot f_q\cdot Y_q) \over {\cal K}_{1p}}\right. \nn &
& \left.+{(k_1\cdot f_p\cdot f_q\cdot k_1) [(k_q\cdot (Y_q-
k_1))(k_p\cdot (Y_p+k_q))-(k_q\cdot Y_q) {\cal K}_{1pq}]\over {\cal
K}_{1pq} {\cal K}_{1q}{\cal K}_{1p}}\right\}~.\eea
The first coefficient can be rewritten as
\bea  C(\{p,q\},\shuffle)=\left\{{(k_1\cdot f_p\cdot X_p)\over {\cal
K}_{1p}} {(k_1\cdot f_q\cdot X_q)\over {\cal K}_{1q}} +{(k_1\cdot
f_p\cdot f_q\cdot k_1)(k_p\cdot (Y_p-k_1))(k_q\cdot X_q) \over {\cal
K}_{1pq} {\cal K}_{1q}{\cal K}_{1p}}\right\}~.~~\label{C2-pq}\eea
When  comparing the factor ${(k_p\cdot (Y_p-k_1))(k_q\cdot X_q)
\over {\cal K}_{1pq} {\cal K}_{1p}}$ in the second term of
$C(\{p,q\},\shuffle)$ with the second term of \eref{G-BCJ-b=2-3}, we
see they are the same. It is not a coincidence and we will see its
reasoning in the next section. The second coefficient
$C(\{q,p\},\shuffle)$ is not relate to the first coefficient by
permutation at the current form, thus we need to do some
manipulations. It is easy to see that the first two terms can be
rewritten as
\bean   {(k_1\cdot f_p\cdot X_p)\over {\cal K}_{1p}} {(k_1\cdot
f_q\cdot Y_q)\over {\cal K}_{1q}}+ {(k_1\cdot f_p\cdot f_q\cdot
k_1)(Y_q\cdot k_q) \over {\cal K}_{1p}{\cal K}_{1q}}~,\eean
where we have used the identity \eref{use-identity-1}
\bea (B\cdot f_p \cdot k_1)(k_q\cdot k_p)=(B\cdot f_p \cdot
k_q)(k_1\cdot k_p)+(k_q\cdot f_p \cdot k_1)(B\cdot
k_p)~.~~\label{use-identity-1} \eea
The identity \eref{use-identity-1} can be easily remembered as
following\footnote{The formula is also like the Schouten identity
$\Spaa{ab}\Spaa{cd}=\Spaa{ac}\Spaa{bd}+\Spaa{ad}\Spaa{cb}$.}: The
two elements $B, k_1$ in $(B\cdot f_p \cdot k_1)$ are exchanged with
the element $k_q$ in $(k_q\cdot k_p)$. Putting  it back,  the second
coefficient becomes
\bea  C(\{q,p\},\shuffle) & = & {(k_1\cdot f_p\cdot X_p)\over {\cal
K}_{1p}} {(k_1\cdot f_q\cdot Y_q)\over {\cal K}_{1q}} + {(k_1\cdot
f_p\cdot f_q\cdot k_1) (k_q\cdot (Y_q- k_1))(k_p\cdot
(Y_p+k_q))\over {\cal K}_{1pq} {\cal K}_{1q}{\cal
K}_{1p}}~~~\label{C2-qp}\eea
Now the permutation relation with $C(\{p,q\},\shuffle)$ is manifest.

Before ending this subsection, we give  some remarks for above
results:
\begin{itemize}

\item (1) First, the two coefficients \eref{C2-pq} and \eref{C2-qp}
are related to each other by a permutation as it should be. This
relation can also be used as a consistent check of our
calculations.

\item (2) Each coefficient contains two terms. The first  term is the
product of two factors ${k_1 f_i X_i\over {\cal K}_{1i}}$ with
$i=p,q$. The same factor  has also appeared in \eref{My-1g-exp}
for the case having just one graviton, and can be interpreted as
"turning a graviton to a gluon in the BCJ-basis"  in a loose
sense. The second term has the factor $(k_1 f_p f_q k_1)$, which
contains the contraction $\eps_p\cdot \eps_q$ as a singal of
the mutual interaction. Thus the physical picture of these two
terms is clear.

\item (3) One of important observations is that the factor $(k_1 f_p
f_q k_1)$ appears naturally in $C(\{p,q\},\shuffle)$, but not in
$C(\{q,p\},\shuffle)$. Only with some manipulations, we can
transfer the form $(k_1 f_p f_q Y_q)$ to $(k_1 f_p f_q k_1)$. As
we will see for the case with three gravitons, the same pattern
will appear again. In other words,  the contraction of
$\eps\cdot \eps$ can always be included inside the form $k_1
f...f k_1$. Observations given in the second and third points
provide information of the building blocks for expansion in the
BCJ-basis when using differential operators.

\item (4) Although it is straightforward, the whole calculation is
nontrivial. An indication is that in middle steps, the BCJ
coefficients \eref{BCJ-exp-1} have been used many times, thus
expressions in middle steps are lengthy and complicated.
However, when summing them up, cancelations happen and the final
result \eref{C2-pq} is very simple. Similar phenomena have been
met many times in the history of  scattering amplitudes. Thus we
believe that there should be a more efficient method where
cancelations in middle steps are automatically avoided.

\end{itemize}
%

\subsection{The case of three gravitons}
Using the recursive expansion \eref{C-single-exp}, the expansion in
the KK-basis  is given by
\bea & & A^{\rm EYM}_{n,3}(1,2,\ldots,n;p,q,r)\nn
&=&
(\epsilon_p\cdot
Y_p)A^{\rm EYM}_{n+1,2}(1,\{2,\ldots,n-1\}\shuffle\{p\},n;q,r)\nn
&&+(\epsilon_p\cdot f_q\cdot Y_q)A_{n+2,1}^{\rm
EYM}(1,\{2,\ldots,n-1\}\shuffle\{q,p\},n;r)\nn
& & +(\epsilon_p\cdot f_r\cdot Y_r)A_{n+2,1}^{\rm
EYM}(1,\{2,\ldots,n-1\}\shuffle\{r,p\},n;q)\nn
&&+(\epsilon_p\cdot f_q\cdot f_r\cdot Y_r)A_{n+3}^{\rm
YM}(1,\{2,\ldots,n-1\}\shuffle\{r,q,p\},n)\nn
& & +(\epsilon_p\cdot f_r\cdot f_q\cdot Y_q)A_{n+3}^{\rm
YM}(1,\{2,\ldots,n-1\}\shuffle\{q,r,p\},n)
.~~~\label{TG3-manifest-1}\eea
By the same trick, i.e., the gauge invariance of $p$, we  can solve
$ A^{\rm EYM}_{n+1,2}(1,p,\{2,3,\ldots,n-1\},n;q,r)$. After putting
it back to \eref{TG3-manifest-1} and simplifying, we reach
\bea & & A^{\rm EYM}_{n,3}(1,2,\ldots,n;p,q,r)\nn
&=&  {(k_1\cdot f_p\cdot
Y_p)\over (k_p\cdot
k_1)}A^{\rm EYM}_{n+1,2}(1,2,\{3,\ldots,n-1\}\shuffle\{p\},n;q,r)+\nn
&&+{(k_1\cdot f_p\cdot f_q\cdot Y_q)\over (k_p\cdot
k_1)}A_{n+2,1}^{\rm
EYM}(1,\{2,\ldots,n-1\}\shuffle\{q,p\},n;r)\nn
& & +{(k_1\cdot f_p\cdot f_r\cdot Y_r)\over (k_p\cdot
k_1)}A_{n+2,1}^{\rm
EYM}(1,\{2,\ldots,n-1\}\shuffle\{r,p\},n;q)\nn
&&+{(k_1\cdot f_p\cdot f_q\cdot f_r\cdot Y_r)\over (k_p\cdot
k_1)}A_{n+3}^{\rm
YM}(1,\{2,\ldots,n-1\}\shuffle\{r,q,p\},n)\nn
& & +{(k_1\cdot f_p\cdot f_r\cdot f_q\cdot Y_q)\over (k_p\cdot
k_1)}A_{n+3}^{\rm YM}(1,\{2,\ldots,n-1\}\shuffle\{q,r,p\},n)
~,~~\label{TG3-manifest-3}\eea
where since only  $f$'s appear, the gauge invariance of all
gravitons is manifest. To expand in the BCJ-basis, i.e.,
\bea & & A^{\rm EYM}_{n,3}(1,2,\ldots,n;p,q,r)\nn
& = & \sum_{\rho\in S_3}\sum_{\shuffle}C(\rho,\shuffle) A^{\rm
YM}_{n+2}(1,2,\{3,\ldots,n-1\}\shuffle \{\rho_1,\rho_2,\rho_3\},n)
 ~~~\label{3g-guess-1}\eea
with $\rho=\{\rho_1,\rho_2,\rho_3\}=\rho\{p,q,r\}$ the permutation
of $\{p,q,r\}$, we need to expand each term in \eref{TG3-manifest-3}
in that basis. Now we do it term by term.

~\\{\bf The first term:} For the first term, we can use the result
\eref{EYM-2g-Exp} directly only paying the attention to the meaning
of $Y_q, Y_r$ since for $q,r$, the $p$ likes a gluon. Thus we have
\bean T_1 & = & {(k_1\cdot f_p\cdot Y_p)\over (k_p\cdot
k_1)}A_{n+3}^{\rm
YM}(1,2,\{\{3,\ldots,n-1\}\shuffle\{p\}\}\shuffle\{r,q\},n)\nn
& & \times \left\{{(k_1\cdot f_r\cdot X_r)\over {\cal K}_{1r}}
{(k_1\cdot f_q\cdot X_q)\over {\cal K}_{1q}} +{(k_1\cdot f_r\cdot
f_q\cdot k_1)(k_r\cdot (X_r-k_1))(k_q\cdot X_q) \over {\cal K}_{1qr}
{\cal K}_{1q}{\cal K}_{1r}}\right\}\nn
& & + {(k_1\cdot f_p\cdot Y_p)\over (k_p\cdot k_1)} A_{n+3}^{\rm
YM}(1,2,\{\{3,\ldots,n-1\}\shuffle\{p\}\}\shuffle\{q,r\},n)  \nn & &
\times \left\{{(k_1\cdot f_r\cdot X_r)\over {\cal K}_{1r}}
{(k_1\cdot f_q\cdot X_q)\over {\cal K}_{1q}} + {(k_1\cdot f_q\cdot
f_r\cdot k_1) (k_q\cdot (X_q- k_1))(k_r\cdot X_r)\over {\cal
K}_{1qr} {\cal K}_{1q}{\cal K}_{1r}}\right\}~.\eean
Now we can find contributions to six coefficients in
\eref{3g-guess-1}. Since we expect they are related to each other by
permutations, we will only write down  contributions to the
coefficient $C(\{p,q,r\})$ for simplicity\footnote{As pointed out in
the previous subsection, for a given choice of fiducial graviton in
the KK-basis expansion, some orderings will have simpler expressions
for coefficients. This is the reason we consider this ordering.}
\bea {C}^{T_1}(\{p,q,r\}) & = &{(k_1\cdot f_p\cdot Y_p)\over
(k_p\cdot k_1)}{(k_1\cdot f_r\cdot (Y_r+k_p+k_q))\over {\cal
K}_{1r}} {(k_1\cdot f_q\cdot (Y_q+k_p))\over {\cal K}_{1q}}\nn & & +
{(k_1\cdot f_p\cdot Y_p)\over (k_p\cdot k_1)}{(k_1\cdot f_q\cdot
f_r\cdot k_1) (k_q\cdot (Y_q+k_p- k_1))(k_r\cdot (Y_r+k_p+k_q))\over
{\cal K}_{1qr} {\cal K}_{1q}{\cal
K}_{1r}}~.~~\label{pqr-coeff-T1}\eea

~\\{\bf The second term:} The  second term can be split to
\bean T_2
& = &  {(k_1\cdot f_p\cdot f_q\cdot Y_q)\over (k_p\cdot
k_1)}A_{n+2,1}^{\rm EYM}(1,2,\{3,\ldots,n-1\}\shuffle\{q,p\},n;r)\nn
&&+{(k_1\cdot f_p\cdot f_q\cdot k_1)\over (k_p\cdot
k_1)}A_{n+2,1}^{\rm
EYM}(1,q,\{2,3,\ldots,n-1\}\shuffle\{p\},n;r)~.\eean
Among these two terms,  using the result \eref{My-1g-exp}, the first
term can be expanded to
\bean T_{2,1} & = & {(k_1\cdot f_p\cdot f_q\cdot Y_q)\over {\cal
K}_{1p}}{(k_1\cdot f_r\cdot X_r)\over {\cal K}_{1r}} A_{n+3}^{\rm
YM}(1,2,\{\{3,\ldots,n-1\}\shuffle\{q,p\}\}\shuffle\{r\},n)~.\eean
It is important to notice that this term does not contribute to the
coefficient ${ C}(\{p,q,r\})$, thus we can forget it at this
moment\footnote{This is also the reason we consider the coefficient
${ C}(\{p,q,r\})$ for simplicity. Although other
coefficients will be similar, but to lead to the pattern observed
from this coefficient, nontrivial algebraic manipulations will be
needed.}. The second term can be expanded to
\bean T_{2,2}& =&{(k_1\cdot f_p\cdot f_q\cdot k_1)\over (k_p\cdot
k_1)}
 {(k_1\cdot f_r\cdot X_r)\over {\cal K}_{1r}}
 A_{n+3}^{\rm
 YM}(1,q,\{2,3,\ldots,n-1\}\shuffle\{p\}\shuffle\{r\},n)\nn
 & = & {(k_1\cdot f_p\cdot f_q\cdot k_1)\over (k_p\cdot
k_1)}
 {(k_1\cdot f_r\cdot (Y_r+k_q+k_p))\over {\cal K}_{1r}}
 A_{n+3}^{\rm
 YM}(1,q,\{2,3,\ldots,n-1\}\shuffle\{p,r\},n)\nn
 & & + {(k_1\cdot f_p\cdot f_q\cdot k_1)\over (k_p\cdot
k_1)}
 {(k_1\cdot f_r\cdot (Y_r+k_q))\over {\cal K}_{1r}}
 A_{n+3}^{\rm
 YM}(1,q,\{2,3,\ldots,n-1\}\shuffle\{r,p\},n)~,\eean
where to distinguish different situations, it is crucial to expand
$X_r$ to $Y_r$. Among these two terms, the first one  needs to be
expanded  further as
\bea T_{2,2}[1]& = &{(k_1\cdot f_p\cdot f_q\cdot k_1)\over {\cal
K}_{1p}}
 {(k_1\cdot f_r\cdot (Y_r+k_q+k_p))\over {\cal K}_{1r}}
 A_{n+3}^{\rm
 YM}(1,q,2,\{3,\ldots,n-1\}\shuffle\{p,r\},n)\nn
 & & + {(k_1\cdot f_p\cdot f_q\cdot k_1)\over {\cal
K}_{1p}}
 {(k_1\cdot f_r\cdot (Y_r+k_q+k_p))\over {\cal K}_{1r}}
 A_{n+3}^{\rm
 YM}(1,q,p,2,\{3,\ldots,n-1\}\shuffle\{r\},n)\nn
 & & + {(k_1\cdot f_p\cdot f_q\cdot k_1)\over {\cal
K}_{1p}}
 {(k_1\cdot f_r\cdot (k_1+k_q+k_p))\over {\cal K}_{1r}}
 A_{n+3}^{\rm
 YM}(1,q,p,r,\{2,3,\ldots,n-1\},n)~.~~\label{C3-T22-mid-1}\eea
 Using the \eref{Fund-BCJ-deform-1}, we can read
out the contribution to the coefficient ${C}(\{p,q,r\})$ from the
first term as
\bea { C}^{T_{2,2}[1,1]}(\{p,q,r\}) & = & - { (k_q\cdot
(Y_q+k_p))\over {\cal K}_{1q}} {(k_1\cdot f_p\cdot f_q\cdot
k_1)\over {\cal K}_{1p}}
 {(k_1\cdot f_r\cdot (Y_r+k_q+k_p))\over {\cal K}_{1r}}~.~~\label{pqr-coeff-T2211}\eea
Using the \eref{G-BCJ-b=2-3}, we can read out the contribution to
the coefficient ${C}(\{p,q,r\})$ from the second term as
\bea {C}^{T_{2,2}[1,2]}(\{p,q,r\}) & = & {(k_q\cdot
k_1+k_p\cdot (Y_p+k_q))(k_q\cdot (Y_q+k_p))\over {\cal K}_{1pq}
{\cal K}_{1q}}  {(k_1\cdot f_p\cdot f_q\cdot k_1)\over {\cal
K}_{1p}}
 {(k_1\cdot f_r\cdot (Y_r+k_q+k_p))\over {\cal K}_{1r}}~.~~~~\label{pqr-coeff-T2212}\eea
Using the \eref{BCJ-pqr-collect}, we can read out the contribution
to the coefficient ${C}_{3}(\{p,q,r\})$ from the third term
as\footnote{Now it is clear that why in \eref{C3-T22-mid-1} we write
$X_r$ out explicitly, since when we use \eref{BCJ-pqr-collect} to
get \eref{pqr-coeff-T2213}, there is another $X_r$ appears. The
meaning of these two $X_r$'s is different and we must make clear
distinction.}
\bea {C}^{T_{2,2}[1,3]}(\{p,q,r\}) & = & - { k_q\cdot
X_q\over {\cal K}_{1q}}  { (k_p\cdot (Y_p-k_1))\over {\cal
K}_{1pq}} { k_r\cdot X_r\over {\cal K}_{1pqr}}  {(k_1\cdot
f_p\cdot f_q\cdot k_1)\over {\cal K}_{1p}}
 {(k_1\cdot f_r\cdot (k_1+k_q+k_p))\over {\cal K}_{1r}}~.~~~~~~\label{pqr-coeff-T2213}\eea
Now we sum up above three coefficients. The sum of the first two is
\bea {C}^{T_{2,2}[1,1]+T_{2,2}[1,2]}(\{p,q,r\})
& = &  { (k_q\cdot X_q)\over {\cal K}_{1q}} {(k_1\cdot f_p\cdot
f_q\cdot k_1)\over {\cal K}_{1p}}
 {(k_1\cdot f_r\cdot X_r)\over {\cal K}_{1r}}{(k_p\cdot (Y_p-k_1))\over {\cal K}_{1pq}}~.\eea
Adding up the third one we get
\bea & & { C}^{T_{2,2}[1]}(\{p,q,r\})  =    {(k_1\cdot
f_p\cdot f_q\cdot k_1)(k_p\cdot (Y_p-k_1))(k_q\cdot (Y_q+k_p))\over
{\cal K}_{1p}{\cal K}_{1q} {\cal K}_{1pq}}\times {(k_1\cdot f_r\cdot
(Y_r+k_p+k_q))\over {\cal K}_{1r}}
 \nn
 & & -{(k_1\cdot
f_p\cdot f_q\cdot k_1)(k_p\cdot (Y_p-k_1))(k_q\cdot (Y_q+k_p))\over
{\cal K}_{1p}{\cal K}_{1q} {\cal K}_{1pq}}\times
 { (k_r\cdot X_r)\over {\cal K}_{1pqr}}{(k_1\cdot f_r\cdot (k_1+k_q+k_p))
 \over {\cal K}_{1r}}~.\eea
For the second term of $T_{2,2}$, we can expand to another three terms
\bean T_{2,2}[2]& = &{(k_1\cdot f_p\cdot f_q\cdot k_1)\over {\cal
K}_{1p}}
 {(k_1\cdot f_r\cdot (Y_r+k_q))\over {\cal K}_{1r}}
 A_{n+3}^{\rm
 YM}(1,q,2,\{3,\ldots,n-1\}\shuffle\{r,p\},n)\nn
 & & + {(k_1\cdot f_p\cdot f_q\cdot k_1)\over {\cal
K}_{1p}}
 {(k_1\cdot f_r\cdot (k_1+k_q))\over {\cal K}_{1r}}
 A_{n+3}^{\rm
 YM}(1,q,r,2,\{3,\ldots,n-1\}\shuffle\{p\},n)\nn
 & & + {(k_1\cdot f_p\cdot f_q\cdot k_1)\over {\cal
K}_{1p}}
 {(k_1\cdot f_r\cdot (k_1+k_q))\over {\cal K}_{1r}}
 A_{n+3}^{\rm
 YM}(1,q,r,p,\{2,3,\ldots,n-1\},n)~.\eean
The first term does not contribute to the coefficient
$C(\{p,q,r\})$. Using \eref{G-BCJ-b=2-3} and \eref{BCJ-pqr-collect},
the  contributions from the second and third terms are
\bea & & {C}^{T_{2,2}[2]}(\{p,q,r\})  = {(k_1\cdot f_p\cdot
f_q\cdot k_1)\over {\cal K}_{1p}}
 {(k_1\cdot f_r\cdot (k_1+k_q))\over {\cal K}_{1r}}{(k_q\cdot (Y_q+k_p- k_1))(k_r\cdot (Y_r+k_p+k_q))\over {\cal K}_{1qr}
{\cal K}_{1q}}\nn
& & -{(k_1\cdot f_p\cdot f_q\cdot k_1)\over {\cal K}_{1p}}
 {(k_1\cdot f_r\cdot (k_1+k_q))\over {\cal K}_{1r}}  { (k_q\cdot (X_q-k_1))\over
{\cal K}_{1q}}  {(k_r\cdot X_r) \over {\cal K}_{1qr}}{
(k_p\cdot (Y_p-k_1)-{\cal K}_{1pqr})\over {\cal K}_{1pqr}}\nn
& = & -{(k_1\cdot f_p\cdot f_q\cdot k_1)\over {\cal K}_{1p}}
 {(k_1\cdot f_r\cdot (k_1+k_q))\over {\cal K}_{1r}}{(k_r\cdot X_r) \over {\cal K}_{1qr}} { (k_q\cdot (X_q-k_1))\over
{\cal K}_{1q}}{(k_p\cdot (Y_p-k_1))\over {\cal K}_{1pqr}}~.\eea
Putting together, we get
\bea & & C^{T_{2,2}}(\{p,q,r\})  =    {(k_1\cdot f_p\cdot
f_q\cdot k_1)(k_p\cdot (Y_p-k_1))(k_q\cdot (Y_q+k_p))\over {\cal
K}_{1p}{\cal K}_{1q} {\cal K}_{1pq}} {(k_1\cdot f_r\cdot
(Y_r+k_p+k_q))\over {\cal K}_{1r}}
 \nn
 & & -{(k_1\cdot
f_p\cdot f_q\cdot k_1)(k_p\cdot (Y_p-k_1))(k_q\cdot (Y_q+k_p))\over
{\cal K}_{1p}{\cal K}_{1q} {\cal K}_{1pq}}
 { (k_r\cdot X_r)\over {\cal K}_{1pqr}}{(k_1\cdot f_r\cdot (k_1+k_q+k_p))
 \over {\cal K}_{1r}}\nn
 & & -{(k_1\cdot f_p\cdot f_q\cdot k_1)(k_p\cdot (Y_p-k_1))(k_q\cdot (X_q-k_1))
 \over {\cal K}_{1p}{\cal K}_{1q}}
 {(k_1\cdot f_r\cdot (k_1+k_q))\over {\cal K}_{1r}}{(k_r\cdot X_r) \over {\cal K}_{1qr}
{\cal K}_{1pqr}}~.\eea

~\\{\bf The third term:} For the third term, we expand further as
\bea T_3
& = & {(k_1\cdot f_p\cdot f_r\cdot Y_r)\over (k_p\cdot
k_1)}A_{n+2,1}^{\rm EYM}(1,2,\{3,\ldots,n-1\}\shuffle\{r,p\},n;q)\nn
& & + {(k_1\cdot f_p\cdot f_r\cdot k_1)\over (k_p\cdot
k_1)}A_{n+2,1}^{\rm EYM}(1,r,\{2,\ldots,n-1\}\shuffle\{p\},n;q)~.\eea
Again, the first term does not contribute to the coefficient  ${C}(\{p,q,r\})$. For the second term, using the result
\eref{My-1g-exp} we get
\bea T_{3,2} & = & {(k_1\cdot f_p\cdot f_r\cdot k_1)\over (k_p\cdot
k_1)} {(k_1\cdot f_q\cdot X_q)\over {\cal K}_{1q}} A_{n+3}^{\rm
YM}(1,r,\{2,\ldots,n-1\}\shuffle\{p\}\shuffle\{q\},n)\nn
& = & {(k_1\cdot f_p\cdot f_r\cdot k_1)\over (k_p\cdot k_1)}
{(k_1\cdot f_q\cdot X_q)\over {\cal K}_{1q}} A_{n+3}^{\rm
YM}(1,r,\{2,\ldots,n-1\}\shuffle\{p,q\},n)\nn
& & + {(k_1\cdot f_p\cdot f_r\cdot k_1)\over (k_p\cdot k_1)}
{(k_1\cdot f_q\cdot X_q)\over {\cal K}_{1q}} A_{n+3}^{\rm
YM}(1,r,\{2,\ldots,n-1\}\shuffle\{q,p\},n)~.\eea
Now we treat term by the term. For the first term, we expand it
further as
\bea T_{3,2}[1]
& = & {(k_1\cdot f_p\cdot f_r\cdot k_1)\over (k_p\cdot k_1)}
{(k_1\cdot f_q\cdot (Y_q+k_r+k_p))\over {\cal K}_{1q}} A_{n+3}^{\rm
YM}(1,r,2,\{3,\ldots,n-1\}\shuffle\{p,q\},n)\nn
& &+ {(k_1\cdot f_p\cdot f_r\cdot k_1)\over (k_p\cdot k_1)}
{(k_1\cdot f_q\cdot (Y_q+k_r+k_p))\over {\cal K}_{1q}} A_{n+3}^{\rm
YM}(1,r,p,2,\{3,\ldots,n-1\}\shuffle\{q\},n)\nn
& & + {(k_1\cdot f_p\cdot f_r\cdot k_1)\over (k_p\cdot k_1)}
{(k_1\cdot f_q\cdot (k_1+k_r+k_p))\over {\cal K}_{1q}} A_{n+3}^{\rm
YM}(1,r,p,q,\{2,\ldots,n-1\},n)~.\eea
Reading out contributions to the coefficient ${C}(\{p,q,r\})$ we get
\bea & & {
C}^{T_{3,2}[1]}(\{p,q,r\})={(k_1\cdot f_p\cdot f_r\cdot k_1)\over (k_p\cdot k_1)}
{(k_1\cdot f_q\cdot (Y_q+k_r+k_p))\over {\cal K}_{1q}} \times {-(k_r\cdot X_r)\over {\cal K}_{1r}}\nn
& & + {(k_1\cdot f_p\cdot f_r\cdot k_1)\over (k_p\cdot k_1)}
{(k_1\cdot f_q\cdot (Y_q+k_r+k_p))\over {\cal K}_{1q}}\times {(k_r\cdot k_1+k_p\cdot (Y_p+k_r))(k_r\cdot (Y_r+k_p+k_q))\over
{\cal K}_{1pr} {\cal K}_{1r}}\nn
& & +{(k_1\cdot f_p\cdot f_r\cdot k_1)\over (k_p\cdot k_1)}
{(k_1\cdot f_q\cdot (k_1+k_r+k_p))\over {\cal K}_{1q}}\times  { -(k_r\cdot ( Y_r+k_p+k_q))(k_q\cdot (Y_q+k_r+k_p))(k_p\cdot (Y_p-k_1))\over {\cal K}_{1pqr}{\cal K}_{1pr} {\cal K}_{1r} }\nn
& = &  {(k_1\cdot f_p\cdot f_r\cdot k_1)\over (k_p\cdot k_1)}
{(k_1\cdot f_q\cdot (Y_q+k_r+k_p))\over {\cal K}_{1q}} {(k_p\cdot (Y_p-k_1))(k_r\cdot X_r)\over
{\cal K}_{1pr} {\cal K}_{1r}}\nn
& & -{(k_1\cdot f_p\cdot f_r\cdot k_1)\over (k_p\cdot k_1)}
{(k_1\cdot f_q\cdot (k_1+k_r+k_p))\over {\cal K}_{1q}}  { (k_r\cdot X_r)(k_q\cdot (Y_q+k_r+k_p))(k_p\cdot (Y_p-k_1))\over {\cal K}_{1pqr}{\cal K}_{1pr} {\cal K}_{1r} }~. \eea
For the second term we expand to
\bea T_{3,2}[2]
& = & {(k_1\cdot f_p\cdot f_r\cdot k_1)\over (k_p\cdot k_1)}
{(k_1\cdot f_q\cdot (Y_q+k_r))\over {\cal K}_{1q}} A_{n+3}^{\rm
YM}(1,r,2,\{3,\ldots,n-1\}\shuffle\{q,p\},n)\nn
& & + {(k_1\cdot f_p\cdot f_r\cdot k_1)\over (k_p\cdot k_1)}
{(k_1\cdot f_q\cdot (k_1+k_r))\over {\cal K}_{1q}} A_{n+3}^{\rm
YM}(1,r,q,2,\{3,\ldots,n-1\}\shuffle\{p\},n)\nn
& & + {(k_1\cdot f_p\cdot f_r\cdot k_1)\over (k_p\cdot k_1)}
{(k_1\cdot f_q\cdot (k_1+k_r))\over {\cal K}_{1q}} A_{n+3}^{\rm
YM}(1,r,q,p,\{2,\ldots,n-1\},n)~.\eea
Reading out contributions to the coefficient ${ C}(\{p,q,r\})$ we
get (the first term does not contribute)
\bea & & {
C}^{T_{3,2}[2]}(\{p,q,r\})=  {(k_1\cdot f_p\cdot f_r\cdot k_1)\over (k_p\cdot k_1)}
{(k_1\cdot f_q\cdot (k_1+k_r))\over {\cal K}_{1q}} \times {(k_r\cdot k_1+k_q\cdot (Y_q+k_p+k_r))(k_r\cdot (Y_r+k_q+k_p))\over
{\cal K}_{1qr} {\cal K}_{1r}}\nn
& & + {(k_1\cdot f_p\cdot f_r\cdot k_1)\over (k_p\cdot k_1)}
{(k_1\cdot f_q\cdot (k_1+k_r))\over {\cal K}_{1q}}\times  { -k_r\cdot X_r\over {\cal
K}_{1r}} \times { -k_q\cdot (X_q-k_1)-{\cal K}_{1qr} \over {\cal
K}_{1qr}} \times{ -k_p\cdot (Y_p-k_1)-{\cal K}_{1pqr}\over {\cal
K}_{1pqr}}\nn
& = & -{(k_1\cdot f_p\cdot f_r\cdot k_1)\over (k_p\cdot k_1)}
{(k_1\cdot f_q\cdot (k_1+k_r))\over {\cal K}_{1q}} { (k_r\cdot X_r)\over {\cal
K}_{1r}}{ (k_q\cdot (X_q-k_1)+{\cal K}_{1qr}) \over {\cal
K}_{1qr}}{ (k_p\cdot (Y_p-k_1))\over {\cal
K}_{1pqr}}~.\eea
Putting two parts together we get
\bea & & {
C}^{T_{3,2}}(\{p,q,r\})=  {(k_1\cdot f_p\cdot f_r\cdot k_1)\over (k_p\cdot k_1)}
{(k_1\cdot f_q\cdot (Y_q+k_r+k_p))\over {\cal K}_{1q}} {(k_p\cdot (Y_p-k_1))(k_r\cdot X_r)\over
{\cal K}_{1pr} {\cal K}_{1r}}\nn
& & -{(k_1\cdot f_p\cdot f_r\cdot k_1)\over (k_p\cdot k_1)}
{(k_1\cdot f_q\cdot (k_1+k_r+k_p))\over {\cal K}_{1q}}  { (k_r\cdot X_r)(k_q\cdot (Y_q+k_r+k_p))(k_p\cdot (Y_p-k_1))\over {\cal K}_{1pqr}{\cal K}_{1pr} {\cal K}_{1r} }\nn
& & -{(k_1\cdot f_p\cdot f_r\cdot k_1)\over (k_p\cdot k_1)}
{(k_1\cdot f_q\cdot (k_1+k_r))\over {\cal K}_{1q}} { (k_r\cdot X_r)\over {\cal
K}_{1r}}{ (k_q\cdot (X_q-k_1)+{\cal K}_{1qr}) \over {\cal
K}_{1qr}}{ (k_p\cdot (Y_p-k_1))\over {\cal
K}_{1pqr}}~. \eea

~\\ {\bf The fourth term:} The fourth term can be expanded to four
terms:
\bea T_4
& =  &{(k_1\cdot f_p\cdot f_q\cdot f_r\cdot Y_r)\over (k_p\cdot
k_1)}A_{n+3}^{\rm YM}(1,2,\{3,\ldots,n-1\}\shuffle\{r,q,p\},n)\nn
& +& {(k_1\cdot f_p\cdot f_q\cdot f_r\cdot k_1)\over (k_p\cdot
k_1)}A_{n+3}^{\rm YM}(1,r,2,\{3,\ldots,n-1\}\shuffle\{q,p\},n)\nn
& + & {(k_1\cdot f_p\cdot f_q\cdot f_r\cdot k_1)\over (k_p\cdot
k_1)}A_{n+3}^{\rm YM}(1,r,q,2,\{3,\ldots,n-1\}\shuffle\{p\},n)\nn
& + & {(k_1\cdot f_p\cdot f_q\cdot f_r\cdot k_1)\over (k_p\cdot
k_1)}A_{n+3}^{\rm YM}(1,r,q,p,\{2,\ldots,n-1\},n)~.\eea
The first and the second terms do not contribute to the coefficient
${ C}(\{p,q,r\})$. The remaining two terms give
\bea & & {
C}^{T_4}(\{p,q,r\})={(k_1\cdot f_p\cdot f_q\cdot f_r\cdot k_1)\over (k_p\cdot
k_1)}\times {(k_r\cdot k_1+k_q\cdot (Y_q+k_r+k_p))(k_r\cdot (Y_r+k_q+k_p))\over
{\cal K}_{1qr} {\cal K}_{1r}} \nn
& & +{(k_1\cdot f_p\cdot f_q\cdot f_r\cdot k_1)\over (k_p\cdot
k_1)}\times { -k_r\cdot X_r\over {\cal
K}_{1r}} \times { -k_q\cdot (X_q-k_1)-{\cal K}_{1qr} \over {\cal
K}_{1qr}} \times{ -k_p\cdot (Y_p-k_1)-{\cal K}_{1pqr}\over {\cal
K}_{1pqr}}\nn
& = & -{(k_1\cdot f_p\cdot f_q\cdot f_r\cdot k_1)\over (k_p\cdot
k_1)} { (k_r\cdot X_r)\over {\cal K}_{1r}} { (k_q\cdot
(X_q-k_1)+{\cal K}_{1qr}) \over {\cal K}_{1qr}}{ (k_p\cdot
(Y_p-k_1))\over {\cal K}_{1pqr}} ~.~~\label{C3[pqr]-T4} \eea

~\\ {\bf The fifth term:} We expand it to four terms
\bea T_5
& = & {(k_1\cdot f_p\cdot f_r\cdot f_q\cdot Y_q)\over (k_p\cdot
k_1)}A_{n+3}^{\rm YM}(1,2,\{3,\ldots,n-1\}\shuffle\{q,r,p\},n)\nn
& +& {(k_1\cdot f_p\cdot f_r\cdot f_q\cdot k_1)\over (k_p\cdot
k_1)}A_{n+3}^{\rm YM}(1,q,2,\{3,\ldots,n-1\}\shuffle\{r,p\},n) \nn
&+ &{(k_1\cdot f_p\cdot f_r\cdot f_q\cdot k_1)\over (k_p\cdot
k_1)}A_{n+3}^{\rm YM}(1,q,r,2,\{3,\ldots,n-1\}\shuffle\{p\},n)\nn
& + & {(k_1\cdot f_p\cdot f_r\cdot f_q\cdot k_1)\over (k_p\cdot
k_1)}A_{n+3}^{\rm YM}(1,q,r,p,\{2,\ldots,n-1\},n)~. \eea
Again the first and the second terms do not contribute to
coefficient ${ C}(\{p,q,r\})$. The remaining two terms give
\bea & & {C}^{T_5}(\{p,q,r\})= {(k_1\cdot f_p\cdot f_r\cdot
f_q\cdot k_1)\over (k_p\cdot k_1)}\times {(k_q\cdot (Y_q+k_p-
k_1))(k_r\cdot (Y_r+k_q+k_p))\over {\cal K}_{1qr} {\cal
K}_{1q}}\nn
& & + {(k_1\cdot f_p\cdot f_r\cdot f_q\cdot k_1)\over (k_p\cdot
k_1)}\times { -(k_q\cdot (X_q-k_1))\over {\cal K}_{1q}} \times
{-k_r\cdot X_r \over {\cal K}_{1qr}} \times{ -k_p\cdot
(Y_p-k_1)-{\cal K}_{1pqr}\over {\cal K}_{1pqr}}\nn
& = & - {(k_1\cdot f_p\cdot f_r\cdot f_q\cdot k_1)\over (k_p\cdot
k_1)}{ (k_q\cdot (X_q-k_1))\over {\cal K}_{1q}}{(k_r\cdot X_r) \over
{\cal K}_{1qr}}{ (k_p\cdot (Y_p-k_1))\over {\cal
K}_{1pqr}}~.~~\label{C3[pqr]-T5} \eea

~\\{\bf In total:} Now we sum up these contributions and simplify
them to {\small
\bea & & {(k_1\cdot f_p\cdot X_p)\over {\cal K}_{1p}}
{(k_1\cdot f_q\cdot X_q)\over {\cal K}_{1q}} {(k_1\cdot
f_r\cdot X_r)\over {\cal K}_{1r}} +{(k_1\cdot f_p\cdot X_p)\over
(k_p\cdot k_1)} {(k_1\cdot f_q\cdot f_r\cdot k_1) (k_q\cdot
(X_q- k_1))(k_r\cdot X_r)\over {\cal K}_{1qr} {\cal K}_{1q}{\cal
K}_{1r}}\nn
& & + {(k_1\cdot f_p\cdot f_q\cdot k_1)(k_p\cdot (X_p-k_1))(k_q\cdot
X_q)\over {\cal K}_{1p}{\cal K}_{1q} {\cal K}_{1pq}} {(k_1\cdot
f_r\cdot X_r)\over {\cal K}_{1r}} +{(k_1\cdot f_p\cdot f_r\cdot
k_1)(k_p\cdot (X_p-k_1))(k_r\cdot X_r)\over {\cal K}_{1p} {\cal
K}_{1pr} {\cal K}_{1r}}{(k_1\cdot f_q\cdot X_q)\over {\cal
K}_{1q}}\nn
& &- {(k_p\cdot (X_p-k_1))(k_r\cdot X_r)\over {\cal K}_{1p} {\cal
K}_{1q} {\cal K}_{1r}{\cal K}_{1pqr}} \left\{ {(k_1\cdot f_p\cdot
f_q\cdot k_1)(k_1\cdot f_r\cdot k_p)(k_q\cdot X_q)\over {\cal
K}_{1pq}}+{(k_1\cdot f_p\cdot f_r\cdot k_1)(k_1\cdot f_q\cdot k_p)
(k_q\cdot (X_q+k_r))\over {\cal K}_{1pr}}\right.\nn
& & + (k_1\cdot f_p\cdot f_q\cdot k_1)(k_1\cdot f_r\cdot k_q)
\left\{ {(k_q\cdot X_q)\over {\cal K}_{1pq}}+ {(k_q\cdot (X_q-k_1))
  \over {\cal K}_{1qr}}\right\}+{(k_1\cdot f_p\cdot f_q\cdot f_r\cdot k_1)(k_1\cdot k_q)
(k_q\cdot (X_q-k_1)+{\cal K}_{1qr}) \over {\cal K}_{1qr}}\nn
& &\left. +(k_1\cdot f_p\cdot f_r\cdot k_1)(k_1\cdot f_q\cdot k_r)
\left\{ { (k_q\cdot (Y_q-k_1))\over {\cal K}_{1pr}}+{ (k_q\cdot
(X_q-k_1)) \over  {\cal K}_{1qr}} \right\}+ {(k_1\cdot f_p\cdot
f_r\cdot f_q\cdot k_1)(k_1\cdot k_r) (k_q\cdot (X_q-k_1))\over {\cal
K}_{1qr}}\right\}\nn
~~~~\label{3g-sum}  \eea
}
We can simplify further, by using
\bea (k_1\cdot f_p\cdot f_q\cdot f_r\cdot k_1)(k_1\cdot k_q)& = &
(k_1\cdot f_p\cdot f_q\cdot k_1)(k_q\cdot f_r\cdot k_1)+(k_1\cdot
f_p\cdot k_q)(k_1\cdot f_q\cdot f_r\cdot k_1)\nn
(k_1\cdot f_p\cdot f_r\cdot f_q\cdot k_1)(k_1\cdot k_r)& = &
(k_1\cdot f_p\cdot f_r\cdot k_1)(k_r\cdot f_q\cdot k_1)+(k_1\cdot
f_p\cdot k_r)(k_1\cdot f_r\cdot f_q\cdot k_1)~~~\label{ff-rel}\eea
to get
\bea  { C}(\{p,q,r\})=A_1+A_2+A_3~~~~\label{3g-nice}\eea
where
\bea A_1& =& {(k_1\cdot f_p\cdot X_p)\over {\cal K}_{1p}} {(k_1\cdot
f_q\cdot X_q)\over {\cal K}_{1q}}{(k_1\cdot f_r\cdot X_r)\over {\cal
K}_{1r}}~,~~~\label{3g-nice-A} \eea
\bea A_2 & = & {(k_1\cdot f_p\cdot X_p)\over
(k_p\cdot k_1)}\times {(k_1\cdot f_q\cdot f_r\cdot k_1) (k_q\cdot
(X_q- k_1))(k_r\cdot X_r)\over {\cal K}_{1qr} {\cal K}_{1q}{\cal
K}_{1r}}\nn
& & + {(k_1\cdot f_p\cdot f_q\cdot k_1)(k_p\cdot (X_p-k_1))(k_q\cdot
X_q)\over {\cal K}_{1p}{\cal K}_{1q} {\cal K}_{1pq}} {(k_1\cdot
f_r\cdot X_r)\over {\cal K}_{1r}}\nn
& & +{(k_1\cdot f_p\cdot f_r\cdot k_1)(k_p\cdot (X_p-k_1))(k_r\cdot
X_r)\over {\cal K}_{1p} {\cal K}_{1pr} {\cal K}_{1r}}{(k_1\cdot
f_q\cdot X_q)\over {\cal K}_{1q}}~,~~~\label{3g-nice-B}\eea
\bea A_3 & = &-{(k_p\cdot (X_p-k_1))(k_r\cdot X_r)\over {\cal K}_{1p} {\cal
K}_{1q} {\cal K}_{1r}{\cal K}_{1pqr}}\nn
& &\times \left\{ (k_1\cdot f_p\cdot
f_q\cdot k_1)(k_1\cdot f_r\cdot k_p){(k_q\cdot X_q)\over {\cal
K}_{1pq}}+(k_1\cdot f_p\cdot f_r\cdot k_1)(k_1\cdot f_q\cdot k_p)
{(k_q\cdot (X_q+k_r))\over {\cal K}_{1pr}}\right.\nn
& & + (k_1\cdot f_p\cdot f_q\cdot k_1)(k_1\cdot f_r\cdot k_q)
 {((k_q\cdot X_q)-{\cal K}_{1pq})\over {\cal K}_{1pq}}+(k_1\cdot f_q\cdot f_r\cdot k_1)(k_1\cdot
f_p\cdot k_q) {(k_q\cdot (X_q-k_1)+{\cal K}_{1qr}) \over {\cal
K}_{1qr}}\nn
& &\left. +(k_1\cdot f_p\cdot f_r\cdot k_1)(k_1\cdot f_q\cdot k_r)
 { (k_q\cdot (Y_q-k_1))\over {\cal K}_{1pr}}+ (k_1\cdot f_q\cdot f_r\cdot k_1) (k_1\cdot
f_p\cdot k_r){(k_q\cdot (X_q-k_1))\over {\cal
K}_{1qr}}\right\}~.~~~\label{3g-nice-C}
  \eea
This is a much nicer expression than the one given by our direct
calculation \eref{3g-sum}, since each $A_i$ has more manifest
physical pattern.

\section{Expansion in the BCJ-basis by differential operators}

In the previous section we have illustrated how to expand sEYM
amplitudes in the  BCJ-basis of YM amplitudes starting from  their
expansion in the KK-basis. Although the whole procedure is
systematical, the algebraic manipulation is not so easy. As we have
emphasized, the strategy  of using differential operators is itself
an independent method, thus in this section, we will show how to do
the calculation.

Before going to calculation details, let us give some general
considerations. A sEYM amplitude
$A^{\text{EYM}}_{n,m}(1,2,\cdots,n;\{h_1,h_2,\cdots,h_m\})$ with
$n\geq 3$ gluons \footnote{The case $n=2$ is special and we will not
discuss it in this paper.} and $m$ gravitons can be  expanded in the
BCJ basis of YM amplitudes as following
\bea
 A^{\text{EYM}}_{n,m}(1,2,\cdots,n;\{h_1,h_2,\cdots,h_m\})=\sum_{\shuffle}\sum_{\rho} C(\shuffle,\rho) A^{\text{YM}}_{n+m}(1,2,\{3,\cdots,n-1\}\shuffle\{\rho_1,\cdots,\rho_m\},n),
~~~\label{Li-gen-exp}\eea
where the color order of $n$ gluons is kept and $\rho$ represents a
permutation of $m$ gravitons,
$\rho\{h_1,h_2,\cdots,h_m\}=\{\rho_1,\rho_2,\cdots,\rho_m\}$. The
coefficients $C(\shuffle,\rho)$ depend on the permutation $\rho$ and
the shuffle $\shuffle$, because gravitons don't have color order and
can occur in all possible orderings.

By the double copy structure \eref{gen-3},  only the $\epsilon$ part
of graviton polarization tensors
$\epsilon^{\mu\nu}=\epsilon^{\mu}\tilde{\epsilon}^{\nu}$ appears in
the coefficients $C$, i.e.,  they are functions of momenta $k_i$'s
of all particles and  polarization vectors $\epsilon_i$ of all
gravitons,
\bea
C(\shuffle,\rho)=C_{\shuffle,\rho}(k_1,\cdots,k_n,k_{h_1},\cdots,k_{h_m},\epsilon_{h_1},\cdots,\epsilon_{h_m}).
\eea
A very important fact of the expansion \eref{Li-gen-exp} is that by
the gauge invariance of EYM amplitudes, when $\epsilon_i$ is
replaced by $k_i$, $A^{\text{EYM}}_{n,m}(\epsilon_i\rightarrow
k_i)=0$ at the left hand side, thus
$C_{\shuffle,\rho}(\epsilon_i\rightarrow k_i)$ must be zero at the
right hand side because each amplitude of the BCJ-basis is
independent to each other. In other words, these coefficients are
also gauge invariant like amplitudes. From experiences in the
previous section we see that these coefficients are functions of
field strength $f^{\mu\nu}$ only, thus are manifestly gauge
invariant.

Another general feature is that coefficients are   multi-linear
functions of polarization vectors $\epsilon_i$. Thus  Lorentz
invariance means  coefficients can only be polynomial functions of
Lorentz invariant contractions $(\epsilon_i\cdot\epsilon_j)$'s and
$(\epsilon_i\cdot k_j)$'s, but can be rational functions of
$(k_i\cdot k_j)$'s. We can separate terms according to contraction
types sketchily as
\bea
C(\shuffle,\rho)= \alpha_0(\epsilon\cdot k)^m + \alpha_1(\epsilon\cdot\epsilon)(\epsilon\cdot k)^{m-2}+ \alpha_2(\epsilon\cdot\epsilon)^2(\epsilon\cdot k)^{m-4}+ \cdots+ \alpha_{[\frac{m}{2}]}(\epsilon\cdot\epsilon)^{[\frac{m}{2}]}(\epsilon\cdot k)^{m-2[\frac{m}{2}]}.
\eea

Above two general considerations are very important since if we
impose such  conditions, i.e., the gauge invariance and
multi-linearity  of polarization vectors $\epsilon_i$, we could
derive the expansion of sEYM amplitudes in the  BCJ-basis naturally
as will be shown in this section. Comparing to expansion in the
KK-basis, building blocks used in this section  are much more
complicated. In the appendix \ref{gauge-invariance}, we will give a
more systematical discussion about the building blocks for the
expansion in the BCJ-basis, although complete understanding is still
not clear for us.

Now we present several examples to show how to  derive the expansion
 of  sEYM amplitudes in the BCJ-basis by differential
operators.

\subsection{The case of one graviton}

Using the knowledge of  building blocks for one polarization vector
(see \eref{1g-building} in the Appendix \ref{gauge-invariance}), we
can expand the sEYM amplitude with one graviton as
\bea A^{\text{EYM}}_{n,1}(1,2,\cdots,n;p) & = &
\sum_{a=2}^{n-1}\frac{(k_1f_pK_a)}{(k_1k_p)}
B_a~.~~\label{1g-BCJ-1-1}\eea
Applying $\mathcal{T}_{jp(j+1)}$ (with $2\le j\le n-1$) on above
expansion \eref{1g-BCJ-1-1}, the left hand side gives
\begin{align}
 \mathcal{L}=\mathcal{T}_{jp(j+1)}A^{\text{EYM}}_{n,1}(1,2,\cdots,n;p)
 =A^{\text{YM}}_{n+1}(1,2,\cdots,j,p,j+1,\cdots,n)~,~~\label{1g-BCJ-1-2}
\end{align}
while the right hand side gives
\bea
 \mathcal{R}=&
 \sum_{a=2}^{n-1}\left\{\mathcal{T}_{jp(j+1)}\frac{(k_1f_pK_a)}{(k_1k_p)}\right\}
B_a=\sum_{i=2}^{n-1} \delta_{ja} B_a=B_j. ~~~\label{1g-BCJ-1-3}\eea
Comparing two sides, we solve
\bea
B_j=A^{\text{YM}}_{n+1}(1,2,\cdots,j,p,j+1,\cdots,n)~.~~\label{1g-BCJ-1-4}\eea
Putting it back, we get the wanted expansion in the BCJ-basis
\begin{align}
 A^{\text{EYM}}_{n,1}(1,2,\cdots,n;p)
 =& \sum_{\shuffle} \frac{(k_1f_pY_p)}{(k_1k_p)} A^{\text{YM}}_{n+1}(1,2,\{3,\cdots,n-1\}
 \shuffle\{p\},n),~~~\label{1g-BCJ-1-5}
\end{align}
where $K_a$ is exactly the $Y_p$ defined before. $Y_p$ has implicit
dependence on the shuffle. We want to emphasize that  using
insertion operators and gauge invariance for the building blocks
only, we have found the expansion of sEYM amplitudes with single
graviton in the BCJ-basis of YM amplitudes without any other
preassumption, such as the KLT relations.

An interesting observation is that among $(n-1)$ independent
insertion operators, we have used only $(n-2)$ of them. Now let us
consider the action of $\mathcal{T}_{jp(j+1)}$ with $j=1$ on
\eref{1g-BCJ-1-5}:
\bea {\cal L} & = &
\mathcal{T}_{1p2}A^{\text{EYM}}_{n,1}(1,2,\cdots,n;p)=
A^{\text{YM}}_{n+1}(1,p,2,\cdots,n) \nn
{\cal R} & =& \sum_{a=2}^{n-1} \left\{
\mathcal{T}_{1p2}\frac{(k_1f_p K_a)}{(k_1k_p)}\right\}
A^{\text{YM}}_{n+1}(1,2,\cdots,a,p,a+1,\cdots,n-1, n)\nn
& = & {1\over (k_p\cdot k_1)}\sum_{a=2}^{n-1} \left\{ \sum_{t=2}^a
(-k_1\cdot k_p)\delta_{t2}-(k_p\cdot
k_t)\right\}A^{\text{YM}}_{n+1}(1,2,\cdots,a,p,a+1,\cdots,n-1, n)\nn
& = & \sum_{a=2}^{n-1} (-k_p\cdot
K_a)A^{\text{YM}}_{n+1}(1,2,\cdots,a,p,a+1,\cdots,n-1,
n)~.~~\label{1g-BCJ-1-6}\eea
Identifying both sides and simplifying, we get
\bea \sum_{\shuffle} (k_p\cdot
X_p)A^{\text{YM}}_{n+1}(1,\{2,\cdots,n-1\}\shuffle\{p\},
n)=0~,~~\label{1g-BCJ-1-7}\eea
which is nothing, but the fundamental BCJ-relation
\eref{generalized-BCJ}. The derivation of such an important relation
using differential operators shows the deep connection between
BCJ-relations and the gauge invariance. Result \eref{1g-BCJ-1-6} can
also be understood from another angle: when identifying ${\cal
L}={\cal R}$, we get the expansion of the left hand side in the
BCJ-basis as reviewed in \eref{BCJ-exp-1}.

\subsection{The case with two gravitons}

Using the results given in  \eref{2g-building-1} in the Appendix
\ref{gauge-invariance}, we can expand the sEYM amplitude with two
gravitons according to the building blocks
\bea & & A^{\text{EYM}}_{n,2}(1,2,\cdots,n;p,q)  =
\sum_{a=2}^{n-1}\sum_{b=2}^{n-1}{(k_1f_pK_a)\over (k_1k_p)}{(k_1f_q
K_b)\over (k_1k_q)} B_{ab}+\sum_{b=2}^{n-1}{(k_1\cdot f_p\cdot
k_q)\over (k_1\cdot k_p)} {(k_1f_q K_b)\over (k_1k_q)}B_{qb}\nn
& & +\sum_{a=2}^{n-1}{(k_1\cdot f_q\cdot k_p)\over (k_1\cdot k_q)}
{(k_1f_p K_a)\over (k_1k_p)}B_{ap}+ {(k_1\cdot f_p\cdot f_q\cdot
k_1)\over (k_1\cdot k_p)
  (k_1\cdot k_q)} D+{(k_1\cdot f_p\cdot
k_q)\over (k_1\cdot k_p)}{(k_1\cdot f_q\cdot k_p)\over (k_1\cdot k_q)} E~.~~\label{2g-BCJ-4-1}\eea
Now we determine these coefficients one by one:
\begin{itemize}

\item {\bf For $E$:} Since the building block contains the index circle
 structure $(k_1\cdot f_q\cdot f_p\cdot k_q)$, i.e.,
\bea (k_1\cdot f_p\cdot k_q)(k_1\cdot f_q\cdot k_p)= (k_1\cdot
f_q\cdot f_p\cdot k_q)(k_1\cdot k_p)- (k_1\cdot f_p\cdot
f_q\cdot k_1)(k_q\cdot k_p)~,\eea
according to
the general argument given in Appendix \ref{cycle-index}, $E=0$.

\item {\bf For $B_{ab}$:} For this one, we consider the actions of $\mathcal{T}_{jp(j+1)}
\mathcal{T}_{mq(m+1)}$ with $j,m=2,...,n-1$. These operators
will annihilate the later three terms in \eref{2g-BCJ-4-1}.
Using the result
\bea \mathcal{T}_{jp(j+1)}\frac{(k_1f_pK_a)}{(k_1k_p)}= \left\{
 \begin{array}{ll} \delta_{ja},~~~ & j,a\geq 2 \\
 -{k_p\cdot K_a\over k_p\cdot k_1},~~~ & j=1,a\geq 2 \end{array} \right.~~~\label{T-action}
\eea
we can solve
\bea B_{ab}= \left\{ \begin{array}{ll}
A^{\text{YM}}_{n+2}(1,2,\cdots,a,p,a+1,\cdots,b,q,b+1,\cdots,n),~~~~
& a<b \\
A^{\text{YM}}_{n+2}(1,2,\cdots,b,q,b+1,\cdots,a,p,a+1,\cdots,n),~~~~
& a>b \\
A^{\text{YM}}_{n+2}(1,2,\cdots,a,p,q,a+1,\cdots,n)+
A^{\text{YM}}_{n+2}(1,2,\cdots,a,q,p,a+1,\cdots,n),~~~~
& a=b \\
\end{array}\right.\eea
Putting them back, \eref{2g-BCJ-4-1} becomes
\bea & & A^{\text{EYM}}_{n,2}(1,2,\cdots,n;p,q)  =
\sum_{\shuffle}{(k_1f_p Y_p)\over (k_1k_p)}{(k_1f_q Y_q)\over
(k_1k_q)}A^{\text{YM}}_{n+2}(1,2,\{3,\cdots,n-1\}\shuffle\{p\}\shuffle\{q\},n)
\nn
& & +\sum_{b=2}^{n-1}{(k_1\cdot f_p\cdot k_q)\over (k_1\cdot
k_p)} {(k_1f_q K_b)\over (k_1k_q)}B_{qb}
+\sum_{a=2}^{n-1}{(k_1\cdot f_q\cdot k_p)\over (k_1\cdot k_q)}
{(k_1f_p K_a)\over (k_1k_p)}B_{ap}+ {(k_1\cdot f_p\cdot f_q\cdot
k_1)\over (k_1\cdot k_p) (k_1\cdot k_q)}
  D~.~~\label{2g-BCJ-4-2-1}\eea

\item {\bf For $B_{qb}$:} For this one, let us apply operators
$ \mathcal{T}_{mq(m+1)}$ with $m=2,...,(n-1)$ first. It will
annihilate the third and fourth terms in \eref{2g-BCJ-4-2-1} and
we get the middle result
\bea & & A^{\text{EYM}}_{n+1,1}(1,2,\cdots,m,q,m+1,\cdots,n;p)
\nn & =& \sum_{\shuffle}{(k_1f_p Y_p)\over
(k_1k_p)}A^{\text{YM}}_{n+2}(1,2,\{3,\cdots,m,q,m+1,\cdots,n-1\}\shuffle\{p\},n)+{(k_1\cdot
f_p\cdot k_q)\over (k_1\cdot k_p)}
B_{qm}~.~~\label{2g-BCJ-4-3-1}\eea
Next, we apply $\mathcal{T}_{mpq}$ and reach
\bea & & A^{\text{YM}}_{n+2}(1,2,\cdots,m,p,q,m+1,\cdots,n) \nn
& =&
\sum_{\shuffle}A^{\text{YM}}_{n+2}(1,2,\cdots,m,\{q,m+1,\cdots,n-1\}\shuffle\{p\},n)-
B_{qm}~.~\label{2g-BCJ-4-3-2}\eea
Thus we can solve
\bea B_{qm} & = &
\sum_{\shuffle}A^{\text{YM}}_{n+2}(1,2,\cdots,m,\{q,m+1,\cdots,n-1\}\shuffle\{p\},n)
-A^{\text{YM}}_{n+2}(1,2,\cdots,m,p,q,m+1,\cdots,n)\nn
& = &
\sum_{\shuffle}A^{\text{YM}}_{n+2}(1,2,\cdots,m,q,\{m+1,\cdots,n-1\}\shuffle\{p\},n)~.
~~\label{2g-BCJ-4-3-3}\eea

\item {\bf For $B_{ap}$:} For this one, let us apply operators $
\mathcal{T}_{mp(m+1)}$ with $m=2,...,(n-1)$ first. It will
annihilate the second and fourth terms in \eref{2g-BCJ-4-2-1}
and we get the middle result
\bea & & A^{\text{EYM}}_{n+1,1}(1,2,\cdots,m,p,m+1,\cdots,n;q)
\nn & =& \sum_{\shuffle}{(k_1f_q Y_q)\over
(k_1k_q)}A^{\text{YM}}_{n+2}(1,2,\{3,\cdots,m,p,m+1,\cdots,n-1\}\shuffle\{q\},n)+{(k_1\cdot
f_q\cdot k_p)\over (k_1\cdot k_q)}
B_{mp}~.~\label{2g-BCJ-4-4-1}\eea
Next, we apply $\mathcal{T}_{mqp}$ and reach
\bea & & A^{\text{YM}}_{n+2}(1,2,\cdots,m,q,p,m+1,\cdots,n) \nn
& =&
\sum_{\shuffle}A^{\text{YM}}_{n+2}(1,2,\cdots,m,\{p,m+1,\cdots,n-1\}\shuffle\{q\},n)-
B_{mp}~.~\label{2g-BCJ-4-4-2}\eea
Thus we can solve
\bea B_{mp} & = &
\sum_{\shuffle}A^{\text{YM}}_{n+2}(1,2,\cdots,m,\{p,m+1,\cdots,n-1\}\shuffle\{q\},n)
-A^{\text{YM}}_{n+2}(1,2,\cdots,m,q,p,m+1,\cdots,n)\nn
& = &
\sum_{\shuffle}A^{\text{YM}}_{n+2}(1,2,\cdots,m,p,\{m+1,\cdots,n-1\}\shuffle\{q\},n)~.
~~\label{2g-BCJ-4-4-3}\eea
Now putting \eref{2g-BCJ-4-4-3} and \eref{2g-BCJ-4-3-3} back to
\eref{2g-BCJ-4-2-1}, we get
\bea & & A^{\text{EYM}}_{n,2}(1,2,\cdots,n;p,q)  =
\sum_{\shuffle}{(k_1f_p Y_p)\over (k_1k_p)}{(k_1f_q Y_q)\over
(k_1k_q)}A^{\text{YM}}_{n+2}(1,2,\{3,\cdots,n-1\}\shuffle\{p\}\shuffle\{q\},n)
\nn
& & +\sum_{\shuffle}{(k_1\cdot f_p\cdot k_q)\over (k_1\cdot
k_p)} {(k_1f_q Y_q)\over
(k_1k_q)}A^{\text{YM}}_{n+2}(1,2,\{3,\cdots,n-1\}\shuffle\{q,p\},n)\nn
& &  +\sum_{\shuffle}{(k_1\cdot f_q\cdot k_p)\over (k_1\cdot
k_q)} {(k_1f_p Y_p)\over
(k_1k_p)}A^{\text{YM}}_{n+2}(1,2,\{3,\cdots,n-1\}\shuffle\{p,q\},n)+
{(k_1\cdot f_p\cdot f_q\cdot k_1)\over (k_1\cdot k_p) (k_1\cdot
k_q)}
  D~.~~\label{2g-BCJ-4-5-1}\eea

\item {\bf For the $D$:} For the last one, we need the action keeping
the last term. One of such a choice is
$\mathcal{T}_{1qp}\mathcal{T}_{1p2}$. Acting on
\eref{2g-BCJ-4-5-1}, we get
\bea {\cal L} & = & A^{\text{YM}}_{n+2}(1,q,p,2,\cdots,n)\nn
{\cal R} & = & \sum_{\shuffle}{(-k_p\cdot Y_p)\over
(k_1k_p)}{(-k_q\cdot (Y_q-k_1))\over
(k_1k_q)}A^{\text{YM}}_{n+2}(1,2,\{3,\cdots,n-1\}\shuffle\{p\}\shuffle\{q\},n)\nn
& & +\sum_{\shuffle}{(-k_p\cdot k_q)\over (k_1\cdot k_p)}
{(-k_q\cdot (Y_q-k_1))\over
(k_1k_q)}A^{\text{YM}}_{n+2}(1,2,\{3,\cdots,n-1\}\shuffle\{q,p\},n)\nn
& &  +\sum_{\shuffle}{-((k_p+k_1)\cdot k_q)\over (k_1\cdot k_q)}
{(-k_p Y_p)\over
(k_1k_p)}A^{\text{YM}}_{n+2}(1,2,\{3,\cdots,n-1\}\shuffle\{p,q\},n)-
{k_q\cdot (k_p+k_1)\over (k_1\cdot k_p) (k_1\cdot k_q)}
  D~~~~~~~~\label{2g-BCJ-4-6-1}\eea
where we have used \eref{T-action}. From it, we can solve that
\bea & & ((k_p+k_1)\cdot k_q) D  = \sum_{\shuffle}(k_p\cdot
X_p)(k_q\cdot(
Y_q-k_1))A^{\text{YM}}_{n+2}(1,2,\{3,\cdots,n-1\}\shuffle\{q,p\},n)\nn
& & + \sum_{\shuffle}(k_q\cdot(k_p+Y_q)) (k_p Y_p)
A^{\text{YM}}_{n+2}(1,2,\{3,\cdots,n-1\}\shuffle\{p,q\},n)\nn
& & -(k_1\cdot k_p)(k_1\cdot
k_q)A^{\text{YM}}_{n+2}(1,q,p,2,\cdots,n)~~\label{2g-BCJ-4-6-2}
\eea
Now using \eref{G-BCJ-b=2-3} and with some algebraic
simplifications, we find
\bea D & = & {(K_q\cdot X_q)(k_p\cdot (Y_p-k_1)) \over {\cal
K}_{1pq} } ~A_{n+1}^{\rm
YM}(1,2,\{3,\ldots,n-1\}\shuffle\{p,q\},n)\nn
&  &+ {(K_p\cdot X_p)(k_q\cdot (Y_q-k_1))\over {\cal K}_{1pq}
}~A^{\rm YM}_{n+2}(1,2, \{3,...,n-1\}\shuffle\{q,p\},n)
~~\label{2g-BCJ-4-6-4}\eea

\end{itemize}
Putting back \eref{2g-BCJ-4-6-4} to \eref{2g-BCJ-4-5-1}, we finally
get
\bea & & A^{\text{EYM}}_{n,2}(1,2,\cdots,n;p,q)  =
\sum_{\shuffle}{(k_1f_p Y_p)\over (k_1k_p)}{(k_1f_q Y_q)\over
(k_1k_q)}A^{\text{YM}}_{n+2}(1,2,\{3,\cdots,n-1\}\shuffle\{p\}\shuffle\{q\},n)
\nn
& & +\sum_{\shuffle}{(k_1\cdot f_p\cdot k_q)\over (k_1\cdot k_p)}
{(k_1f_q Y_q)\over
(k_1k_q)}A^{\text{YM}}_{n+2}(1,2,\{3,\cdots,n-1\}\shuffle\{q,p\},n)\nn
& &  +\sum_{\shuffle}{(k_1\cdot f_q\cdot k_p)\over (k_1\cdot k_q)}
{(k_1f_p Y_p)\over
(k_1k_p)}A^{\text{YM}}_{n+2}(1,2,\{3,\cdots,n-1\}\shuffle\{p,q\},n)\nn
& & + {(k_1\cdot f_p\cdot f_q\cdot k_1)\over (k_1\cdot k_p)
(k_1\cdot k_q)}\left\{{(K_q\cdot X_q)(k_p\cdot (Y_p-k_1)) \over
{\cal K}_{1pq} } ~A_{n+1}^{\rm
YM}(1,2,\{3,\ldots,n-1\}\shuffle\{p,q\},n)\right. \nn & & \left.+
{(K_p\cdot X_p)(k_q\cdot (Y_q-k_1))\over {\cal K}_{1pq} }~A^{\rm
YM}_{n+2}(1,2, \{3,...,n-1\}\shuffle\{q,p\},n)\right\}
  ~,~~\label{2g-BCJ-4-7-1}\eea
which is nothing, but the result \eref{EYM-2g-Exp}. We want to
emphasize again that using the differential operators on the gauge
invariant building blocks, we reach the expansion in the BCJ-basis
naturally without any further preassumption. In the derivation, the
symmetry between $p,q$ is automatically kept.

Before ending this part, let us give another remark. To solve $D$ we
have used the differential operator
$\mathcal{T}_{1qp}\mathcal{T}_{1p2}$ and the result
\eref{G-BCJ-b=2-3}. However, we can use another operator
$\mathcal{T}_{1q2}\mathcal{T}_{qp2}$ to find $D$. If we act it on
both sides of \eref{2g-BCJ-4-5-1}, we will get the same ${\cal L}$
and different ${\cal R}$ in \eref{2g-BCJ-4-6-1}. Using these two
different expressions of ${\cal R}$, we can solve $D$. Putting the
$D$ back, we get the ${\cal L}$,  which is nothing, but the
expression \eref{G-BCJ-b=2-3}. In other words, we have derived the
expansion of YM amplitudes to its BCJ-basis for this special case,
just like the derivation of the fundamental BCJ-relations in the
previous subsection \eref{1g-BCJ-1-7}. From the new angle, the BCJ
relations are consistent conditions of these differential operators.
It is similar to the observation that consistent conditions for
different KLT relations
\cite{Kawai:1985xq,Bern:1998ug,BjerrumBohr:2009rd,Stieberger:2009hq,
BjerrumBohr:2010ta,BjerrumBohr:2010zb,BjerrumBohr:2010yc} will imply
these BCJ relations.

\subsubsection{Another derivation}

The starting point of the previous derivation  is the expansion
\eref{2g-BCJ-4-1} of $A^{\text{EYM}}_{n,2}$ using building blocks of
two gravitons. Now we consider another derivation, which has the
recursive structure like the expansion in the KK-basis. We can first
view $A_{n,2}^{\text{EYM}}$ as the polynomial functions of
$\epsilon_p$ and write its manifestly gauge invariant expansion by
building blocks of $\epsilon_p$ as
\begin{align}
A^{\text{EYM}}_{n,2}(1,\cdots,n;p,q)=\sum_{a=2}^{n-1}
\frac{(k_1f_pK_a)}{(k_1k_p)}B_a + \frac{(k_1f_pk_q)}{(k_1k_p)}B_q
+\frac{(k_1f_pf_qk_1)}{(k_1k_p)(k_1k_q)} E_{pq},~~~~\label{Li-BCJ-1}
\end{align}
with $B_a$ and $B_q$ being polynomials of $\epsilon_q$. Now we  use
differential operators to determine these unknown variables one by
one:
\begin{itemize}

\item (1) First, we use insertion operators $\mathcal{T}_{ip(i+1)}$ with
$2\le i\le n-1$ to determine $B_a$. After applying these
operators, we get
\begin{align}
 A^{\text{EYM}}_{n+1,1}(1,\cdots,i,p,i+1,n;q)
=& \sum_{a=2}^{n-1} \mathcal{T}_{ip(i+1)}\frac{(k_1f_pK_a)}{(k_1k_p)}B_a = B_i, \notag
\end{align}

\item (2) For the remaining two variables, there are no differential
operators involving only graviton $p$ to select them one by one,
and we need to expand it further according to $\epsilon_q$ as
\begin{align}
B_q= \sum_{b=2}^{n-1} \frac{(k_1f_qK_b)}{(k_1k_q)} B_{qb} +\frac{(k_1f_qk_p)}{(k_1k_q)}B_{qp}.
\end{align}
Now
\begin{align}
A^{\text{EYM}}_{n,2}(1,\cdots,n;p,q)=&\sum_{\shuffle} \frac{(k_1f_pY_p)}{(k_1k_p)}A^{\text{EYM}}_{n+1,1}(1,2,\{3,\cdots,n-1\}\shuffle\{p\},n;q) \notag\\
  &+\sum_{b=2}^{n-1} \frac{(k_1f_pk_q)}{(k_1k_p)}\frac{(k_1f_qK_b)}{(k_1k_q)} B_{qb} +\frac{(k_1f_pk_q)}{(k_1k_p)}\frac{(k_1f_qk_p)}{(k_1k_q)}B_{qp}+\frac{(k_1f_pf_qk_1)}{(k_1k_p)(k_1k_q)} E_{pq}.
\end{align}
Because of the index circle structure, we have immediately
$B_{qp}=0$.

\item (3) Now we apply insertion operators
$\mathcal{T}_{iq(i+1)}\mathcal{T}_{qp(i+1)}$ to the above
equation to determine $B_{qb}$. We get
\begin{align}
 A^{\text{YM}}_{n+2}(1,\cdots,i,q,p,i+1,\cdots,n)=& -\sum_{\shuffle}
  A_{n+2}(1,\cdots,i,q,i+1,\{i+2,\cdots,n-1\}\shuffle\{p\},n)+B_{qb}~. \notag
\end{align}
From it, we  can  solved
\begin{align}
B_{qb}=\sum_{\shuffle} A_{n+2}(1,\cdots,i,q,\{i+1,\cdots,n-1\}\shuffle\{p\},n). \notag
\end{align}
Thus we can write
\bea & & A^{\text{EYM}}_{n,2}(1,\cdots,n;p,q)= \sum_{\shuffle}
 \frac{(k_1f_pY_p)}{(k_1k_p)}A^{\text{EYM}}_{n+1,1}(1,2,\{3,\cdots,n-1\}
 \shuffle\{p\},n;q)\nn
  &&+\sum_{\shuffle}
\frac{(k_1f_pk_q)}{(k_1k_p)}\frac{(k_1f_qY_q)}{(k_1k_q)}
A_{n+2}(1,2,\{3,\cdots,n-1\}\shuffle\{q,p\},n)+\frac{(k_1f_pf_qk_1)}{(k_1k_p)(k_1k_q)}
E_{pq}.  \eea

\item (4) The determination of $E_{pq}$ will be exactly the same as  the
determination of $D$ in \eref{2g-BCJ-4-6-4} and we will not
repeat the calculation.

\end{itemize}

Assembling all pieces together, we get
\begin{align}
A^{\text{EYM}}_{n,2}(1,\cdots,n;p,q)=
  &\sum_{\shuffle} \frac{(k_1f_pX_p)}{(k_1k_p)}\frac{(k_1f_qX_q)}{(k_1k_q)} A_{n+2}(1,2,\{3,\cdots,n-1\}\shuffle\{p\}\shuffle\{q\},n) \notag\\
  &+\frac{(k_1f_pf_qk_1)}{(k_1k_p)(k_1k_q)}  \frac{[k_p(Y_p-k_1)](k_qX_q)}{\mathcal{K}_{1pq}} A^{\text{YM}}_{n+2}(1,2,\{3,\cdots,n-1\}\shuffle\{p,q\},n) \notag\\
  &+\frac{(k_1f_pf_qk_1)}{(k_1k_p)(k_1k_q)} \frac{[k_q(Y_q-k_1)](k_pX_p)}{\mathcal{K}_{1pq}} A^{\text{YM}}_{n+2}(1,2,\{3,\cdots,n-1\}\shuffle\{q,p\},n).
\end{align}

\subsection{The case with three gravitons}
In this subsection, we consider a little more complicated case,
i.e., the sEYM amplitudes with three gravitons. We will use the
recursive structure to express the expansion.

First we write down the expansion according to the manifestly gauge
invariant building blocks  of $\epsilon_p$ as
\bea A^{\text{EYM}}_{n,3}(1,2,\cdots,n;p,q,r)&= &\sum_{a=2}^{n-1}
\frac{(k_1f_pK_a)}{(k_1k_p)}
B_a+\frac{(k_1f_pk_q)}{(k_1k_p)}B_p+\frac{(k_1f_pk_r)}{(k_1k_p)}B_r
\nn & & +\frac{(k_1f_pf_qk_1)}
{(k_1k_p)(k_1k_q)}D_{pq}+\frac{(k_1f_pf_rk_1)}{(k_1k_p)(k_1k_r)}D_{pr},~~~\label{Li-BCJ-3g-1}
\eea
with $B$'s being polynomials of $\epsilon_q$ and $\epsilon_r$,
$D_{pq}$ of $\epsilon_r$ and $D_{pr}$ of $\epsilon_q$. Now we solve
these coefficients step by step:
\begin{itemize}

\item (1) It is clear that applying insertion operators
$\mathcal{T}_{ip(i+1)}$ with $i=2,\cdots,n-1$ directly, we can
solve
\begin{align}
\mathcal{T}_{ip(i+1)}A^{\text{EYM}}_{n,3}(1,2,\cdots,n;p,q,r)=& A^{\text{EYM}}_{n+1,2}(1,\cdots,i,p,i+1,\cdots,n;q,r)
=& \sum_{a=2}^{n-1} \mathcal{T}_{ip(i+1)}\frac{(k_1f_pK_a)}{(k_1k_p)} B_a=B_i, \notag
\end{align}

\item (2)
To determine $B_p$, we need to expand it further as
\begin{align}
\frac{(k_1f_pk_q)}{(k_1k_p)}B_q=& \frac{(k_1f_pk_q)}{(k_1k_p)} \Big\{ \sum_{b=2}^{n-1} \frac{(k_1f_qK_b)}{(k_1k_q)} B_{qb}
+\frac{(k_1f_qk_p)}{(k_1k_q)}B_{qp}+\frac{(k_1f_qk_r)}{(k_1k_q)}B_{qr}+\frac{(k_1f_qf_rk_1)}{(k_1k_q)(k_1k_r)}D_{qqr} \Big\}.
\end{align}
Because of the index circle structure, $B_{qp}=0$. Next we
compute coefficients $B_{qb}$ and $B_{qr}$, and  leave $D_{qqr}$
to later. Applying the insertion operator
$\mathcal{T}_{iq(i+1)}\mathcal{T}_{qp(i+1)}$, we have
\begin{align}
A^{\text{EYM}}_{n+2,1}(1,\cdots,i,q,p,i+1,\cdots,n;r)
=& -\sum_{\shuffle} A_{n+2,1}(1,\cdots,i,q,i+1,\{i+2,\cdots,n-1\}\shuffle\{p\},n;r)+B_{qi}, \notag
\end{align}
from which we solve
\begin{align}
B_{qi}=\sum_{\shuffle} A_{n+2,1}(1,\cdots,i,q,\{i+1,\cdots,n-1\}\shuffle\{p\},n;r).
\end{align}
To determine $B_{qr}$, we need to expand it further
\begin{align}
\frac{(k_1f_pk_q)}{(k_1k_p)}\frac{(k_1f_qk_r)}{(k_1k_q)}B_{qr}
=& \frac{(k_1f_pk_q)}{(k_1k_p)}\frac{(k_1f_qk_r)}{(k_1k_q)} \Big\{ \sum_{c=2}^{n-1}\frac{(k_1f_rK_c)}{(k_1k_r)}B_{qrc}+\frac{(k_1f_rk_p)}{(k_1k_r)}B_{qrp}+\frac{(k_1f_rk_q)}{(k_1k_r)}B_{qrq} \Big\}. \notag
\end{align}
Again because of the index circle structure,
$B_{qrp}=B_{qrq}=0$. We apply the insertion operator
$\mathcal{T}_{iq(i+1)}\mathcal{T}_{rq(i+1)}\mathcal{T}_{qp(i+1)}$
to
 get
\begin{align}
A_{n+3}(1,\cdots,i,r,q,p,i+1,\cdots,n)
=& -\sum_{\shuffle} A_{n+3}(1,\cdots,i,r,q,i+1,\{i+2,\cdots,n-1\}\shuffle\{p\},n) \notag\\
&-\sum_{\shuffle} A_{n+3}(1,\cdots,i,r,i+1,\{i+2,\cdots,n-1\}\shuffle\{q,p\},n)+B_{qri}, \notag
\end{align}
so  $B_{qrc}$ is solved to be
\begin{align}
B_{qri}=\sum_{\shuffle} A_{n+3}(1,\cdots,i,r,\{i+1,\cdots,n-1\}\shuffle\{q,p\},n).
\end{align}

\item (3)
As for the coefficient $B_r$, because of the symmetry between
the $q$ and $r$, we can take the exactly same steps as before
and get similar results
\begin{align}
\frac{(k_1f_pk_r)}{(k_1k_p)}B_r=
&\sum_{\shuffle} \frac{(k_1f_pk_r)}{(k_1k_p)} \frac{(k_1f_rY_r)}{(k_1k_r)} A_{n+2,1}(1,2,\{3,\cdots,n-1\}\shuffle\{r,p\},n;q) \notag\\
&+\sum_{\shuffle} \frac{(k_1f_pk_r)}{(k_1k_p)} \frac{(k_1f_rk_q)}{(k_1k_r)} \frac{(k_1f_qY_q)}{(k_1k_r)} A_{n+3}(1,2,\{3,\cdots,n-1\}\shuffle\{q,r,p\},n) \notag\\
&+\frac{(k_1f_pk_r)}{(k_1k_p)} \frac{(k_1f_rf_qk_1)}{(k_1k_q)(k_1k_r)}D_{rqr},  \notag
\end{align}

\item (4) Up to now, we have
\begin{align}
A^{\text{EYM}}_{n,3}(1,2,\cdots,n;p,q,r)=&\sum_{\shuffle} \frac{(k_1f_pY_p)}{(k_1k_p)} A^{\text{EYM}}_{n+1,2}(1,2,\{3,\cdots,n-1\}\shuffle\{p\},n;q,r) \notag\\
&+\sum_{\shuffle} \frac{(k_1f_pk_q)}{(k_1k_p)} \frac{(k_1f_qY_q)}{(k_1k_q)} A_{n+2,1}(1,2,\{3,\cdots,n-1\}\shuffle\{q,p\},n;r) \notag\\
&+\sum_{\shuffle} \frac{(k_1f_pk_q)}{(k_1k_p)} \frac{(k_1f_qk_r)}{(k_1k_q)} \frac{(k_1f_rY_r)}{(k_1k_r)} A_{n+3}(1,2,\{3,\cdots,n-1\}\shuffle\{r,q,p\},n)  \notag\\
&+\sum_{\shuffle} \frac{(k_1f_pk_r)}{(k_1k_p)} \frac{(k_1f_rY_r)}{(k_1k_r)} A_{n+2,1}(1,2,\{3,\cdots,n-1\}\shuffle\{r,p\},n;q) \notag\\
&+\sum_{\shuffle} \frac{(k_1f_pk_r)}{(k_1k_p)} \frac{(k_1f_rk_q)}{(k_1k_r)} \frac{(k_1f_qY_q)}{(k_1k_r)} A_{n+3}(1,2,\{3,\cdots,n-1\}\shuffle\{q,r,p\},n) \notag\\
&+\frac{(k_1f_pk_r)}{(k_1k_p)} \frac{(k_1f_rf_qk_1)}{(k_1k_q)(k_1k_r)}D_{rqr}
+\frac{(k_1f_pk_q)}{(k_1k_p)} \frac{(k_1f_qf_rk_1)}{(k_1k_q)(k_1k_r)}D_{qqr} \notag\\
&+\frac{(k_1f_pf_qk_1)}{(k_1k_p)(k_1k_q)}D_{pq}+\frac{(k_1f_pf_rk_1)}{(k_1k_p)(k_1k_r)}D_{pr}.
\end{align}

\item (5) Now we consider these left $D$'s. First we  expand $D_{pq}$ as
\begin{align}
\frac{(k_1f_pf_qk_1)}{(k_1k_p)(k_1k_q)}D_{pq}
=& \frac{(k_1f_pf_qk_1)}{(k_1k_p)(k_1k_q)} \Big\{ \sum_{c=2}^{n-1} \frac{(k_1f_rK_c)}{(k_1k_r)} D_{pqc}+\frac{(k_1f_rk_p)}{(k_1k_r)} D_{pqp}+\frac{(k_1f_rk_q)}{(k_1k_r)} D_{pqq} \Big\},
\end{align}
Now  applying insertion operators
$\mathcal{T}_{ir(i+1)}\mathcal{T}_{1q2}\mathcal{T}_{qp2}$ with
$2\le i\le n-1$ we get
\begin{align}
 &A_{n+3}(1,q,p,2,\cdots,i,r,i+1,\cdots,n) \notag\\
=& -\sum_{\shuffle}A_{n+3}(1,q,2,\{3,\cdots,i,r,i+1,\cdots,n-1\}\shuffle\{p\},n) \notag\\
    &-\sum_{\shuffle}\frac{(k_qY_q)}{(k_1k_q)} A_{n+3}(1,2,\{3,\cdots,i,r,i+1,\cdots,n-1\}\shuffle\{q,p\},n) \notag\\
    &-\sum_{\shuffle}\frac{(k_rY_r)}{(k_1k_r)} A_{n+3}(1,2,\cdots,i,r,\{i+1,\cdots,n-1\}\shuffle\{q,p\},n)
    +\frac{1}{(k_1k_q)}D_{pqi}. \notag
\end{align}
Solving it in the BCJ-basis of YM amplitudes, we find
\begin{align}
D_{pqi} =& \frac{[k_q(X_q-k_1)](k_pX_p)}{\mathcal{K}_{1pq}} A_{n+3}(1,2,\{3,\cdots,i,r,i+1,\cdots,n-1\}\shuffle\{q,p\},n) \notag\\
    &+\frac{[k_p(X_p-k_1)](k_qX_q)}{\mathcal{K}_{1pq}} A_{n+3}(1,2,\{3,\cdots,i,r,i+1,\cdots,n-1\}\shuffle\{p,q\},n), \notag
\end{align}
Similarly by the symmetry between $q$ and $r$, after expanding
$D_{pr}$ and applying
$\mathcal{T}_{iq(i+1)}\mathcal{T}_{1r2}\mathcal{T}_{rp2}$, we
can write
\begin{align}
&\frac{(k_1f_pf_rk_1)}{(k_1k_p)(k_1k_r)}D_{pr} \notag\\
=&  \sum_{\shuffle} \frac{(k_1f_pf_rk_1)}{(k_1k_p)(k_1k_r)} \frac{(k_1f_qY_q)}{(k_1k_q)} \Big\{ \frac{[k_r(X_r-k_1)](k_pX_p)}{\mathcal{K}_{1pr}} A_{n+3}(1,2,\{3,\cdots,n-1\}\shuffle\{q\}\shuffle\{r,p\},n) \notag\\
    &+\frac{[k_p(X_p-k_1)](k_rX_r)}{\mathcal{K}_{1pr}} A_{n+3}(1,2,\{3,\cdots,n-1\}\shuffle\{q\}\shuffle\{p,r\},n) \Big\}  \notag\\
& + \frac{(k_1f_pf_rk_1)}{(k_1k_p)(k_1k_r)} \frac{(k_1f_qk_p)}{(k_1k_q)} D_{prp}+ \frac{(k_1f_pf_rk_1)}{(k_1k_p)(k_1k_r)} \frac{(k_1f_qk_r)}{(k_1k_q)} D_{prr}.
\end{align}
Up to now the expansion of $A^{\text{EYM}}_{n,3}$ is given by
\begin{align}
&A^{\text{EYM}}_{n,3}(1,2,\cdots,n;p,q,r) \notag\\
=&\sum_{\shuffle} \frac{(k_1f_pY_p)}{(k_1k_p)} A^{\text{EYM}}_{n+1,2}(1,2,\{3,\cdots,n-1\}\shuffle\{p\},n;q,r) \notag\\
&+\sum_{\shuffle} \frac{(k_1f_pk_q)}{(k_1k_p)} \frac{(k_1f_qY_q)}{(k_1k_q)} A_{n+2,1}(1,2,\{3,\cdots,n-1\}\shuffle\{q,p\},n;r) \notag\\
&+\sum_{\shuffle} \frac{(k_1f_pk_q)}{(k_1k_p)} \frac{(k_1f_qk_r)}{(k_1k_q)} \frac{(k_1f_rY_r)}{(k_1k_r)} A_{n+3}(1,2,\{3,\cdots,n-1\}\shuffle\{r,q,p\},n)  \notag\\
&+\sum_{\shuffle} \frac{(k_1f_pk_r)}{(k_1k_p)} \frac{(k_1f_rY_r)}{(k_1k_r)} A_{n+2,1}(1,2,\{3,\cdots,n-1\}\shuffle\{r,p\},n;q) \notag\\
&+\sum_{\shuffle} \frac{(k_1f_pk_r)}{(k_1k_p)} \frac{(k_1f_rk_q)}{(k_1k_r)} \frac{(k_1f_qY_q)}{(k_1k_r)} A_{n+3}(1,2,\{3,\cdots,n-1\}\shuffle\{q,r,p\},n) \notag\\
&+\sum_{\shuffle} \frac{(k_1f_pf_qk_1)}{(k_1k_p)(k_1k_q)} \frac{(k_1f_rY_r)}{(k_1k_r)} \Big\{ \frac{[k_q(X_q-k_1)](k_pX_p)}{\mathcal{K}_{1pq}} A_{n+3}(1,2,\{3,\cdots,n-1\}\shuffle\{r\}\shuffle\{q,p\},n) \notag\\
    &+\frac{[k_p(X_p-k_1)](k_qX_q)}{\mathcal{K}_{1pq}} A_{n+3}(1,2,\{3,\cdots,n-1\}\shuffle\{r\}\shuffle\{p,q\},n) \Big\}  \notag\\
 &+\sum_{\shuffle} \frac{(k_1f_pf_rk_1)}{(k_1k_p)(k_1k_r)} \frac{(k_1f_qY_q)}{(k_1k_q)} \Big\{ \frac{[k_r(X_r-k_1)](k_pX_p)}{\mathcal{K}_{1pr}} A_{n+3}(1,2,\{3,\cdots,n-1\}\shuffle\{q\}\shuffle\{r,p\},n) \notag\\
    &+\frac{[k_p(X_p-k_1)](k_rX_r)}{\mathcal{K}_{1pr}} A_{n+3}(1,2,\{3,\cdots,n-1\}\shuffle\{q\}\shuffle\{p,r\},n) \Big\}  \notag\\
&+\frac{(k_1f_pk_r)}{(k_1k_p)} \frac{(k_1f_rf_qk_1)}{(k_1k_q)(k_1k_r)}D_{rqr}
+\frac{(k_1f_pk_q)}{(k_1k_p)} \frac{(k_1f_qf_rk_1)}{(k_1k_q)(k_1k_r)}D_{qqr} + \frac{(k_1f_pf_qk_1)}{(k_1k_p)(k_1k_q)} \frac{(k_1f_rk_p)}{(k_1k_r)} D_{pqp} \notag\\
&+ \frac{(k_1f_pf_qk_1)}{(k_1k_p)(k_1k_q)} \frac{(k_1f_rk_q)}{(k_1k_r)} D_{pqq}
 + \frac{(k_1f_pf_rk_1)}{(k_1k_p)(k_1k_r)} \frac{(k_1f_qk_p)}{(k_1k_q)} D_{prp}+
 \frac{(k_1f_pf_rk_1)}{(k_1k_p)(k_1k_r)} \frac{(k_1f_qk_r)}{(k_1k_q)} D_{prr}. \label{3expansion3}
\end{align}

\item (6) Finally we need to determine the six left
$D$'s. We  rewrite these six terms as
\begin{align}
 C_1
 =&a_1\frac{(k_1f_pk_q)}{\mathcal{K}_{1p}}\frac{(k_1f_qf_rk_1)}{\mathcal{K}_{1q}\mathcal{K}_{1r}}
 + a_2\frac{(k_1f_pk_r)}{\mathcal{K}_{1p}}\frac{(k_1f_rf_qk_1)}{\mathcal{K}_{1q}\mathcal{K}_{1r}}
 + a_3\frac{(k_1f_qk_p)}{\mathcal{K}_{1q}}\frac{(k_1f_pf_rk_1)}{\mathcal{K}_{1p}\mathcal{K}_{1r}} \notag\\
 &+ a_4\frac{(k_1f_qk_r)}{\mathcal{K}_{1q}}\frac{(k_1f_rf_pk_1)}{\mathcal{K}_{1p}\mathcal{K}_{1r}}
 +a_5\frac{(k_1f_rk_p)}{\mathcal{K}_{1r}}\frac{(k_1f_pf_qk_1)}{\mathcal{K}_{1p}\mathcal{K}_{1q}}
 + a_6\frac{(k_1f_rk_q)}{\mathcal{K}_{1r}}\frac{(k_1f_qf_pk_1)}{\mathcal{K}_{1p}\mathcal{K}_{1q}} \notag\\
 =& \sum_{\rho} a(\rho) \frac{(k_1f_{\rho_3}k_{\rho_2})}{\mathcal{K}_{1\rho_3}}\frac{(k_1f_{\rho_2}f_{\rho_1}k_1)}{\mathcal{K}_{1\rho_2}\mathcal{K}_{1\rho_1}},  \notag
\end{align}
where $\rho$ is a permutation of $\{p,q,r\}$, and the other
known terms in the expansion of $A^{\text{EYM}}_{n,3}$ are
denoted by $C_0$. To determine them, we need to use the
operators
$\mathcal{T}_{1\{\rho\}2}\equiv\mathcal{T}_{1\rho_1\rho_2\rho_32}=\mathcal{T}_{\rho_2\rho_32}\mathcal{T}_{\rho_1\rho_22}\mathcal{T}_{1\rho_12}$,
whose effect is to insert $\{p,q,r\}$ between $1$ and $2$
consecutively with the ordering $(\rho_1,\rho_2,\rho_3)$.
Applying $\mathcal{T}_{1\{\rho'\}2}$ to \eref{3expansion3}, the
left hand side gives
\bea  & & \mathcal{T}_{1\{\rho'\}2}
  A^{\text{EYM}}_{n,3}(1,2,\cdots,n;p,q,r) =
 A^{\text{YM}}_{n+3}(1,\rho'_1,\rho'_2,\rho'_3,2,\cdots,n) \nn %
 &=& \sum_{\rho}\sum_{\shuffle}
 \mathcal{C}^{(\rho')}(\rho,\shuffle)
 A^{\text{YM}}_{n+3}(1,2,\{3,\cdots,n-1\}\shuffle\{\rho_1,\rho_2,\rho_3\},n),
\eea
 where the expansions of general color ordered YM amplitudes to
their BCJ-basis have been used (see \eref{BCJ-pqr-collect}). The
right hand side gives
\bea \mathcal{T}_{1\{\rho'\}2} (C_0+C_1) =
 \mathcal{T}_{1\{\rho'\}2} C_0+\mathcal{T}_{1\{\rho'\}2}
 \sum_{\rho} a(\rho)
 \frac{(k_1f_{\rho_3}k_{\rho_2})}{\mathcal{K}_{1\rho_3}}\frac{(k_1f_{\rho_2}f_{\rho_1}k_1)}{\mathcal{K}_{1\rho_2}\mathcal{K}_{1\rho_1}}=\mathcal{T}_{1\{\rho'\}2}
 C_0+\frac{a(\rho')}{(k_1k_{\rho_1})},~~ \eea
 where we have used the result
$\mathcal{T}_{1\{\rho'\}2}\frac{(k_1f_{\rho_3}k_{\rho_2})}{\mathcal{K}_{1\rho_3}}\frac{(k_1f_{\rho_2}f_{\rho_1}k_1)}{\mathcal{K}_{1\rho_2}\mathcal{K}_{1\rho_1}}=\delta_{\rho,\rho'}
\frac{1}{(k_1k_{\rho_1})}$.
Comparing two sides we get
\bea
 a({\rho'})=-(k_1k_{\rho_1}) \mathcal{T}_{1\{\rho'\}2}C_0+
(k_1k_{\rho_1}) \sum_{\rho}\sum_{\shuffle}
\mathcal{C}^{(\rho')}(\rho,\shuffle)
A^{\text{YM}}_{n+3}(1,2,\{3,\cdots,n-1\}\shuffle\{\rho_1,\rho_2,\rho_3\},n).~~~~~~
\label{2coe3} \eea
Having the general formula for $a(\rho')$'s,  we can calculate
their expressions explicitly. To do so, first we rewrite $C_0$
as
\begin{align}
C_0=\sum_{\rho}\sum_{\shuffle} C_0(\rho,\shuffle) A^{\text{YM}}_{n+3}(1,2,\{3,\cdots,n-1\}\shuffle\{\rho_1,\rho_2,\rho_3\},n), \notag
\end{align}
with \bea & &
  C_0(\rho,\shuffle)=\frac{(k_1f_{\rho_1}X_{\rho_1})}{\mathcal{K}_{1\rho_1}}
 \frac{(k_1f_{\rho_2}X_{\rho_2})}{\mathcal{K}_{1\rho_2}}\frac{(k_1f_{\rho_3}X_{\rho_3})}{\mathcal{K}_{1\rho_3}}
+\frac{[k_{\rho_2}(X_{\rho_2}-k_1)](k_{\rho_3}X_{\rho_3})}{\mathcal{K}_{1\rho_2\rho_3}}\frac{(k_1f_{\rho_1}Y_{\rho_1})}{\mathcal{K}_{1\rho_1}}
    \frac{(k_1f_{\rho_2}f_{\rho_3}k_1)}{\mathcal{K}_{1\rho_2}\mathcal{K}_{1\rho_3}}
    \nn
 &&+\frac{[k_{\rho_1}(X_{\rho_1}-k_1)](k_{\rho_3}X_{\rho_3})}{\mathcal{K}_{1\rho_1\rho_3}}\frac{(k_1f_{\rho_2}Y_{\rho_2})}{\mathcal{K}_{1\rho_2}}
    \frac{(k_1f_{\rho_1}f_{\rho_3}k_1)}{\mathcal{K}_{1\rho_1}\mathcal{K}_{1\rho_3}}
    +\frac{[k_{\rho_1}(X_{\rho_1}-k_1)](k_{\rho_2}X_{\rho_2})}{\mathcal{K}_{1\rho_1\rho_2}}\frac{(k_1f_{\rho_3}Y_{\rho_3})}{\mathcal{K}_{1{\rho_3}}}
    \frac{(k_1f_{\rho_1}f_{\rho_2}k_1)}{\mathcal{K}_{1\rho_1}\mathcal{K}_{1\rho_2}},~~~~~~~
    \notag
\eea then $a(\rho')$ becomes
\begin{align}
 a({\rho'})=& \sum_{\rho}\sum_{\shuffle} \Big\{ -(k_1k_{\rho_1}) \mathcal{T}_{1\{\rho'\}2}C_0(\rho,\shuffle)+ (k_1k_{\rho_1}) \mathcal{C}^{(\rho')}(\rho,\shuffle) \Big\} A^{\text{YM}}_{n+3}(1,2,\{3,\cdots,n-1\}\shuffle\{\rho_1,\rho_2,\rho_3\},n) \notag\\
 =& \sum_{\rho}\sum_{\shuffle} a(\rho',\rho) A^{\text{YM}}_{n+3}(1,2,\{3,\cdots,n-1\}\shuffle\{\rho_1,\rho_2,\rho_3\},n),
\end{align}
with
\begin{align}
a(\rho',\rho)=-(k_1k_{\rho_1}) \mathcal{T}_{1\{\rho'\}2}C_0(\rho,\shuffle)+ (k_1k_{\rho_1})
 \mathcal{C}^{(\rho')}(\rho,\shuffle).~~~\label{LI-rel}
\end{align}
After substituting $a(\rho',\rho)$ into (\ref{3expansion3}), we
 get the final expansion of $A^{\text{EYM}}_{n,3}$ amplitudes in
 the BCJ-basis
\begin{align}
A^{\text{EYM}}_{n,3}
=\sum_{\rho}\sum_{\shuffle} C(\rho,\shuffle) A^{\text{YM}}_{n+3}(1,2,\{3,\cdots,n-1\}\shuffle\{\rho_1,\rho_2,\rho_3\},n)
\end{align}
with
\begin{align}
C(\rho,\shuffle)=C_0(\rho,\shuffle)+\sum_{\rho'} a(\rho',\rho).~~~\label{Li-3g-diff}
\end{align}

\end{itemize}

Although \eref{Li-3g-diff} is the final result, to compare with
\eref{3g-nice} given in the  previous section,  we give the explicit
expression of coefficients with the ordering $\rho=(p,q,r)$.
 Because $C_0(\rho,\shuffle)$ has been
given, we only need to calculate $a(\rho',\rho)$. We have left the
calculation details in the Appendix \ref{3gdetail} and just quote
the result of $C(\{pqr\},\shuffle)$
\bea
 & & C(\{pqr\},\shuffle)
 =C_0(\{pqr\},\shuffle)+C_1(\{pqr\},\shuffle) \nn
& =&
\frac{(k_1f_pX_p)}{\mathcal{K}_{1p}}\frac{(k_1f_qX_q)}{\mathcal{K}_{1q}}\frac{(k_1f_rX_r)}{\mathcal{K}_{1r}}
  +\frac{[k_q(X_q-k_1)](k_rX_r)}{\mathcal{K}_{1qr}}\frac{(k_1f_pX_p)}
  {\mathcal{K}_{1p}}\frac{(k_1f_qf_rk_1)}{\mathcal{K}_{1q}\mathcal{K}_{1r}}
  \nn
 &&+ \frac{[k_p(X_p-k_1)](k_rX_r)}{\mathcal{K}_{1pr}}\frac{(k_1f_qX_q)}{\mathcal{K}_{1q}}
 \frac{(k_1f_pf_rk_1)}{\mathcal{K}_{1p}\mathcal{K}_{1r}} +\frac{[k_{p}(X_p-k_1)]
 (k_qX_q)}{\mathcal{K}_{1pq}}\frac{(k_1f_rX_r)}{\mathcal{K}_{1r}}\frac{(k_1f_pf_qk_1)}
 {\mathcal{K}_{1p}\mathcal{K}_{1q}}
  \nn
 & &-\frac{(k_rX_r)[k_q(X_q-k_1)+\mathcal{K}_{1qr}][k_p(Y_p-k_1)]}{\mathcal{K}_{1qr}\mathcal{K}_{1pqr}}\frac{(k_1f_pk_q)}{\mathcal{K}_{1p}}\frac{(k_1f_qf_rk_1)}{\mathcal{K}_{1q}\mathcal{K}_{1r}}  \notag\\
 && - \frac{[k_q(X_q-k_1)](k_rX_r)[k_p(Y_p-k_1)]}{\mathcal{K}_{1qr}\mathcal{K}_{1pqr}}\frac{(k_1f_pk_r)}{\mathcal{K}_{1p}}\frac{(k_1f_rf_qk_1)}{\mathcal{K}_{1q}\mathcal{K}_{1r}} \notag\\
 && - \frac{(k_rX_r)[k_p(Y_p-k_1)][k_q(X_q+k_r)]}{\mathcal{K}_{1pr}\mathcal{K}_{1pqr}}\frac{(k_1f_qk_p)}{\mathcal{K}_{1q}}\frac{(k_1f_pf_rk_1)}{\mathcal{K}_{1p}\mathcal{K}_{1r}}  \notag\\
 && - \frac{[k_p(Y_p-k_1)](k_rX_r)[k_1(Y_q-k_1)]}{\mathcal{K}_{1pr}\mathcal{K}_{1pqr}}\frac{(k_1f_qk_r)}{\mathcal{K}_{1q}}\frac{(k_1f_rf_pk_1)}{\mathcal{K}_{1p}\mathcal{K}_{1r}} \notag\\
 && - \frac{(k_qX_q)[k_p(Y_p-k_1)](k_rX_r)}{\mathcal{K}_{1pq}\mathcal{K}_{1pqr}}\frac{(k_1f_rk_p)}{\mathcal{K}_{1r}}\frac{(k_1f_pf_qk_1)}{\mathcal{K}_{1p}\mathcal{K}_{1q}}  \notag\\
 && - \frac{[k_p(Y_p-k_1)][(k_qX_q)-\mathcal{K}_{1pq}](k_rX_r)}{\mathcal{K}_{1pq}\mathcal{K}_{1pqr}}\frac{(k_1f_rk_q)}{\mathcal{K}_{1r}}\frac{(k_1f_qf_pk_1)}{\mathcal{K}_{1p}\mathcal{K}_{1q}},
\eea
which is the same as the one given in the last section.

Similar to the previous two subsections, we see that to reach
$A^{\text{YM}}_{n+3}(1,\rho_1,\rho_2,\rho_3,2,\cdots,n)$ with the
fixed ordering $\rho$,  there are  five different insertion
operators, i.e.,
\bea & &
\mathcal{T}_{1\rho_12}\mathcal{T}_{\rho_1\rho_32}\mathcal{T}_{\rho_1\rho_2\rho_3},
~~~\mathcal{T}_{1\rho_12}\mathcal{T}_{\rho_1\rho_2
2}\mathcal{T}_{\rho_2\rho_3 2}\nn
& &
\mathcal{T}_{1\rho_22}\mathcal{T}_{1\rho_1\rho_2}\mathcal{T}_{\rho_2\rho_32}\nn
& &
\mathcal{T}_{1\rho_32}\mathcal{T}_{1\rho_1\rho_3}\mathcal{T}_{\rho_1\rho_2\rho_3},
~~~\mathcal{T}_{1\rho_32}\mathcal{T}_{1\rho_2\rho_3}\mathcal{T}_{1\rho_1\rho_2}\eea
which can be used to  act on the left hand side of
\eref{3expansion3}. With six orderings of $\rho$, we will get $29$
equations after identifying the resulted right hand sides of
\eref{3expansion3}. Although we have not explicitly checked, we
believe using them we can solve these six unknown $D$ coefficients,
thus we find the BCJ expansions \eref{BCJ-exp-1} as the consistent
conditions for differential operators.


\section{Conclusion and discussion}
\label{conclusion}

The double copy, and the related KLT relation, imply that amplitudes
of one theory can be expanded in terms of the amplitudes of another
theory as in \eref{gen-5}. Although this is, in principle, a simple
fact, directly using the KLT form \eref{gen-4} to find expansion
coefficients is a very nontrivial task despite many efforts  devoted
to find  alternative ways.

In this paper, we have proposed a new method to efficiently find
expansion coefficients  by using the differential operators
introduced in \cite{Cheung:2017ems}. Moreover, we have actually
achieved more than our initial goal. Using  the proper building
blocks, we have shown that the differential operators, together with
the web connections established in \cite{Cheung:2017ems},  naturally
lead  to the expansion of sEYM amplitude in the KK-basis or
BCJ-basis without any extra assumptions. Furthermore,  with the use
of these building blocks, we have translated the problem into a set
of linear equations, thus greatly reducing its complexity.

For the expansion in the  KK-basis, the expansion coefficients are
polynomial and the natural building blocks are $\eps\cdot \eps$ and
$\eps\cdot k$. With this simplification, the recursive expansion of
sEYM amplitudes to YM amplitudes \eref{eq:singletrace} has been
reproduced by the new method. As a further demonstration of the
efficacy of the new techniques, we have also discussed the expansion
of gravity theory and Born-Infeld theory in the KK-basis.

For the expansion in the BCJ-basis, which has not been dealt with
much in previous works, finding the expansion coefficients is a much
more difficult task. The technical challenge lies in that a proper
understanding of the building blocks, for arbitrary number of
gravitons, is still missing. As  explained above, the building
blocks $\eps\cdot \eps$ and $\eps\cdot k$ in the KK-basis expansion
are too large. Even if we constrain ourselves to the gauge invariant
combinations, such as $(k_1 f X)$ and $(kf...fk)$, they are still
too large. As we have argued, the index cycle structure should not
appear. Furthermore, as given in \eref{ff-rel}, not all allowed
gauge invariant building blocks are independent of each other. Thus
a proper understanding of these building blocks becomes one of the most
important problems.


Both expansions are worth studying. For the expansion in the
KK-basis, since all physical poles are included in the basis,
expansion coefficients can be arranged to be polynomials. Thus such
an expansion is more suitable when considering the analytical
structure of amplitudes. However, in this manner, the manifest gauge
invariance of gravitons is lost. On the other hand, for the
expansion in the BCJ-basis, the gauge invariance of all gravitons is
manifest. The price to pay is that some physical poles are moved to
the coefficients.

Due to  the  difficulty related to gauge invariant building blocks,
as pointed out before, we could not give the complete solution for
the expansion of sEYM amplitudes in the BCJ-basis of YM amplitudes
at this moment. Thus a clear understanding of the building blocks
will be an important future problem. Our examples of the previous
section, especially the recursive construction,  may provide us with
some guidance in this issue. Our results here show that differential
operators introduced in \cite{Cheung:2017ems} provide a much wider
set of  applications (such as the three generalized relations
presented in \eref{color-rev}, \eref{pho-dec-gen} and
\eref{gener-KK-rela}) and have a deeper meaning (such as the
relation between expansion coefficients and BCJ expansion
coefficients given in \eref{LI-rel}) than previously thought.
Therefore, further studies are needed in order to exploit their full
potential, such as, for instance, their applications to soft and
collinear limits.

\section*{Acknowledgments}

This work is supported by Qiu-Shi Funding and Chinese NSF funding
under Grant No.11575156, No.11935013 and No.11805163, as well as NSF funding of
Jiangsu Province under Grant No.BK20180897.

\appendix

\section{Terms with index circle structure}\label{cycle-index}

When we try to derive   the recursive expansion of sEYM amplitudes,
we encounter some terms like $(\epsilon_{h_1}\cdot f_{h_j}\cdot
f_{h_{j_2}}\cdot k_{h_1})B_{h_jh_{j_2}h_1}$ and
$(\epsilon_{h_1}\cdot f_{h_j}\cdot f_{h_{j_2}}\cdot
k_{h_j})B_{h_jh_{j_2}h_j}$, where $f=k\eps-\eps k$. If we expand $f$
in these corresponding building blocks completely, such as
$(\epsilon_{h_1}\cdot f_{h_j}\cdot f_{h_{j_2}}\cdot k_{h_1})$ and
$(\epsilon_{h_1}\cdot f_{h_j}\cdot f_{h_{j_2}}\cdot k_{h_j})$, we
will get terms containing factors like  $(\epsilon_{i_1}\cdot
k_{i_2})(\epsilon_{i_2}\cdot k_{i_3})\cdots(\epsilon_{i_{s}}\cdot
k_{i_1})$, where all $i_t$ are gravitons. We will call such a
structure the "index cycle structure". In this appendix we will show
that all building blocks with the index cycle structure will not
appear in the expansion.

To prove above claim, we need to consider some special  combinations
of insertion operators, such as the
$\mathcal{T}_{ah_{i_1}h_{i_2}}$$\mathcal{T}_{ah_{i_2}h_{i_3}}\cdots
\mathcal{T}_{ah_{i_s}h_{i_1}}$, where $a$ represents a gluon and
$h_i$'s represent  gravitons. Acting it on both sides of
 the expansion of sEYM amplitude $A^{\text{EYM}}_{n,m}$,
on the right hand side only the term
$(\epsilon_{h_{i_1}}f_{h_{i_2}}\cdots
f_{h_{i_s}}k_{h_{i_1}})B_{h_{i_1}\cdots h_{i_s}h_{i_1}}$ is not
annihilated, while at the left hand side we have
\begin{align}
\mathcal{T}_{ah_{i_1}h_{i_2}}\mathcal{T}_{ah_{i_2}h_{i_3}}\cdots \mathcal{T}_{ah_{i_s}h_{i_1}}A^{\text{EYM}}_{n,m}
=& \int d\mu \text{PT}(1,2,\cdots,n) (\mathcal{T}_{ah_{i_1}h_{i_2}}
\mathcal{T}_{ah_{i_2}h_{i_3}}\cdots \mathcal{T}_{ah_{i_s}h_{i_1}}
\text{Pf}\Psi_{H_m}) \text{Pf}'\Psi.~~~~\label{cycle-zero}
\end{align}
Thus to fix unknown $B_{h_{i_1}\cdots h_{i_s}h_{i_1}}$  we just need
to work out the effect of these operators on $\text{Pf}\Psi_{H_m}$.

According to \cite{Lam:2016tlk,He:2016iqi}, $\text{Pf}\Psi_{H_m}$
can be expanded as the sum of permutations like
\begin{align}
\text{Pf}\Psi_{H_m}=\sum_{1\le i_1\le i_2\le \cdots\le i_m\le n\atop i_1+i_2+\cdots+i_m=n}
(-1)^{n-m}P_{i_1i_2\cdots i_m},
\end{align}
where the sum is organized by the unique cycle decomposition of
these permutations. When the length of a cycle is one, it is given
by $\Psi_{(h_i)}=-\sum_{b\ne h_i}\frac{\epsilon_{h_i}\cdot
k_b}{\sigma_{h_ib}}$. When the length of a cycle is bigger than one,
it's given by $\Psi_{(h_{i_1}\cdots
h_{i_r})}=\frac{\text{tr}(f_{h_{i_1}}\cdots
f_{h_{i_r}})}{2\sigma_{h_{i_1}h_{i_2}}\cdots
\sigma_{h_{i_r}h_{i_1}}}$. For example,
\begin{align}
\text{Pf}\Psi_3
=& P_{111}-P_{12}+P_3
= \Psi_{(1)}\Psi_{(2)}\Psi_{(3)}-\Psi_{(1)}\Psi_{(23)}-\Psi_{(2)}\Psi_{(13)}-\Psi_{(3)}\Psi_{(12)}+\Psi_{(123)}+\Psi_{(132)}.
\end{align}

Having the expression of $\text{Pf}\Psi_{H_m}$,  we  present two
simple examples to show why \eref{cycle-zero} is zero at the first,
and then rigorous proof follows. Let us start with the simplest
case, i.e., the one with only two gravitons. Without loss of
generality, let us consider the operator
$\mathcal{T}_{ah_1h_2}\mathcal{T}_{ah_2h_1}$. Putting it to
\eref{cycle-zero}, we have
\bea & &  \mathcal{T}_{ah_1h_2}\mathcal{T}_{ah_2h_1}
\text{Pf}\Psi_{H_m} =\mathcal{T}_{ah_1h_2}\mathcal{T}_{ah_2h_1}
\sum_{1\le i_1\le i_2\le \cdots\le i_m\le n\atop
i_1+i_2+\cdots+i_m=n} (-1)^{n-m}P_{i_1i_2\cdots i_m} \nn
& =& \mathcal{T}_{ah_1h_2}\mathcal{T}_{ah_2h_1}\Big\{
\Psi_{(1)}\Psi_{(2)} \Psi_{H_{m-2}}-\Psi_{(12)}
\Psi_{H_{m-2}}+\Psi_{(1)}\Psi_{(2\cdots)}(\cdots)
+\Psi_{(2)}\Psi_{(1\cdots)}(\cdots) +\Psi_{(1..2...)}(\cdots) \Big\}
\nn
& =& \mathcal{T}_{ah_1h_2}\mathcal{T}_{ah_2h_1}\Big\{
\Psi_{(1)}\Psi_{(2)} -\Psi_{(12)}
\Big\}\text{Pf}\Psi_{H_{m-2}}.~~~~\label{cycle-exp-1} \eea
In \eref{cycle-exp-1}, $\Psi_{(2\cdots)}$ represents any cycle
containing $h_2$ and other gravitons than $h_1$, and similar
understanding for $\Psi_{(1\cdots)}$ and $\Psi_{(1\cdots 2\cdots)}$.
All of them are annihilated by at least one of insertion operators.
Thus among all terms in the sum of $\text{Pf}\Psi$, only the first
two give nonzero contributions. Carrying it out explicitly, we get
\begin{align}
\mathcal{T}_{ah_1h_2}\mathcal{T}_{ah_2h_1} \text{Pf}\Psi_{H_m}
=& \mathcal{T}_{ah_1h_2}\mathcal{T}_{ah_2h_1}\Big\{ \Psi_{(1)}\Psi_{(2)} -\Psi_{(12)}
\Big\}\text{Pf}\Psi_{H_{m-2}}\notag\\
=& \Big\{ \frac{\sigma_{h_2a}}{\sigma_{h_1a}\sigma_{h_1h_2}}
\frac{\sigma_{h_1a}}{\sigma_{h_2h_1}\sigma_{h_2a}} -\frac{1}{\sigma_{h_1h_2}
\sigma_{h_2h_1}} \Big\} \text{Pf}\Psi_{H_{m-2}}
=0.
\end{align}

The next simple case is the one with three gravitons, for example,
the string $(h_1, h_2, h_3)$. By similar reason, the action of
$\mathcal{T}_{ah_1h_2}\mathcal{T}_{ah_2h_3}\mathcal{T}_{ah_3h_1}$ is
\begin{align}
&\mathcal{T}_{ah_1h_2}\mathcal{T}_{ah_2h_3}\mathcal{T}_{ah_3h_1} \text{Pf}\Psi_{H_m} \notag\\
=&\mathcal{T}_{ah_1h_2}\mathcal{T}_{ah_2h_3}\mathcal{T}_{ah_3h_1} \Big\{
 \Psi_{(h_1)}\Psi_{(h_2)}\Psi_{(h_3)}
 -\Psi_{(h_1)}\Psi_{(h_2h_3)}-\Psi_{(h_2)}\Psi_{(h_1h_3)}-\Psi_{(h_3)}\Psi_{(h_1h_2)} \notag\\
 &+\Psi_{(h_1h_2h_3)}+\Psi_{(h_1h_3h_2)}
 \big\} \text{Pf}\Psi_{H_{m-3}},
\end{align}
with $\text{Pf}\Psi_{H_{m-2}}$ the Pfaffian of other gravitons.
 Carrying it out explicitly,
first we note that
\begin{align}
&\mathcal{T}_{ah_1h_2}\Psi_{(h_1)}\mathcal{T}_{ah_2h_3}\mathcal{T}_{ah_3h_1}\Psi_{(h_2h_3)}
= \frac{\sigma_{h_2a}}{\sigma_{h_1h_2}\sigma_{h_1a}}   \mathcal{T}_{ah_2h_3}
\left\{\mathcal{T}_{ah_3h_1} \frac{[(\epsilon_{h_2}f_{h_3}k_{h_2})-(k_{h_2}f_{h_3}
\epsilon_{h_2})]}{2\sigma_{h_2h_3}\sigma_{h_3h_2}}\right\}
=0,
\end{align}
 because $\Psi_{(h_2h_3)}$ doesn't contain
$(\epsilon_{h_3}k_{h_1})$ and $(\epsilon_{h_3}k_a)$, thus
$\mathcal{T}_{ah_3h_1}\Psi_{(h_2h_3)}=0$.  Similarly we have
\begin{align}
\mathcal{T}_{ah_2h_3}\Psi_{(h_2)}\mathcal{T}_{ah_1h_2}\mathcal{T}_{ah_3h_1}\Psi_{(h_1h_3)}
=0, \qquad
\mathcal{T}_{ah_3h_1}\Psi_{(h_3)}\mathcal{T}_{ah_1h_2}\mathcal{T}_{ah_2h_3}\Psi_{(h_1h_2)}
=0.
\end{align}
For the remaining three terms, we have
\bea & &
\mathcal{T}_{ah_1h_2}\mathcal{T}_{ah_2h_3}\mathcal{T}_{ah_3h_1}
\text{Pf}\Psi_{H_m}\nn
& =&\Big\{ \frac{\sigma_{h_2a}}{\sigma_{h_1h_2}\sigma_{h_1a}}
\frac{\sigma_{h_3a}}{\sigma_{h_2h_3}\sigma_{h_2a}}
\frac{\sigma_{h_1a}}{\sigma_{h_3h_1}\sigma_{h_3a}} -
\frac{1}{2\sigma_{h_1h_2}\sigma_{h_2h_3}\sigma_{h_3h_1}}
+\frac{1}{2\sigma_{h_1h_3}\sigma_{h_3h_2}\sigma_{h_2h_1}} \Big\}
\text{Pf}\Psi_{H_{m-3}} \nn
& =& \Big\{
 \frac{1}{\sigma_{h_1h_2}\sigma_{h_2h_3}\sigma_{h_3h_1}}
 -\frac{1}{2\sigma_{h_1h_2}\sigma_{h_2h_3}\sigma_{h_3h_1}}+\frac{1}{2\sigma_{h_1h_3}\sigma_{h_3h_2}\sigma_{h_2h_1}}
 \Big\} \text{Pf}\Psi_{H_{m-3}}
= 0. \eea

From the previous two simple examples, we can see how to generalize
to strings of  $s$ gravitons, for example, the $(h_1,h_2,...,h_s)$.
With the action of insertion operators, we have
\begin{align}
\mathcal{T}_{ah_{1}h_{2}}\mathcal{T}_{ah_{2}h_{3}}\cdots \mathcal{T}_{ah_{s}h_{1}} \text{Pf}\Psi_{H_m}
=& \mathcal{T}_{ah_{1}h_{2}}\mathcal{T}_{ah_{2}h_{3}}\cdots \mathcal{T}_{ah_{s}h_{1}} \sum_{1\le i_1\le i_2\le \cdots\le i_m\le n\atop i_1+i_2+\cdots+i_m=n} (-1)^{n-m}P_{i_1i_2\cdots i_m} \notag\\
=& \mathcal{T}_{ah_{1}h_{2}}\mathcal{T}_{ah_{2}h_{3}}\cdots \mathcal{T}_{ah_{s}h_{1}} \Big\{
\sum_{1\le j_1\le j_2\le \cdots\le j_l\le s\atop j_1+j_2+\cdots+j_l=s} (-1)^{s-1}
P_{j_1j_2\cdots j_l} \Big\} \text{Pf}\Psi_{H_{m-s}},
\end{align}
where $\text{Pf}\Psi_{H_{m-s}}$ is the Pfaffian of  other gravitons.
Because of the particular circle order $(h_1,h_2,\cdots,h_s)$, the
action of insertion operator $\mathcal{T}_{ah_{j}h_{j+1}}$ is not
zero when and only when there is the factor $\eps_{h_j}\cdot k_a$ or
$\eps_{h_j}\cdot k_{h_{j+1}}$. Thus only terms, which are either the
multiplication of $s$'s length one cycles or the length $s$ cycles
having the same or reversing  order, will contribute. With the
simplification, we have
\begin{align}
&\mathcal{T}_{ah_{1}h_{2}}\mathcal{T}_{ah_{2}h_{3}}\cdots \mathcal{T}_{ah_{s}h_{1}} \text{Pf}\Psi_{H_m} \notag\\
=& \mathcal{T}_{ah_{1}h_{2}}\mathcal{T}_{ah_{2}h_{3}}\cdots \mathcal{T}_{ah_{s}h_{1}} \Big\{
\Psi_{h_1}\Psi_{h_2}\cdots\Psi_{h_s}+(-1)^{s-1} \Psi_{(h_1h_2\cdots h_s)}+(-1)^{s-1} \Psi_{(h_sh_{s-1}\cdots h_1)} \Big\} \Psi_{H_{m-s}} \notag\\
=& \Big\{ \frac{\sigma_{h_2a}}{\sigma_{h_1a}\sigma_{h_1h_2}} \frac{\sigma_{h_3a}}{\sigma_{h_2a}\sigma_{h_2h_3}}\cdots \frac{\sigma_{h_1a}}{\sigma_{h_sa}\sigma_{h_sh_1}}+(-1)^{s-1}(-1)^s\frac{1}{2\sigma_{h_1h_2}\sigma_{h_2h_3}\cdots\sigma_{h_sh_1}} \notag\\
 &+(-1)^{s-1}\frac{1}{2\sigma_{h_sh_{s-1}}\sigma_{h_{s-1}h_{s-2}}\cdots\sigma_{h_1h_s}} \Big\}\Psi_{H_{m-s}} \notag\\
=& \Big\{ \frac{1}{\sigma_{h_1h_2}\sigma_{h_2h_3}\cdots\sigma_{h_sh_1}}-\frac{1}{\sigma_{h_1h_2}\sigma_{h_2h_3}\cdots\sigma_{h_sh_1}} \Big\} \Psi_{H_{m-s}} \notag\\
=& 0.
\end{align}
So finally we have proved that any term with the building block
having index circle structure  will vanish.

\section{ General discussions of manifestly gauge invariant functions
}\label{gauge-invariance}

In the appendix, we make some general discussions of manifestly
gauge invariant functions, which are important for the construction
of building blocks when expanding in the BCJ basis. For amplitudes
involving gauge particles, they must satisfy some basic
requirements: Lorentz invariance, on-shell condition, momentum
conservation, transversality \footnote{Here, we follow the name from
\cite{Boels:2017gyc}, it means $\epsilon_i\cdot k_i=0$.}, gauge
invariance, etc. With those requirements, properties of amplitudes
have been thoroughly discussed  in
\cite{Barreiro:2013dpa,Boels:2016xhc,Boels:2017gyc,Arkani-Hamed:2016rak,Bern:2017tuc}.
Especially, in \cite{Boels:2016xhc} starting from the fact that
amplitudes are multi-linear functions of $\eps$'s the authors are
able to write down a set of linear equations coming from gauge
invariance condition. After solving them,  they can derive some
important consequence. Going further in \cite{Arkani-Hamed:2016rak}
the authors prove that using only gauge invariance, locality and
minimal power counting, one can determine amplitudes uniquely.

Since gauge invariance is a so important and strong constraint, it
is better to construct functions which are manifestly gauge
invariant by sacrificing other properties \cite{Bern:2017tuc}. By
Feynman rules amplitudes are multi-linear functions of polarization
vectors. Gauge invariant condition means that when replacing
$\epsilon_i^{\mu}$ by $k_i^{\mu}$ for each $i$, amplitudes vanish. A
simple and naive combination satisfying above condition is
$(\epsilon_i^{\mu}-k_i^{\mu})$, which is wrong since the  dimension
of two terms does not match. Correcting with dimension, one leads to
the form $(\epsilon_i^{\mu}k_j^{\nu}-\epsilon_r^{\nu}k_i^{\mu})$.
Imposing gauge invariant condition for each $i$, we end up the
combination $\epsilon_i^{\mu}k_i^{\nu}-k_i^{\mu}\epsilon_i^{\nu}$,
which is nothing, but the familiar field strength
$f_i^{\mu\nu}=k_i^{\mu}\epsilon_i^{\nu}-\epsilon_i^{\mu}k_i^{\nu}$.
Above argument seems to indicate that any manifestly gauge invariant
function could be rewritten  as the function of $f^{\mu\nu}$.
Although we could not give complete proof about this statement,  for
some special cases, such as the sEYM amplitudes studied in this
paper, we will show it now.

Actually, we can go further.  Lorentz invariance requires that
$f^{\mu\nu}$ must be contracted by a momentum or another
$f^{\mu\nu}$, thus eventually we will reach two types of
contractions: Type-I with the form $(k_a\cdot f_b\cdots f_c\cdot
k_d)$ and Type II with the form $\text{tr}(f_a\cdot f_b\cdots
f_c\cdot f_d)\equiv(f_{a\nu}^\mu f_{b\rho}^\nu\cdots
f_{c\beta}^\alpha f_{d\mu}^\beta)$. Lorentz and gauge invariant
building blocks should be functions of both types,  but in this
paper we only encounter the Type-I contractions\footnote{In
\cite{Bern:2017tuc}, four point amplitude of gauge theory has been
rewritten to a form using the Type-II contractions. }, so we just
discuss some properties of them.

For the Type-I contraction, we use the number of $f$'s appearing in
the term to characterize them (called the $f$-degree), thus we have
$d_f=m-2$, where $d_f$ is the $f$-degree while $m$ is the mass
dimension. Some examples are $(k_af_bk_c)$ of degree one and
$(k_if_jf_lk_s)$ of degree two. Two properties of Type-I
contractions can be easily proved.  The first is
\begin{align}
 (k_{i_1}f_{i_2}\cdots f_{i_{m-1}}k_{i_m})=(-1)^{m-2}(k_{i_m}f_{i_{m-1}}\cdots f_{i_2}k_{i_1}).
\end{align}
The second is
\begin{align}
 &(k_{i_1}f_{i_2}\cdots f_{i_{a-1}}f_{i_{a}}f_{i_{a+1}}\cdots f_{i_{m-1}}k_{i_m})(k_jk_{i_a}) \notag\\
 =& (k_{i_1}f_{i_2}\cdots f_{i_{a-1}}k_{i_a})(k_jf_{i_{a}}f_{i_{a+1}}
 \cdots f_{i_{m-1}}k_{i_m})+(k_{i_1}f_{i_2}\cdots f_{i_{a-1}}
 f_{i_{a}}k_j)(k_{i_a}f_{i_{a+1}}\cdots f_{i_{m-1}}k_{i_m})~, \label{ftermformula}
\end{align}
which is just the identity \eref{use-identity-1}. The formula
(\ref{ftermformula}) tells us that any Type-I contraction with
higher degrees can always be decomposed to the sum of terms as the
product of Type-I contractions with  lower degrees. This
decomposition terminates at Type-I contractions of degree one and
degree two, which will be called fundamental. Relations
\eref{ftermformula} give the first indication why the building
blocks for BCJ-basis are much more complicated.

Having above general discussions, let us move to the case related to
sEYM amplitudes.

\subsection{Having only one polarization vector}

Now we consider the simplest example, i.e., a general function of
$(n+1)$ momenta and only one polarization vector. As the linear
functions of $\eps$,  the functions can be generally written as
\begin{align}
 F=F(k_1,\cdots,k_n,k_p,\epsilon_p)=\sum_{i=1}^{n-1}\alpha_i(\epsilon_p\cdot k_i),
\end{align}
where $\alpha_i$'s are  unknown functions of $(k_j\cdot k_l)$. Here
we have used momentum conservation to eliminate the momentum $k_n$,
so all remaining $(\epsilon_p\cdot k_i)$'s are independent. Gauge
invariant condition leads to
\begin{align}
 F(\epsilon_p\rightarrow k_p)=\sum_{i=1}^{n-1}\alpha_i(k_p\cdot k_i)=0, \label{1polarization}
\end{align}
Solving
\begin{align}
 \alpha_1=-\sum_{i=2}^{n-1}\alpha_i \frac{(k_p\cdot k_i)}{(k_p\cdot k_1)},
\end{align}
from the equation and putting it back, we get
\bea
 F & =& \sum_{i=2}^{n-1}\alpha_i(\epsilon_p\cdot k_i)-(\epsilon_p\cdot k_1)
 [-\sum_{i=2}^{n-1}\alpha_i \frac{(k_p\cdot k_i)}{(k_p\cdot k_1)}]
 =\sum_{i=2}^{n-1}\alpha_i \frac{(k_1f_pk_i)}{{\cal
 K}_{1p}}~~~\label{1g-building}
\eea
with
$(k_1f_pk_i)=k_{1\mu}(k_p^{\mu}\epsilon_p^{\nu}-\epsilon_p^{\mu}
k_p^{\nu})k_{i\nu}$ and ${\cal K}_{1p}=(k_p\cdot k_1)$. Thus we have
shown the appearance of Type-I contractions  in this
case\footnote{We should note that the number of momenta should be
greater than $3$, since when $n+1=3$, the function will vanish
because of the special kinematics of three particles.}. Result
\eref{1g-building} tells us that the basis of  gauge invariant
building blocks of single polarization vector is given by
$(k_1f_pk_i)$ (or $\frac{(k_1f_pk_i)}{{\cal
 K}_{1p}}$) with $i=2,...,n-1$. A equivalent, but more convenient
basis can be taken as $(k_1f_p K_i)$ (or $\frac{(k_1f_pK_i)}{{\cal
 K}_{1p}}$) with $i=2,...,n-1$ and $K_i=\sum_{t=1}^i k_t$.

\subsection{Having two polarization vectors}

Now we consider the case with $(n+2)$ momenta and two polarization
vectors. The general form of the function $F$ is
\begin{align}
 F=F(k_1,\cdots,k_n,k_p,k_q,\epsilon_p,\epsilon_q)=\alpha(\epsilon_p\cdot \epsilon_q)+\sum_{i=1}^{n-1,q}\sum_{j=1}^{n-1,p}\beta_{ij}(\epsilon_p\cdot k_i)(\epsilon_q\cdot k_j),
\end{align}
where we have used the momentum conservation to eliminate the
contractions $(\epsilon_p\cdot k_{n})$ and $(\epsilon_q\cdot
k_{n})$. Gauge invariant condition of $p$ leads to
\begin{align}
 F(\epsilon_p\rightarrow k_p)
 =& \alpha(k_p\cdot \epsilon_q)+\sum_{i=1}^{n-1,q}\sum_{j=1}^{n-1,p}\beta_{ij}(k_p\cdot k_i)(\epsilon_q\cdot k_j) \notag\\
 =& (\epsilon_q\cdot k_p)[\alpha+\sum_{i=1}^{n-1,q}\beta_{ip}(k_p\cdot k_i)]+\sum_{j=1}^{n-1}[\sum_{i=1}^{n-1,q}\beta_{ij}(k_p\cdot k_i)](\epsilon_q\cdot k_j) \notag\\
 =& 0.
\end{align}
By the independence of $(\epsilon_q\cdot k_p)$ and $(\epsilon_q\cdot k_j)$'s,
we get the first set of equations
\begin{align}
 \left\{ \begin{array}{ll}
   \alpha+\sum_{i=1}^{n-1,q}\beta_{ip}(k_p\cdot k_i)= 0  \notag\\
   \sum_{i=1}^{n-1,q}\beta_{ij}(k_p\cdot k_i)= 0 & \text{for $j=1,2,\cdots,n-1$}.
   \end{array} \right.
\end{align}
Similarly the gauge invariant condition of $q$ gives us another set
of equations
\begin{align}
 \left\{ \begin{array}{ll}
   \alpha+\sum_{j=1}^{n-1,p}\beta_{qj}(k_q\cdot k_j)= 0  \notag\\
   \sum_{j=1}^{n-1,p}\beta_{ij}(k_q\cdot k_j)= 0 & \text{for $i=1,2,\cdots,n-1$}.
   \end{array} \right.
\end{align}
Now we  solve these equations. Using the first set of equations we
get
\begin{align}
   \beta_{1p}=& -\alpha\frac{1}{(k_p\cdot k_1)}-\sum_{i=2}^{n-1,q}\beta_{ip} \frac{(k_p\cdot k_i)}{(k_p\cdot k_1)} \notag\\
 \beta_{1j}=& -\sum_{i=2}^{n-1,q}\beta_{ij} \frac{(k_p\cdot k_i)}{(k_p\cdot k_1)} \qquad \text{for $j=1,2,\cdots,n-1$},
 \end{align}
while using  the second set of equations we get
\begin{align}
 \beta_{q1}=& -\alpha \frac{1}{(k_1\cdot k_q)}-\sum_{j=2}^{n-1,p}\beta_{qj}\frac{(k_q\cdot k_j)}{(k_1\cdot k_q)} \notag\\
 \beta_{i1}=& -\sum_{j=2}^{n-1,p}\beta_{ij}\frac{(k_q\cdot k_j)}{(k_1\cdot k_q)} \qquad \text{for $i=1,2,\cdots,n-1$}.
\end{align}
Putting  these results back and doing some algebraic manipulations,
we reach
\bea
  F=&-\alpha \frac{(k_1\cdot f_p\cdot f_q\cdot k_1)} {(k_1\cdot k_p)
  (k_1\cdot k_q)}+\sum_{i=2}^{n-1,q}\sum_{j=2}^{n-1,p}\beta_{ij}
  \frac{(k_1\cdot f_p\cdot k_i)} {(k_1\cdot k_p)}\frac{(k_1\cdot f_q\cdot k_j)}
  {(k_1\cdot k_q)},~~~~\label{appendix-2g}
\eea
which is manifestly gauge invariant with only field strength
appearing at the cost of introducing poles $(k_1\cdot k_p)$ and
$(k_1\cdot k_q)$\footnote{There is a choice when solving $\beta$'s.
Different choices lead to different poles.}. Above result is
applicable when $(n+2)>4$. For the special case $n=2$, the  $F$ is
\begin{align}
  F=&-\alpha \frac{(k_1\cdot f_p\cdot f_q\cdot k_1)} {(k_1\cdot k_p)(k_1\cdot k_q)}+\beta_{qp} \frac{(k_1\cdot f_p\cdot k_q)} {(k_1\cdot k_p)}\frac{(k_1\cdot f_q\cdot k_p)}{(k_1\cdot k_q)}.
\end{align}
Result \eref{appendix-2g} shows that with two polarization vectors,
the basis of gauge invariant building blocks contains following
three types:
\bea {\rm type-I}: &~~~ & \frac{(k_1\cdot f_p\cdot k_i)} {(k_1\cdot
k_p)}\frac{(k_1\cdot f_q\cdot k_j)}
  {(k_1\cdot k_q)},~~~i,j=2,...,n-2 \nn
{\rm  type-II}: &~~~ & \frac{(k_1\cdot f_p\cdot k_q)} {(k_1\cdot
k_p)}\frac{(k_1\cdot f_q\cdot k_j)}
  {(k_1\cdot k_q)},~~~\frac{(k_1\cdot f_p\cdot k_i)} {(k_1\cdot
k_p)}\frac{(k_1\cdot f_q\cdot k_p)}
  {(k_1\cdot k_q)},~~~i,j=2,...,n-2 \nn
  {\rm type-III}: & ~~~& \frac{(k_1\cdot f_p\cdot f_q\cdot k_1)} {(k_1\cdot k_p)
  (k_1\cdot k_q)}~~~~\label{2g-building-1} \eea
We can also replace $k_i\to K_i$ to get equivalent basis.

\subsection{Having three polarization vectors}

Now we consider the case with three polarization vectors. The general form of $F$ can be written as
\begin{align}
 F =& \alpha^{ijl}(\epsilon_{n+1}\cdot k_i)(\epsilon_{n+2}\cdot k_j)(\epsilon_{n+3}\cdot k_l) \notag\\
   +& \beta_1^{i}(\epsilon_{n+1}\cdot k_i)(\epsilon_{n+2}\cdot \epsilon_{n+3}) +\beta_2^{j}(\epsilon_{n+2}\cdot k_j)(\epsilon_{n+1}\cdot\epsilon_{n+3})
        +\beta_3^l (\epsilon_{n+3}\cdot k_l)(\epsilon_{n+1}\cdot\epsilon_{n+2})
\end{align}
where the summations of $i,j,l$ are implicit for simplicity, and
their ranges are $i\in \{1,2,\cdots,n-1,n+2,n+3\}$,
$j\in\{1,2,\cdots,n-1,n+1,n+3\}$ and $l\in
\{1,2,\cdots,n-1,n+1,n+2\}$. Again we have used momentum
conservation to eliminate $k_n$.

Using the  gauge invariant conditions for
$\epsilon_{n+1}$,$\epsilon_{n+2}$ and $\epsilon_{n+3}$, we can get
three sets of linear equations. The first set is
\begin{align}
 &\sum_{i\ne n,n+1}\beta^i_1(k_{n+1}\cdot k_i)=0, \notag\\
 &\sum_{i\ne n,n+1}\alpha^{i(n+1)(n+1)}(k_{n+1}\cdot k_i)+\beta^{n+1}_2+\beta^{n+1}_3 =0, \notag\\
 &\sum_{i\ne n,n+1}\alpha^{ij(n+1)}(k_{n+1}\cdot k_i)+\beta^{j}_2 =0, \qquad \text{for $j\ne n,n+1,n+2$}, \notag\\
 &\sum_{i\ne n,n+1}\alpha^{i(n+1)l}(k_{n+1}\cdot k_i)+\beta^{l}_3 =0, \qquad \text{for $l\ne n,n+1,n+3$}, \notag\\
 &\sum_{i\ne n,n+1}\alpha^{ijl}(k_{n+1}\cdot k_i)=0,\qquad \text{for $j\ne n,n+1,n+2$ and $l\ne n,n+1,n+3$}.
\end{align}
The second set is
\begin{align}
 &\sum_{j\ne n,n+2}\beta^j_2(k_{n+2}\cdot k_j)=0, \notag\\
 &\sum_{j\ne n,n+2}\alpha^{(n+2)j(n+2)}(k_{n+2}\cdot k_j)+\beta^{n+2}_1+\beta^{n+2}_3 =0, \notag\\
 &\sum_{j\ne n,n+2}\alpha^{ij(n+2)}(k_{n+2}\cdot k_j)+\beta^{i}_1 =0, \qquad \text{for $i\ne n,n+1,n+2$}, \notag\\
 &\sum_{j\ne n,n+2}\alpha^{(n+2)jl}(k_{n+2}\cdot k_j)+\beta^{l}_3 =0, \qquad \text{for $l\ne n,n+2,n+3$}, \notag\\
 &\sum_{j\ne n,n+2}\alpha^{ijl}(k_{n+2}\cdot k_j)=0, \qquad \text{for $i\ne n,n+1,n+2$ and $l\ne n,n+2,n+3$}.
\end{align}
The third set is
\begin{align}
 &\sum_{l\ne n,n+3}\beta^l_3(k_{n+3}\cdot k_l)=0, \notag\\
 &\sum_{l\ne n,n+3}\alpha^{(n+3)(n+3)l}(k_{n+3}\cdot k_l)+\beta^{n+3}_1+\beta^{n+3}_2 =0, \notag\\
 &\sum_{l\ne n,n+3}\alpha^{i(n+3)l}(k_{n+3}\cdot k_l)+\beta^{i}_1 =0, \qquad \text{for $i\ne n,n+1,n+3$}, \notag\\
 &\sum_{l\ne n,n+3}\alpha^{(n+3)jl}(k_{n+3}\cdot k_l)+\beta^{j}_2 =0, \qquad \text{for $j\ne n,n+2,n+3$}, \notag\\
 &\sum_{l\ne n,n+3}\alpha^{ijl}(k_{n+3}\cdot k_l)=0, \qquad \text{for $i\ne n,n+1,n+3$ and $j\ne n,n+2,n+3$}.
\end{align}
For simplicity, in the sum  we just indicate the unallowed  indexes,
for example, $i\ne n,n+1$ means the sum range is
$i=1,2,\cdots,n-1,n+2,n+3$. From these three sets of equations we
can solve some $\beta$ variables and then put them back to get the
manifestly gauge invariant form of $F$ as
\begin{align}
 F=& \sum_{i\ne 1,n,n+1}\sum_{j\ne 1,n,n+2}\sum_{l\ne 1,n,n+3}\alpha^{ijl} \frac{(k_1\cdot f_{n+1}\cdot k_i)}{(k_1\cdot k_{n+1})}
      \frac{(k_1\cdot f_{n+2}\cdot k_j)}{(k_1\cdot k_{n+2})} \frac{(k_1\cdot f_{n+3}\cdot k_l)}{(k_1\cdot k_{n+3})} \notag\\
  &- \sum_{i\ne 1,n,n+1}\beta_1^i \frac{(k_1\cdot f_{n+1}\cdot k_i)}{(k_1\cdot k_{n+1})}
     \frac{(k_1\cdot f_{n+2}\cdot f_{n+3}\cdot k_1)}{(k_1\cdot k_{n+2})(k_1\cdot k_{n+3})} \notag\\
  &- \sum_{j\ne 1,n,n+2}\beta_2^j \frac{(k_1\cdot f_{n+2}\cdot k_j)}{(k_1\cdot k_{n+2})}
     \frac{(k_1\cdot f_{n+1}\cdot f_{n+3}\cdot k_1)}{(k_1\cdot k_{n+1})(k_1\cdot k_{n+3})} \notag\\
  &- \sum_{l\ne 1,n,n+3}\beta_3^l \frac{(k_1\cdot f_{n+3}\cdot k_l)}{(k_1\cdot k_{n+3})}
     \frac{(k_1\cdot f_{n+1}\cdot f_{n+1}\cdot k_1)}{(k_1\cdot k_{n+1})(k_1\cdot k_{n+2})}.
     ~~~\label{appendix-3g}
\end{align}
As before, the special role of $k_1$ comes from our particular
choice when solving equations. Above form is applicable when
$n+3>4$. When $n+3=4$, $F$ is simplified to
\begin{align}
 F =& \alpha^{ijl}(\epsilon_{2}\cdot k_i)(\epsilon_{3}\cdot k_j)(\epsilon_{4}\cdot k_l) \notag\\
   +& \beta_1^{i}(\epsilon_{2}\cdot k_i)(\epsilon_{3}\cdot \epsilon_{4}) +\beta_2^{j}(\epsilon_{3}\cdot k_j)(\epsilon_{2}\cdot\epsilon_{4})
        +\beta_3^l (\epsilon_{4}\cdot k_l)(\epsilon_{2}\cdot\epsilon_{3})
\end{align}
with $i=3,4$, $j=2,4$ and $l=2,3$. Using  gauge invariant conditions
to get three sets of equations and solving them, we find
\begin{align}
F=& \alpha^{342} \frac{(k_4\cdot f_{2}\cdot k_3)}{(k_4\cdot k_{2})}
      \frac{(k_2\cdot f_{3}\cdot k_4)}{(k_2\cdot k_{3})} \frac{(k_3\cdot f_{4}\cdot k_2)}{(k_3\cdot k_{4})} \notag\\
  &- \beta_1^3 \frac{(k_4\cdot f_{2}\cdot k_3)}{(k_4\cdot k_{2})}
     \frac{(k_2\cdot f_{3}\cdot f_{4}\cdot k_3)}{(k_2\cdot k_{3})(k_3\cdot k_{4})} \notag\\
  &- \beta_2^4 \frac{(k_2\cdot f_{3}\cdot k_4)}{(k_2\cdot k_{3})}
     \frac{(k_4\cdot f_{2}\cdot f_{4}\cdot k_3)}{(k_4\cdot k_{2})(k_3\cdot k_{4})} \notag\\
  &- \beta_3^2 \frac{(k_3\cdot f_{4}\cdot k_2)}{(k_3\cdot k_{4})}
     \frac{(k_4\cdot f_{2}\cdot f_{3}\cdot k_2)}{(k_4\cdot k_{2})(k_3\cdot k_{2})},
\end{align}

Like the case in previous subsection, result \eref{appendix-3g}
gives us also the basis of gauge invariant building blocks for three
polarization vectors. Since it is straightforward, we will not write
down them explicitly.

Although in this paper we will only use results up to three
polarization vectors, above calculations show that the manifestly
gauge invariant form of $F$ does have some patterns, and we
conjecture that these patterns will also appear even with more than
three polarization vectors. But there must be some new things happen
when the number of momenta is equal to that of polarization vectors.
For example, when there are only four gluons involved, the
manifestly gauge invariant form of $F$ has been given in
\cite{Bern:2017tuc} consisting of the Type-II contractions, while
the Type-I contractions don't appear. We will discuss these in the
future work.

\section{Some calculation details using differential
operators}\label{3gdetail}

In this Appendix,  we present the explicit calculations for unknown
variables $a(\rho',\rho)$'s appearing in $C(\{pqr\},\shuffle)$
\eref{Li-3g-diff}. Since coefficients are related by permutations,
we could fix the ordering $\rho$ to be $\rho=\{p,q,r\}$ and use
different orderings $\rho'$'s to denote different $a$'s:
\begin{itemize}

\item (1) $a(\{p,q,r\})$: We have
\begin{align}
 a(\{p,q,r\})
 =& \mathcal{K}_{1p}[\mathcal{C}^{(pqr)}(\{p,q,r\},\shuffle)-\mathcal{T}_{1pqr2}C_0(\{p,q,r\},\shuffle)] \notag\\
 =& \mathcal{K}_{1p}[\mathcal{C}^{(pqr)}(\{p,q,r\},\shuffle)-\mathcal{T}_{1p2}\frac{(k_1f_pX_p)}{\mathcal{K}_{1p}}\mathcal{T}_{pq2}\frac{(k_1f_qX_q)}{\mathcal{K}_{1q}}\mathcal{T}_{qr2}\frac{(k_1f_rX_r)}{\mathcal{K}_{1r}} \notag\\
  &-\frac{[k_p(Y_p-k_1)](k_qX_q)}{\mathcal{K}_{1pq}}\mathcal{T}_{qr2}\frac{(k_1f_rX_r)}{\mathcal{K}_{1r}}\mathcal{T}_{pq2}\mathcal{T}_{1p2}\frac{(k_1f_qf_pk_1)}{\mathcal{K}_{1q}\mathcal{K}_{1p}}] \notag\\
 =& \mathcal{K}_{1p}\mathcal{C}^{(pqr)}(\{p,q,r\},\shuffle) = -\frac{[k_p(Y_p-k_1)][(k_qX_q)-\mathcal{K}_{1pq}](k_rX_r)}{\mathcal{K}_{1pq}\mathcal{K}_{1pqr}}
\end{align}

\item (2) $a(\{p,r,q\})$: We have
\begin{align}
a(\{p,r,q\})
=& \mathcal{K}_{1p}[\mathcal{C}^{(prq)}(\{p,q,r\},\shuffle)-\mathcal{T}_{1prq2}C_0(\{p,q,r\},\shuffle)] \notag\\
=&  \mathcal{K}_{1p}[\mathcal{C}^{(prq)}(\{p,q,r\},\shuffle)-\mathcal{T}_{1p2}\frac{(k_1f_pX_p)}{\mathcal{K}_{1p}}\mathcal{T}_{pr2}\frac{(k_1f_rX_r)}{\mathcal{K}_{1r}}\mathcal{T}_{rq2}\frac{(k_1f_qX_q)}{\mathcal{K}_{1q}} \notag\\
  &-\frac{[k_p(Y_p-k_1)](k_rX_r)}{\mathcal{K}_{1pr}}\mathcal{T}_{rq2}\frac{(k_1f_qX_q)}{\mathcal{K}_{1q}}\mathcal{T}_{pr2}\mathcal{T}_{1p2}\frac{(k_1f_rf_pk_1)}{\mathcal{K}_{1r}\mathcal{K}_{1p}}] \notag\\
 =& \mathcal{K}_{1p}[\mathcal{C}^{(prq)}(\{p,q,r\},\shuffle)- \frac{[k_p(Y_p-k_1)](k_rX_r)}{\mathcal{K}_{1p}\mathcal{K}_{1pr}} ] \notag\\
 =& \mathcal{K}_{1p}[-\frac{[k_p(Y_p-k_1)](k_rX_r)[k_1(Y_q-k_1)+\mathcal{K}_{1pqr}]}{\mathcal{K}_{1p}\mathcal{K}_{1pr}\mathcal{K}_{1pqr}}+ \frac{[k_p(Y_p-k_1)](k_rX_r)}{\mathcal{K}_{1p}\mathcal{K}_{1pr}} ] \notag\\
 =& -\frac{[k_p(Y_p-k_1)](k_rX_r)[k_1(Y_q-k_1)]}{\mathcal{K}_{1pr}\mathcal{K}_{1pqr}}
\end{align}

\item (3) $a(\{q,p,r\})$: We have
\begin{align}
 a(\{q,p,r\})
 =& \mathcal{K}_{1q}[\mathcal{C}^{(qpr)}(\{p,q,r\},\shuffle)-\mathcal{T}_{1qpr2}C_0(\{p,q,r\},\shuffle)] \notag\\
 =& \mathcal{K}_{1q}[\mathcal{C}^{(qpr)}(\{p,q,r\},\shuffle)-\mathcal{T}_{pr2}\frac{(k_1f_rX_r)}{\mathcal{K}_{1r}} \mathcal{T}_{qp2}\frac{(k_1f_pY_p)}{\mathcal{K}_{1p}} \mathcal{T}_{1q2}\frac{(k_1f_qX_q)}{\mathcal{K}_{1q}} \notag\\
  & -\mathcal{T}_{pr2}\frac{(k_1f_rX_r)}{\mathcal{K}_{1r}} \mathcal{T}_{qp2}\mathcal{T}_{1q2} \frac{(k_1f_pf_qk_1)}{\mathcal{K}_{1p}\mathcal{K}_{1q}} \frac{[k_p(Y_p-k_1)](k_qX_q)}{\mathcal{K}_{1pq}}] \notag\\
 =& \mathcal{K}_{1q}\mathcal{C}^{(qpr)}(\{p,q,r\},\shuffle) = -\frac{(k_qX_q)[k_p(Y_p-k_1)](k_rX_r)}{\mathcal{K}_{1pq}\mathcal{K}_{1pqr}}
\end{align}

\item (4) $a(\{q,r,p\})$: We have
\begin{align}
 a(\{q,r,p\})
 =& \mathcal{K}_{1q}[\mathcal{C}^{(qrp)}(\{p,q,r\},\shuffle)-\mathcal{T}_{1qrp2}C_0(\{p,q,r\},\shuffle)] \notag\\
 =& \mathcal{K}_{1q}[-\frac{[k_q(X_q-k_1)](k_rX_r)[k_p(Y_p-k_1)+\mathcal{K}_{1pqr}]}{\mathcal{K}_{1q}\mathcal{K}_{1qr}\mathcal{K}_{1pqr}}+\frac{[k_q(X_q-k_1)](k_rX_r)}{\mathcal{K}_{1q}\mathcal{K}_{1qr}}] \notag\\
 =& -\frac{[k_q(X_q-k_1)](k_rX_r)[k_p(Y_p-k_1)]}{\mathcal{K}_{1qr}\mathcal{K}_{1pqr}}
\end{align}

\item (5) $a(\{r,q,p\})$: We have
\begin{align}
 a(\{r,q,p\})
 =& \mathcal{K}_{1r}[\mathcal{C}^{(rqp)}(\{p,q,r\},\shuffle)-\mathcal{T}_{1rqp2}C_0(\{p,q,r\},\shuffle)] \notag\\
 =& \mathcal{K}_{1r}[-\frac{(k_rX_r)[k_q(X_q-k_1)+\mathcal{K}_{1qr}][k_p(Y_p-k_1)+\mathcal{K}_{1pqr}]}{\mathcal{K}_{1r}\mathcal{K}_{1qr}\mathcal{K}_{1pqr}} + \frac{(k_rX_r)[k_q(X_q-k_1)+\mathcal{K}_{1qr}]}{\mathcal{K}_{1r}\mathcal{K}_{1qr}}] \notag\\
 =& -\frac{(k_rX_r)[k_q(X_q-k_1)+\mathcal{K}_{1qr}][k_p(Y_p-k_1)]}{\mathcal{K}_{1qr}\mathcal{K}_{1pqr}}
\end{align}

\item (6) $a(\{r,p,q\})$: We have
\begin{align}
 a(\{r,p,q\})
 =& \mathcal{K}_{1r}[\mathcal{C}^{(rpq)}(\{p,q,r\},\shuffle)-\mathcal{T}_{1rpq2} C_0(\{p,q,r\},\shuffle)] \notag\\
 =& \mathcal{K}_{1r}\mathcal{C}^{(rpq)}(\{p,q,r\},\shuffle) = -\frac{(k_rX_r)[k_p(Y_p-k_1)][k_q(X_q+k_r)]}{\mathcal{K}_{1pr}\mathcal{K}_{1pqr}}
\end{align}
In all above calculations, we have used the results
\eref{BCJ-pqr-collect}, which are the expansion of color ordered
YM amplitudes to its BCJ basis.
\end{itemize}
%


\end{document}